\documentclass[10pt]{article}
\usepackage{amsmath}
\usepackage{amssymb}
\usepackage{amsfonts}
\usepackage{array}
\usepackage{caption}
\usepackage{comment}
\usepackage{graphicx}
\usepackage[a4paper, portrait, margin=2cm]{geometry}
\usepackage{hyperref}
\usepackage{indentfirst} 
\usepackage{lineno,hyperref}
\usepackage{mathrsfs}
\usepackage{multirow}
\usepackage{ragged2e}
\usepackage{rotating}
\usepackage{siunitx}
\usepackage{soul}
\usepackage{subcaption}
\usepackage[T1]{fontenc}
\usepackage{tabularx}
\usepackage[normalem]{ulem}
\usepackage{url}
\usepackage{xcolor}
\usepackage{verbatim}

\newcommand{\bsym}[1]{\boldsymbol{#1}}

\newcommand{\ds}[1]{{\color{orange} \bf #1}}
\newcommand{\pe}[1]{{\color{teal} \bf #1}}

\newcommand{\ex}[0]{\hat{\bsym{e}}_x}

\def\arxiv{1}

\begin{document}
	
\centering \Huge Benchmark of no-slip boundary conditions for meshless Lattice Boltzmann Method in Stokes flow.

\vspace{0.75cm}
\centering
\large Dawid Strzelczyk$^{*,1,2}$, Pavel Eichler$^{3}$, Maciej Matyka$^{1,2}$ \normalsize

\vspace{0.25cm}

\noindent\textit{$^1$Faculty of Physics and Astronomy, University of Wrocław, pl. Maxa Borna 9, 50-204 Wrocław, Poland}

\noindent\textit{$^2$Jo\v{z}ef Stefan Institute, Scientific Computing Laboratory, Jamova cesta 39, 1000 Ljubljana, Slovenia}

\noindent\textit{$^3$Department of Software Engineering, Faculty of Nuclear Sciences and Physical Engineering, Czech Technical University in Prague, Trojanova 13, Prague, 12000, Czech Republic}

\vspace{0.25cm}
\RaggedRight
$^*$ \texttt{dawid.strzelczyk@ijs.si}
\linebreak

\justifying

\noindent \textbf{Abstract} 
The introduction of the approximated streaming to the standard Lattice Boltzmann Method (LBM) decouples the space and velocity discretizations. Although many no-slip implementations knwon from LBM can be directly adopted to the models with approximated streaming, their exact behavior in the off-lattice setting remains unexplored. In this work, we investigate the meshfree off-lattice D2Q9 model equipped with several no-slip implementations -- non-equilibrium extrapolation, simple and interpolated bounceback, and moment-based boundaries -- applied to a Stokes flow around a cylinder. We compare the pressure, velocity and velocity gradient fields obtained with various no-slips, on polar and hybrid polar-scattered discretizations. We find that non-equilibrium extrapolation performs the best of all the studied no-slip realizations, giving smooth stresses and zero velocity on the solid walls. Other no-slips suffer from oscillations and/or discontinuities in the hydrodynamic fields. Hybrid discretizations are found to give higher errors than the polar grid, but improve the stability of the solution. Finally, we simulate the deformation of an elastic ring under the stresses obtained with the studied no-slips to highlight the importance of the sensible choice of the no-slip realization in more complex problems.
\vspace{0.25cm}

\vspace{1cm}

\section{Introduction}

Over the years, a plethora of numerical approaches have been introduced to solve the Boltzmann Transport Equation (BTE). For the systems close to equilibrium, one usually discretizes the velocities using Hermite polynomials, accompanied by the minimal (or close to the minimal) set of the discrete velocities that guarantee the correct evolution of the considered hydrodynamic moments of the velocity distribution function (VDF)~\cite{Shan2006}. This leads to the discrete velocities Boltzmann equation (DVBE). The discretization of space and time in DVBE can be achieved with one of the known numerical schemes for the advection equation with sources. One of the most popular choices, which results in the celebrated Lattice Boltzmann Method~\cite{Succi2018,Guo2013,Kruger2017}, is to integrate the DVBE along the characteristics, and exploit the fact that the microscopic velocities are constant in space and time, to solve the advective part of DVBE exactly. This, however, couples the space and velocity discretization such that only the velocity sets based on shapes that tesselate the space can be used (e.g. squares in 2D, or cubes in 3D). What follows immediately is that the space discretization must be constructed accordingly (leading to, e.g., square or cubic grids). This leads to the necessity of a special treatment of complex boundaries with, e.g., more elaborate implementations of boundary conditions~\cite{Sanjeevi2018,Bouzidi2001,Ginzburg2003}, non-trivial discretization refinement~\cite{Lyu2023}, and less control over the order of approximation in space. An especially undesired effect related to non-trivial shapes of the flow domain is the reduction of the order of convergence of the solution when the problem boundaries are not aligned with the discrete lattice vectors or their shape changes in time~\cite{Molins2021}.

To overcome those limitations, a number of approaches was introduced that do not require the exact compliance between the velocity and space discretizations, called off-grid LBM (OLBM). In general, they exploit the same velocity discretization as LBM, and employ for the solution of DVBE one of the methods for the approximation of the value of VDF or its gradients/fluxes in space, e.g. polynomial interpolation~\cite{He1997,He1996,He1997c,He1997a}, finite differences~\cite{
Hejranfar2014a}, finite volumes~\cite{Ubertini2004,Nannelli1992,Misztal2015a}, finite elements~\cite{Zadehgol2014a,Min2011,Lee2003,Misztal2015,Matin2017}, or spectral methods~\cite{Hejranfar2015}. The applications of such numerical schemes, at present, consist of a number of model flow problems, like lid driven cavity~\cite{Hejranfar2015,Lee2003,Min2011,Zadehgol2014a,Lin2019,Lee2003,Hejranfar2015,Hejranfar2014a}, cylinder flow~\cite{Min2011,Lee2003,He1997,He1997a,Ubertini2004,Zadehgol2014a,Lin2019,Lee2003,Hejranfar2014a}, Taylor-Green vortices~\cite{Hejranfar2015,Hejranfar2015}, shear layers~\cite{Hejranfar2015,Misztal2015a,Hejranfar2015}, channel flows of various kinds~\cite{Nannelli1992,He1996,He1997c,Misztal2015a,Lee2003}, and also more complex setups, e.g., flow through porous media~\cite{Misztal2015a,Misztal2015,Zadehgol2014a}, flow around an airfoil~\cite{Lin2019,Hejranfar2014a}, or multiphase flows~\cite{Matin2017}. Giving up the exact streaming comes with more freedom as to the placement of the computational nodes in space, as, e.g., unstructured grids can be used, but also introduces challenges like additional numerical errors due to the approximation or higher computational and memory cost of the method.


Whichever approach for the numerical solution of the DVBE one takes, the used space discretization may fall into one of the two categories -- structured or unstructured. Positions of the points in structured discretizations are placed in space according to some know, regular pattern, e.g. on a rectangular lattice or aligned with a polar coordinate system. Such discretizations allow for a quick calculation of the points' positions and a straightforward construction of interpolants or discrete approximations of linear operators~\cite{Bertin2002}. However, their utility is often limited to simple geometries, and local refinement of the discretization or displacement of the boundaries pose a challenge in such models. On the other hand, the unstructured discretizations give much more freedom as to the placement of the discretization points, since their positions are coupled to one another to a much smaller extent (e.g. within a single finite element). When dealing with complex geometries of the flow boundaries, like those in porous media~\cite{Naqvi2026} or aerial vehicles~\cite{Akkurt2022}, boundaries changing their shape in time, e.g. in circulatory system~\cite{MALVE2019}, reactive flows with dissolution~\cite{Molins2021}, or unbounded flows over obstacles~\cite{PM2023108074}, one usually is bound to use unstructured approach for the space discretization. Nevertheless, the use of unstructured discretizations comes with an additional cost of storing and accessing the connectivity data, which increases the memory and computational demands of the method, and motivates the use of hybrid, structured-unstructured, discretizations within one problem.

The approximation methods capable of operating naturally on both structured and unstructured discretizations are the so called meshless (or meshfree) methods~\cite{Liu2005}. They do not require any information about the connectivity between the discretization nodes, thus allowing for the approximation to be performed on a set of scattered nodes. This feature makes the discretizaion (and also -- the dynamic re-discretization) much faster, flexible, and requires less human input to improve the quality of once generated discretization. An especially attractive member of the meshfree methods family is the so-called Radial Basis Functions-Generated Finite Differences (RBF-FD), generalizing the standard finite difference method to non-regular node arrangements. It has been successfully used to study a range of fluid dynamical problems~\cite{Zamolo2019,Rot2024}, including the solution of DVBE~\cite{Lin2019,Strzelczyk2024a,Strzelczyk2024,Musavi2016,Maidenberg2026}. Even though operating on scattered nodes is handled naturally in RBF-FD, there are certain advantages of using regular nodes arrangement for this method, at least in the part of the flow domain. The scattered part of the discretization is usually confined to the vicinity of irregular boundaries, unfit for structured nodes layouts. In the bulk of the fluid, on the other hand, one can use regular discretization and standard finite differences or Lagrange interpolation, which require smaller stencils than the ones used in RBF-FD, and thus reduce the computational and memory demands of the numerical model. Additionally, even if RBF-FD approximation is used in the whole domain of a hybrid problem, the approximation weights can be computed only once and reused for all stencils consisting of only regular nodes. Several works has already exploited the use of hybrid discretization in a single flow domain~\cite{Javed2013,Ding2004,Rot2024} and there is still ongoing work aimed at assessing the stability and accuracy of RBF-FD on regular, scattered and hybrid discretizations, see, e.g.,~\cite{KOLARPOZUN2026} for hyperbolic problems.

Even though in OLBM one is usually interested in obtaining the hydrodynamic solutions by means of solving the BTE/DVBE, the macroscopic boundary conditions for hydrodynamic moments need to be implemented via the VDF. To do so, a number of strategies has been presented in the context of LBM, and a common choice in OLBM simulations is to adopt those implementations as directly as possible, especially when it comes to imposing the no-slip walls. In general, the boundary conditions in LBM can be classified into two groups, depending on which populations at a boundary node are replaced when the boundary condition is imposed. Approaches like simple or interpolated bounceback~\cite{Bouzidi2001}, non-equilibrium bounceback~\cite{Zou1997}, moments-based boundary conditions~\cite{Krastins2020,Eichler2024}, or multireflection boundaries~\cite{Ginzburg2003} identify the populations flowing into the fluid from the solid (called the {\itshape missing} or {\itshape unknown} populations) and replace only those. On the other hand, implementations like non-equilibrium extrapolation~\cite{Guo2002_nee}, regularized boundary condition~\cite{Latt2007}, or the one using the finite differences to assess velocity gradients on walls~\cite{Skordos1993} replace all of the populations in a boundary node. The use of many LBM-specific realizations of the boundary conditions in OLBM is often pretty straightforward, because interpolation, or some other kind of approximation in space, needed in, e.g., interpolated bounceback, is naturally available in OLBM, while the determination of the missing populations can be done using the local representation of the boundary. Unfortunately, in the off-grid setups, the discrete distributions are usually not streamed exactly between the boundary nodes and because of this, the decoupling of the space and velocity discretizations affects also how the shape of the boundary is represented in the model, see, e.g., the pressure fields on the spherical obstacles in our previous work~\cite{Strzelczyk2024a}. Many of the currently available works on OLBM make an arbitrary choice as to which realization of a given boundary condition to use, e.g., simple~\cite{Lin2019} or non-equilibrium~\cite{Musavi2016} bounceback, imposing equilibrium distributions~\cite{Musavi2016}, non-equilibrium extrapolation~\cite{Maidenberg2026,Hu2024}, using the strain-rate tensor to approximate the off-equilibrium part~\cite{Pribec2021}, but little is known about the performance of various no-slip implementations studied in a rigorous setting allowing for a fair comparison between them.

The motivation for this work is to study the accuracy and stability of the meshless RBF-FD Lattice Boltzmann solver equipped with five LBM-specific realizations of the no-slip boundaries, adapted to the off-grid setup. As the benchmark problem, we consider a low-Reynolds number flow around a two-dimensional cylinder. We perform simulations for several viscosity values, to compare the stability limits of each no-slip type. We investigate the profiles of the pressure, velocity, and its gradient on the solid walls, as well as the overall drag force. Finally, we show how the differences between the stresses obtained with each no-slip can affect the shape of the cylinder once it is allowed to deform elastically. This work fills the gap in knowledge about the fundamental properties of meshless LBM schemes and provides guidelines for a more sensible choice of no-slip realizations in meshless LBM simulations.

\section{Methods}

\subsection{Definition of the considered Stokes problem}

\begin{figure}
    \centering
    \includegraphics[width=0.6\linewidth]{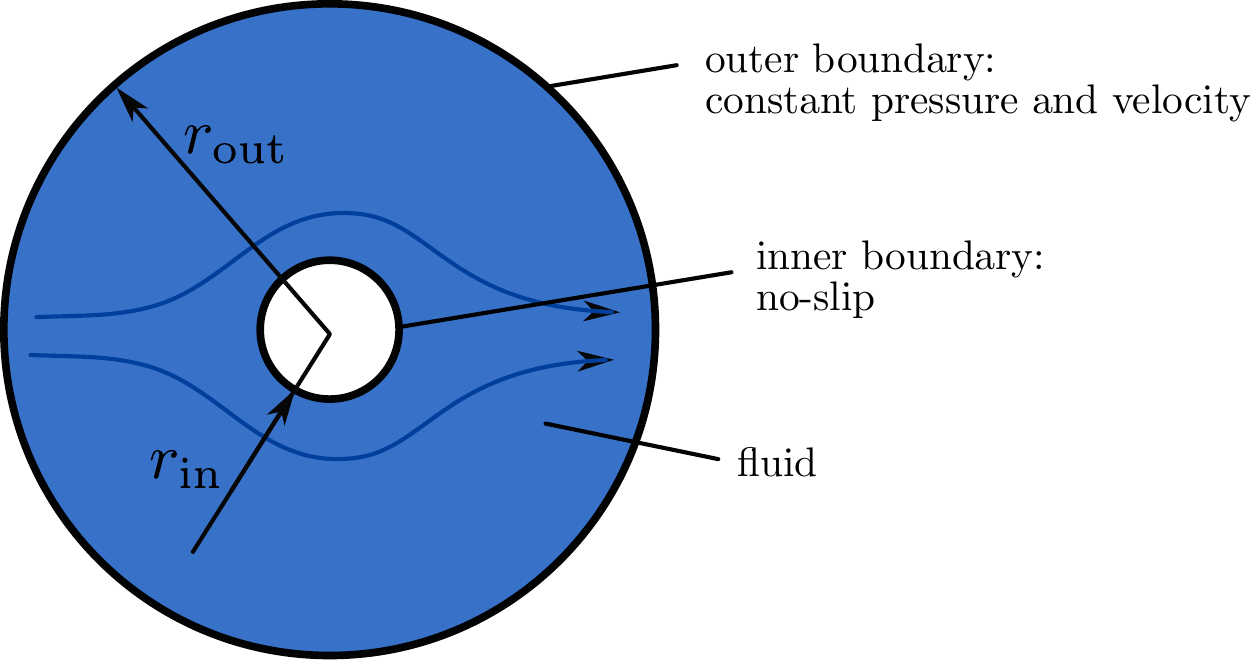}
    \caption{Schematic representation of the considered Stokes problem.}
    \label{fig:stokes_problem}
\end{figure}

We consider a two dimensional flow around a cylinder of radius $r_\text{in}$ placed in a circular domain of radius $r_\text{out}$, concentric with it (see Fig.~\ref{fig:stokes_problem}. For vanishingly small Reynolds number $Re = 2V_\infty r_\text{in}/\nu$, as long as the outer boundary has a finite radius, it is possible to analytically derive a closed-form solution to the steady-state Stokes equation subject to Dirichlet boundary conditions a the outer boundary and no-slip condition on the inner boundary, i.e.,
\begin{subequations}\label{eq:stokes_eq}
	\renewcommand{\arraystretch}{2}
	\begin{align}
		-\nabla P + \mu\nabla^2\bsym{V} &= \bsym{0}, \\
		P=P_\infty, \> \bsym{V}=V_\infty\ex & \quad\text{for }  |\bsym{x}| = r_\text{out} > 0, \\
		\bsym{V}=\bsym{0} & \quad\text{for }  |\bsym{x}| = r_\text{in}, \quad 0<r_\text{in}<r_\text{out}.
	\end{align}
\end{subequations}
where $\mu=1$ is fluid's dynamic viscosity, $\bsym{n}$ is the local normal vector on the boundary pointing inside the fluid, $\ex$ is the unit vector aligned with the $x$-coordinate of the Cartesian coordinate system, and $\bsym{x}$ is the position vector and $\bsym{V}$ is the velocity vector. In our computations we keep $Re = 0.01$.

We delegate the full analytical solution procedure to Appendix~\ref{sec:stokes_analytical_solution}, here we limit ourselves to presenting its final form in polar coordinates
\begin{subequations}\label{eq:stokes_analytical}
	\renewcommand{\arraystretch}{1.5}
	\begin{align}
        V_r &= \cos{\theta} \left( Ar^2 + B\ln{r} + C + Dr^{-2} \right), \\
		V_\theta &= -\sin{\theta} \left( 3Ar^2 + B(1+\ln{r}) + C - Dr^{-2} \right), \\
		P(r,\theta) &= P_\infty + \nu \cos{\theta} \displaystyle\int\limits_R^r d\xi \mathcal{R}(\xi), \quad \mathcal{R}(\xi) = 2B\xi^{-2} + 8A\label{eq:pressure_final}
	\end{align}
\end{subequations}
where $r =: |\bsym{x}|$ and $\theta$ are the radial and angular coordinates and $A,B,C,D \in \mathbb{R}$ are constants dependent on $r_\text{out}$, $r_\text{in}$, $P_\infty$, and $V_\infty$.

\subsection{Meshless approximation}\label{sec:meshless}

In this work, we use the radial basis functions generated finite differences method (RBFFD) to interpolate a function given at $N$ points $\bsym{x}_i \in \mathbb{R}^d$ with $d=2$. This technique is capable of achieving high-order approximation on scattered discretizations, i.e., it does not require the generation of volumes (or elements), but operates on non-connected sets of points. Despite of the freedom in positioning the discretization points, one can arrange them in a regular manner, e.g., on a square or polar grid. It has been previously used with success to address fluid flow problems both with LBM-like~\cite{Strzelczyk2025,Strzelczyk2024,Lin2019,Maidenberg2025} and direct Navier-Stokes~\cite{Rot2024,Javed2013,Shankar2015} solvers. In contrast to, e.g., finite elements method, where the approximation is carried out within a single element consisting of a number of nodes, in RBFFD one defines the so-called {\itshape stencils} $S$, which are subsets of the set off all discretization points. A common choice is to take $N_L$ points closest (in the sense of the Euclidean norm) to the point where the approximation is to take place (called the {\itshape target point}). At the target point $\bsym{x}$, one looks for the approximation to $\mathcal{L}f$ as a linear combination of the function's values at the $N_L$ stencil points
\begin{equation}\label{eq:rbffd_general_fd}
    \mathcal{L}f(\bsym{x}) \approx \sum\limits_{i=1}^{N_L} w_i^\mathcal{L}(\bsym{x}) f(\bsym{x}_{S(i)})
\end{equation}
where $\mathcal{L}$ is some linear operator, $S(i)$ is the index of the $i$-th member of the stencil $S$, and $w_i^\mathcal{L}(\bsym{x})$ is the weight associated with the value of $\mathcal{L}f$ at point $\bsym{x}_{S(i)}$ during the evaluation at point $\bsym{x}$. The operator $\mathcal{L}$ can in particular be a derivative operator, e.g., $\partial/\partial x$ or an identity operator $\mathcal{L}f = f$ (when one speaks of interpolation). We will further omit the superscript $\mathcal{L}$ and will keep in mind that a certain operator is being considered. To obtain the weights $w$, one can approximate the function $f$ as a linear combination of $N_L$ radial basis functions centered at the stencil points
\begin{equation}\label{eq:rbffd_basis_representation}
    f(\bsym{x}) = \sum\limits_{j=1}^{N_L} \gamma_j \phi_{S(j)}(\bsym{x})
\end{equation}
where
\begin{equation}
    \phi_j(\bsym{x}) =: \phi\left(\dfrac{||\bsym{x}-\bsym{x}_j||_2}{\delta}\right)
\end{equation}
is the radial (basis) function (RBF) centered at node $\bsym{x}_j$ and $\delta$ is the local scaling factor (to be specified later) . Substituting Eq.~\eqref{eq:rbffd_basis_representation} into Eq.~\eqref{eq:rbffd_general_fd} one obtains the following system of linear equations
\begin{equation}\label{eq:rbffd_linear_system}
    \Phi \cdot \bsym{w}(\bsym{x}) = \bsym{\phi}(\bsym{x})
\end{equation}
where
\begin{equation}\label{eq:rbffd_interpolation_matrix}
    \{\Phi\}_{ij} =: \phi_{S(i)}(\bsym{x}_{S(j)}) = \{\Phi\}_{ji}
\end{equation}
is the interpolation matrix dependent only on the relative positions of the stencil members,
\begin{equation}\label{eq:rbffd_values_vector}
    \bsym{\phi}(\bsym{x}) =: [\phi_{S(0)}(\bsym{x}), \> \dots, \> \phi_{S(N_L)}(\bsym{x})]^T
\end{equation}
is the vector of values of the basis functions at the target node, and
\begin{equation}\label{eq:rbffd_weights_vector}
    \bsym{w}(\bsym{x}) =: [w_0(\bsym{x}), \> \dots, \> w_{N_L}(\bsym{x})]^T
\end{equation}
is the vector of the sought-for weights. In this work we use cubic RBFs
\begin{equation}
    \phi_j(\bsym{x}) = \left(\dfrac{||\bsym{x}-\bsym{x}_j||_2}{\delta}\right)^{3/2}
\end{equation}
as they do not require tuning of any parameters (in contrast to, e.g., Gaussians~\cite{Lehto2017}). As the interpolation matrices constructed with such RBFs need not be positive-definite, one has to be augment them with monomials of order $m$ for the linear system to solvable uniquely and in a stable manner~\cite{Flyer2016}. Thus, we expand the system in Eq.~\eqref{eq:rbffd_linear_system} with $N_p$ monomials
\begin{equation}\label{eq:rbffd_linear_system_expanded}
    \begin{bmatrix}
        \Phi & P\\
        P^T & 0
    \end{bmatrix}
    \cdot
    \left\{
    \begin{array}{c}
         \bsym{w}(\bsym{x})  \\
         \bsym{\lambda} 
    \end{array}
    \right\}
    =
    \left\{
    \begin{array}{c}
         \bsym{\phi}(\bsym{x})  \\
         \bsym{p}(\bsym{x}) 
    \end{array}
    \right\}
\end{equation}
where
\begin{equation}
    P =
    \begin{bmatrix}
        p_1(\bsym{x},S(1)) & \dots & p_{N_p}(\bsym{x},S(1)) \\
        \vdots & \ddots & \vdots \\
        p_1(\bsym{x},S(N_L)) & \dots & p_{N_p}(\bsym{x},S(N_L)) \\
    \end{bmatrix}
    , \quad
    \bsym{p}(\bsym{x})
    =
    \left\{
    \begin{array}{c}
          p_1(\bsym{x}) \\
          \vdots \\
          p_{N_p}(\bsym{x}) \\
    \end{array}
    \right\}.
\end{equation}
In particular, we use the rescaled form of the monomials centered at point $\bsym{x}_c$
\begin{equation}
    p_l(\bsym{x},\bsym{x}_c) =: p_l\left(\dfrac{\bsym{x}-\bsym{x}_c}{\delta}\right),
\end{equation}
where the central point and the scale $\delta$ will be defined later. We use all the monomials of order $m=2$ and lower, i.e., the monomial basis is $\{1,x,y,x^2,xy,y^2\}$ and the number of monomials in it is
\begin{equation}
    N_p =
    \left(
    \begin{array}{c}
         m+d  \\
         m 
    \end{array}
    \right).
\end{equation}
The system from Eq.~\eqref{eq:rbffd_linear_system_expanded} is solved for weights $\bsym{w}$, and the monomial weights $\bsym{\lambda}$ are treated as Lagrange multipliers and are not used for the approximation further. The extension of the above procedure to the approximation of any linear operator acting on the unknown function $f$ does not pose greater difficulties~\cite{Strzelczyk2025}. The stencil size needs to be no smaller than the number of the monomials, $N_L \ge N_P$, with the recommended value $N_L \ge 2N_P$. Thus, while the monomial augmentation stabilizes the interpolation and provides higher order of convergence~\cite{Flyer2016}, it increases the computational cost of the method. Table~\ref{tab:rbffd_interpolation_params} summarizes the interpolation setup used in this work. We note that the used monomials are defined using the Cartesian coordinates.

\begin{table}[!ht]
    \centering
    \begin{tabular}{lcc}
        \multirow{2}{*}{\textbf{Parameter}} & \textbf{Symbol} & \multirow{2}{*}{\textbf{Value}}\\
        & \textbf{(if applies)} &\\
        \hline
        RBF type & -- & cubic \\
        Stencil size & $N_L$ & 21 \\
        Monomials order & $m$ & 2 \\
        Monomials set & -- & $\{1,x,y,x^2,xy,y^2\}$ \\
        Number of monomials & $N_P$ & 6
    \end{tabular}
    \caption{Setup of the RBFFD interpolation used in this work.}
    \label{tab:rbffd_interpolation_params}
\end{table}

\subsection{Structured and hybrid discretization}\label{sec:discretization}

\begin{figure}[ht!]
	\centering
    \includegraphics[height=.5\linewidth]{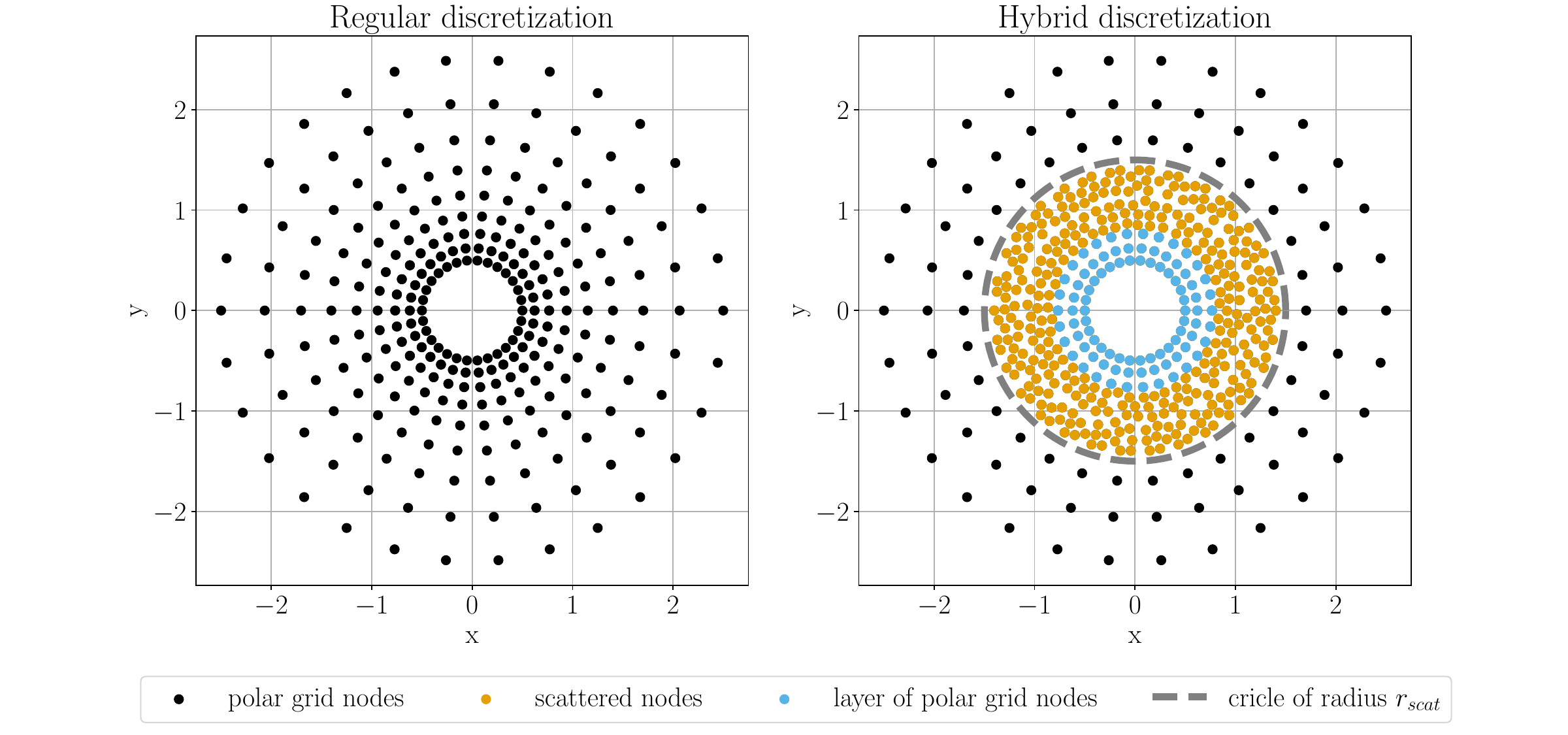}
	\caption{Examples of regular and hybrid discretizations of the considered problem domain. The radius of the outer boundary and the boundary between the scattered and regular region are chosen to ensure the clarity of the Figure and are different from the ones actually used in the simulations.}
	\label{fig:CYLINDER_points}
\end{figure}

In this study, for the discretization of the cylinder case, we use polar (called {\itshape regular} further in the text) or hybrid discretization (Fig.~\ref{fig:CYLINDER_points} for and exemplary visualization). The regular discretization is created to resemble the one from the work of Lin and others~\cite{Lin2019} as closely as possible. Similarly to them, we set the radius of the cylinder equal to $r_\text{in}=0.5$ and the radius of the outer boundary equal to $r_\text{out}=55$. Concerning the angular coordinates $\theta_i$, $i=1,\dots,N_\theta$ of the points positions, we use a constant spacing $\Delta \theta = 2\pi/N_{\theta}$, where $N_\theta$ is the number of points discretizing the cylinder's surface. On the other hand, from Fig.~5 in~\cite{Lin2019}, one sees that the the grid parameter along the radial coordinate scales approximately proportionally to the distance from the cylinder, thus, in our discretization we use the following recursive relation to generate the series of the radial coordinates $r_i$ of the points
\begin{equation}
	r_{i+1} = r_i (1+\Delta r), \quad r_\text{out} = r_\text{in}
\end{equation}
where the constant $\Delta r = 0.03$ for $N_{\theta}=180$ and is scaled proportionally to $\Delta \theta$ to keep the ratio of the angular and radial spacing constant for each point in the domain. The $r_i$ series generation is terminated such that none of the points $r_i,\> i=1,\dots,N_r$ falls beyond the outer boundary radius, $r_\text{out}$, i.e. $r_{N_r} < r_\text{out}$. Then, the whole series is rescaled according to
\begin{equation}
	r_i \rightarrow (r_i-r_\text{out}) \dfrac{r_\text{out}-r}{r_{N_r} - r} + r_\text{out}
\end{equation}
in order to ensure that the outermost points are placed exactly on the radius $r_\text{out}$.

The hybrid discretization with $N_\text{layers}$ layers of structured adjacent to the cylinder is generated by starting with an polar grid defined above, and first removing points $(r_i,\theta_i)$ for which $r_i < r_\text{scat} = 5$ and $i > N_\text{layers}=3$. Then, the empty space is populated with a scattered nodes using the Poisson disk sampling algorithm~\cite{Slak2019}. The target internodal spacing scales linearly between $r \Delta \theta$ at the cylinder's surface to $r_\text{scat} \Delta \theta$ at $r_\text{scat}$.

\subsection{Meshless IS-LBM}

\begin{figure}[!ht]
    \centering
    \includegraphics[width=0.75\linewidth]{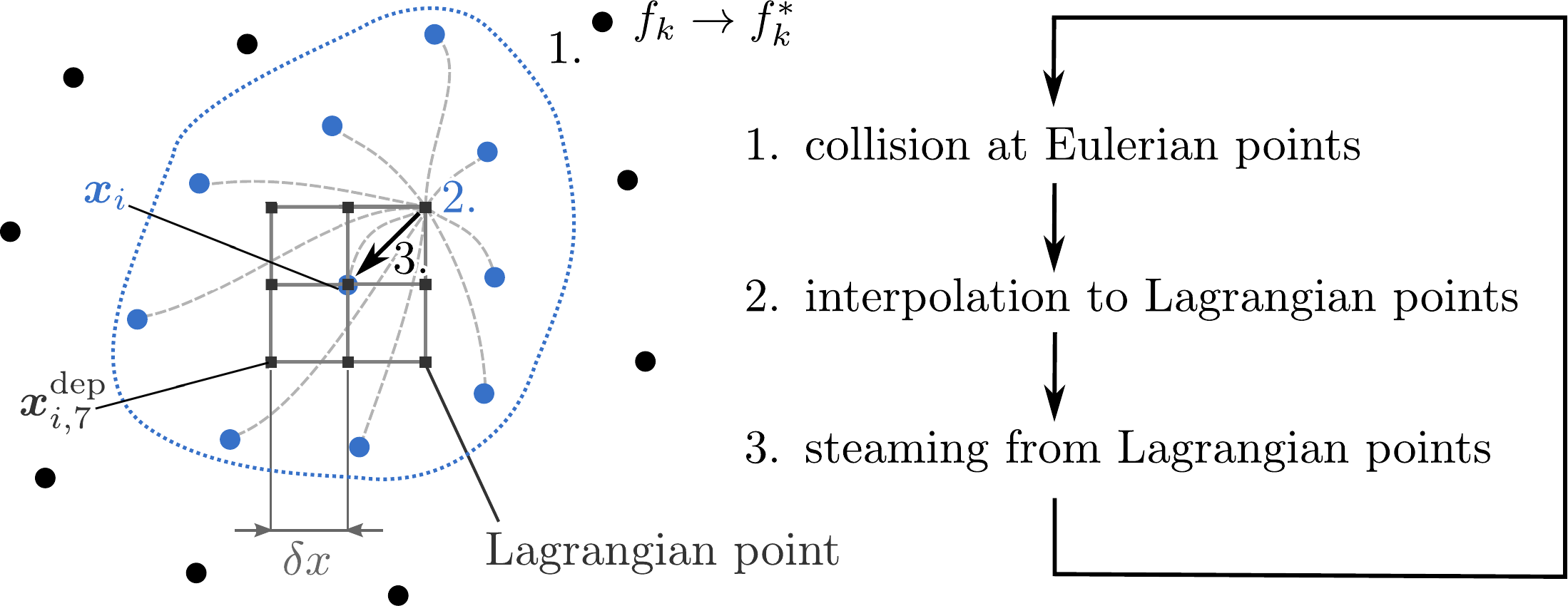}
    \caption{Graphical interpretation of the semi-Lagrangian meshfree LBM model considered in this work. Blue points constitute the interpolation stencil of the central node, $\bsym{x}_i$. The gray square centered at $\bsym{x}_i$ represents the departure (Lagrnagian) points from which the post-collision VDFs $k_k^*$ are streamed to the centlan node. For instance, point $\bsym{x}^\text{dep}_{i,7}$ is the departure node for $f^*_7$.}
    \label{fig:olbm_basics}
\end{figure}

In the present work we solve the most popular version of DVBE with BGK collision~\cite{Bhatnagar1954}, discretized along characteristics using trapezoid rule, with shifted velocity distribution functions (VDF) and relaxation time to recover the explicitness of the scheme
\begin{equation}
	f_k(t+1,\bsym{x}_i) = f_k(t,\bsym{x}_{i,k}^\text{dep}) - \dfrac{1}{\tau}
	\left(
		f_k(t,\bsym{x}_{i,k}^\text{dep}) - f_k^\text{eq}(t,\bsym{x}_{i,k}^\text{dep})
	\right)
	=
	f_k^*(t,\bsym{x}_{i,k}^\text{dep})
\end{equation}
where $f_k(t,\bsym{x}_i)$ denotes the $k$-th VDF at discretization point $\bsym{x}_i$ at timestep $t$. We use the D2Q9 model for velocity discretization, where the discrete lattice vectors are
\begin{equation}
	\begin{tabular}{ll}
		$\bsym{e}_0 = [0,0]$ \hspace{1cm}& $\bsym{e}_5 = [1,1]$ \\
		$\bsym{e}_1 = [1,0]$ & $\bsym{e}_6 = [-1,1]$ \\
		$\bsym{e}_2 = [0,1]$ & $\bsym{e}_7 = [-1,-1]$ \\
		$\bsym{e}_3 = [-1,0]$ & $\bsym{e}_8 = [1,-1]$ \\
		$\bsym{e}_4 = [0,-1]$ &  \\
	\end{tabular}
\end{equation}
and second-order discretization of the compressible equilibrium VDF $f_k^\text{eq}$
\begin{equation}
	f_k^\text{eq} = \omega_k \rho
	\left(
		1 + \dfrac{\bsym{V} \cdot \bsym{e}_k}{c_s^2} +
		\dfrac{(\bsym{V} \cdot \bsym{e}_k)^2}{2c_s^4} -
		\dfrac{\bsym{V}^2}{2c_s^2}
	\right)
	=
	\omega_k \rho
	\left(
	1 + S_k[\bsym{V}]
	\right)
\end{equation}
where $c_s^2 = 1/3$ is the lattice speed of sound and the mass density and velocity are the quadratures of VDF in velocity space
\begin{equation}\label{eq:moments}
	\begin{array}{rcl}
		\rho &=& \displaystyle\sum\limits_{q=0}^{8} f_k,\\
		
		\vspace{.1cm}\\

		\bsym{V} &=& \dfrac{1}{\rho}\displaystyle\sum\limits_{q=0}^{8} \bsym{e}_kf_k.
	\end{array}
\end{equation}
$\tau$ is the non-dimensional relaxation time. To approximate the values of VDF at the departure points $\bsym{x}_{i,k}^\text{dep} =: \bsym{x}_i - \bsym{e}_k\delta x$, which are generally missing from the discretization, we use the RBFFD interpolation described in Section~\ref{sec:meshless}. $\delta x$ is the streaming distance and we use the value of $\delta x=0.06$ for $N_\theta=45$ and scale it proportionally with the internodal distance when the space discretization is refined. We define the interpolation stencils as $N_L$ points lying closest to each discretization node $\bsym{x}_i$, called the stencil's center. We use one interpolant for the approximation at all departure nodes of $\bsym{x}_i$ apart from $\bsym{x}_{i,0}^\text{dep} \equiv \bsym{x}_i$, for which we do not interpolate $f_0$, rather leaving it intact during the streaming (see Fig.~\ref{fig:olbm_basics} for the graphical interpretation of the considered semi-Lagrangian meshfree LBM). We can now define the scaling factor $\delta$ of the used RBFs and monomials, to be the distance between the point $\bsym{x}_i$ and its closest neighbor within its stencil. The scaling factor is the same for RBFs and monomial within each stencil. The vector $\bsym{x}_c$ by which the monomials in each stencil are translated, is taken to be the stencil's center $\bsym{x}_i$.

To improve accuracy, we use the well-conditioned representation of the discrete VDFs, $\tilde{f_k} =: f_k - \omega_k$~\cite{Skordos1993}. We implicitly assume unit mean density. In the preliminary simulations, we found this step necessary to achieve the convergence in time at the desired level when the approximate streaming was used. In turn, the well-conditioned mass density is defined as
\begin{equation}\label{eq:moments_well}
	\begin{array}{rcl}
		\tilde{\rho} &=& \displaystyle\sum\limits_{q=0}^{8} \tilde{f}_k = \rho-1,\\
		
		\vspace{.1cm}\\
		
		\bsym{V} &=& \dfrac{1}{\tilde{\rho}+1}\displaystyle\sum\limits_{q=0}^{8} \bsym{e}_k \tilde{f}_k.
	\end{array}
\end{equation}
and the well conditioned equilibrium distribution becomes
\begin{equation}\label{eq:feq_well_conditioned}
	\tilde{f_k}^\text{eq} =: f_k^\text{eq} - \omega_k = \omega_k\tilde{\rho} + \omega_k (\tilde{\rho}+1) S_k[\bsym{V}].
\end{equation}
We note that the summation for the momentum velocity does not change due to the symmetry of the lattice weights, i.e., $\omega_k = \omega_{k'}$. Apart from the above changes, switching to well-conditioned variables renders the considered DBE the same.

\subsection{Implementation of boundary conditions in LBM and IS-LBM}\label{sec:lbm_BCs}

All the no-slip boundary implementations used in this work are adopted and typical for the standard LBM. Due to the fact that in meshfree LBM one does not have to stick to a square grid, two major differences with respect the lattice-based model arise, namely
\begin{enumerate}
	\item no links are cut by the boundary since the boundary nodes are placed exactly on it. Thus, one needs to define a way to identify the missing populations (i.e., those that are streamed to the boundary nodes from the solid). We will refer to the set of the unknown populations at a given boundary node as $UK$ and to the set of the known ones as $K$,
	\item the nearest bulk neighbor of a boundary node can be placed along the boundary local normal vector, which allows for implementing the Neumann-type conditions practically without the use of approximation.
\end{enumerate}
Whenever necessary in this Section, each of the differences will be properly addressed when introducing particular no-slip implementation.

\subsubsection{Imposing equilibrium populations}

A common choice to realize the far-field Dirichlet boundaries for the pressure $P_\infty(\bsym{x}_B)$ and velocity $\bsym{V}_\infty(\bsym{x}_B)$ is to use equilibrium values parametrized by those macroscopic moments for all populations in the far-field boundary nodes~\cite{He1997,He1997a,Lin2019,Wu2009}
\begin{equation}
    f_k(t+1;\bsym{x}_B) = f^\text{eq}_k(\rho_\infty(\bsym{x}_B),\bsym{V}_\infty(\bsym{x}_B)).
\end{equation}
This is a justified choice since in the far field, the gradient of the density and velocity should tend to zero (uniform, undisturbed flow), and also should the non-equilibrium part of VDF. In the well-conditioned representation, this boundary implementation takes the form as in Eq.~\eqref{eq:feq_well_conditioned}.

To make our results comparable to those of other authors, we keep the ratio between the radius of the outer and the inner boundary similar to other works, i.e., $r_\text{out}/r_\text{in}=110$. We note that while it is a common choice to set uniform density/pressure and velocity at the far-field boundary and Stokes problem does have a solution for a finite $r_\text{out}$ then, it introduces a spurious source of momentum to the system and its applicability to model real systems without such source is limited to large enough $r_\text{out}/r_\text{in}$ ratios.

\subsubsection{Non-equilibrium extrapolation (NEE)}

Originally introduced by Guo and Shi~\cite{Guo2002_nee} for boundaries aligned with discrete lattice vectors, and extended in~\cite{Guo2002} for curved boundaries. This no-slip implementation replaces all the populations of a boundary node $\bsym{x}_B$. The new values are decomposed into the equilibrium part parametrized with the desired macroscopic velocity and density copied from the nearest bulk neighbor $\bsym{x}_N$, and the non-equilibrium part, also copied from $\bsym{x}_N$
\begin{equation}\label{eq:nee_rule}
	f_k(t+1;\bsym{x}_B) = f^\text{eq}_k(\rho(t+1;\bsym{x}_N),\bsym{0}) + f^\text{neq}_k(t+1;\bsym{x}_N),
\end{equation}
where $f_k^\text{neq} =: f_k - f_k^\text{eq}$. Note that the values copied from the neighbor are both taken after the streaming is done. We derive the analytical rate of convergence of this no-slip in the off-grid setting in Section 1 of the Supplementary Materials. When using the well-conditioned variables, Eq.~\eqref{eq:nee_rule} changes to
\begin{equation}\label{eq:nee_rule_well}
	\tilde{f}_k(t+1;\bsym{x}_B) = \tilde{\rho}(t+1;\bsym{x}_N)\omega_k +
	\tilde{f}_k(t+1;\bsym{x}_N) -
	\tilde{f}^\text{eq}_k(\tilde{\rho}(t+1;\bsym{x}_N),\bsym{0}).
\end{equation}

\subsubsection{Simple bounceback (SBB)}

\begin{figure}[!ht]
    \centering
    \includegraphics[width=0.75\linewidth]{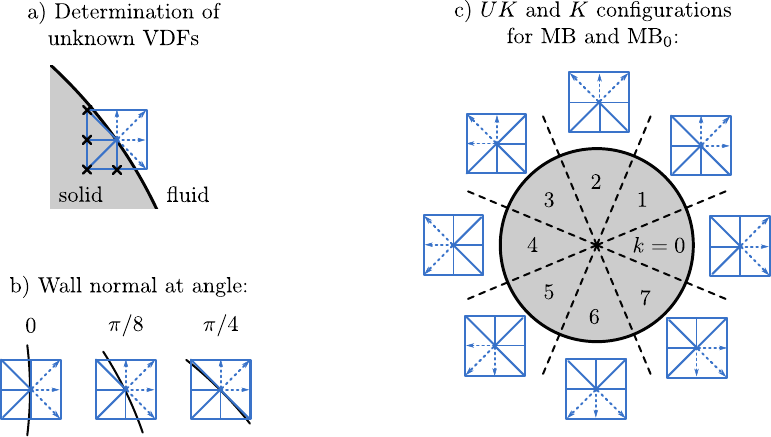}
    \caption{Graphical interpretation of the choice of the sets of the populations known ($K$) and unknown ($UK$) during the imposition of the no-slip boundary: a) the populations streamed from the departure nodes lying inside the cylinder are considered unknown (denoted with dotted-line arrows), b) exemplary $U$ and $UK$ configurations at characteristic angles, c) the sets of $U$ and $UK$ chosen for MB and MB$_0$ no-slips, along with the angle sectors corresponding to each configuration.}
    \label{fig:UK_and_K}
\end{figure}

Originating from lattice gas automata~\cite{CORNUBERT1991241}, this no-slip replaces the unknown post-streaming populations with the opposite-sign post-collision populations from the same timestep at the same node
\begin{equation}\label{eq:sbb_rule}
	\text{for } k\in UK: \> f_k(t+1;\bsym{x}_B) = f^*_{k'}(t;\bsym{x}_B)
\end{equation}
where $k'$ denotes the population going in the direction opposite to $f_k$, i.e., $\bsym{e}_k = -\bsym{e}_{k'}$. We define the set of the unknown populations $UK$ as those whose departure points lie inside the cylinder (see Fig.~\ref{fig:UK_and_K}, subplot a) for the graphical interpretation)
\begin{equation}\label{eq:unknown_vdf_set_rule}
	UK = \left\{ k \> : \> |\bsym{x}_B+\bsym{e}_{k'}\delta x| < r_\text{in} \right\}.
\end{equation}
Due to the symmetry of the lattice weights, with well-conditioned VDFs the bounceback rule stays unchanged
\begin{equation}\label{eq:sbb_rule}
	\text{for } k\in UK: \> \tilde{f}_k(t+1;\bsym{x}_B) = \tilde{f}^*_{k'}(t;\bsym{x}_B).
\end{equation}

\subsubsection{Interpolated bounceback (IBB)}

Suggested by Bouzidi and others~\cite{Bouzidi2001}, this realization of no-slip allows for fine-tuning of the boundary position by assuming that the boundary link is cut in some proportion $q\in [0,1]$. In LBM, it replaces the unknown populations with the opposite-direction post-collision populations starting from the departure points in the bulk, whose position is determined by the cut ratio $q$. The value of VDF at the departure node is reconstructed using linear interpolation between the boundary node $\bsym{x}_B$ and its nearest bulk neighbor $\bsym{x}_N$ along the streaming direction
\begin{equation}\label{eq:ibb_rule_general}
	f_k(t+1,\bsym{x}_B) =
	\begin{cases}
		2qf_{k'}^*(t,\bsym{x}_B) + (1-2q)f_{k'}^*(t,\bsym{x}_N), &\text{for } q<1/2\\
		\dfrac{1}{2q}f_{k'}^*(t,\bsym{x}_B) - \dfrac{(1-2q)}{2q}f_{k'}^*(t,\bsym{x}_N), &\text{for } q \ge 1/2\\
	\end{cases}.
\end{equation}
In off-grid setting, boundary nodes lying exactly on the boundary give $q=0$ always, and thus the recipe simplifies to
\begin{equation}\label{eq:ibb_rule_general}
	\text{for } k\in UK: \> f_k(t+1,\bsym{x}_B) =
	f_{k'}^*(t,\bsym{x}_B+\bsym{e}_k\delta x).
\end{equation}
We use Eq.~\eqref{eq:unknown_vdf_set_rule} to determine the set of the unknown populations. Here, again, due to the symmetry of the lattice weights, with well-conditioned VDFs the bounceback rule stays the same
\begin{equation}\label{eq:sbb_rule}
	\text{for } k\in UK: \> \tilde{f}_k(t+1;\bsym{x}_B) = \tilde{f}^*_{k'}(t;\bsym{x}_B+\bsym{e}_k\delta x).
\end{equation}

\subsubsection{Moments-based no-slips (MB and MB$_0$}\label{sssec:moments_bc}

The moment-based boundary conditions were originally proposed by Krastins and others~\cite{Krastins2020}, later extended to D3Q27 model by Eichler and others~\cite{Eichler2024}. Unlike previous work, the meshless approache allows for an unknown number of VDFs at the boundary. This fact presents certain challenges when implementing and deriving the formulation of moment-based boundary conditions.

Let $UK = \{i_1, i_2, \cdots, i_m\}$ and $K = \{j_1, j_2, \cdots, j_n\}$, $m + n = 9$, are the sets containing indexes of unknown and known populations in the boundary node $\bsym{x}_B$, respectively. We denote $\bsym{f}_{UK} = \left[f_{i_1}, f_{i_2}, \cdots, f_{i_m} \right]^T$ and $\bsym{f}_{K} = \left[ f_{j_1}, f_{j_2}, \cdots, f_{j_n}\right]^T$ vectors of uknonw and known populations, respectively. Then, the unknown populations can be solved as
\begin{equation}\label{eq:moment_based_rule}
    \bsym{f}_{UK} = \mathbb{M} \cdot \bsym{f}_{K} + \mathbb{A}_m\bsym{m}_{UK}^\text{eq},
\end{equation}
where the particular form of the matrix $\mathbb{M}$ and the vector $\bsym{m}$ consisting of raw moments is constructed on the basis of a system of equations for the calculation of raw moments using expressions of unknown populations. When expressing unknown distributions, we use the same strategy as in \cite{Eichler2024} - that is, we try to use equations with the lowest possible order of raw moments. Furthermore, all moments are approximated by their equilibrium.

\if\arxiv0
    \pe{TODO: Would it be a good idea to add comments explaining how this is implemented—that is, are all the system matrices stored?}
    
    \pe{TODO: Maybe we can mention the generator in python? But if yes, it should be published somewhere.} -- \ds{I refrenced it at the end of this subsection, ok? Will need to remember to include it in the repo.}
\fi

When the well-conditioned populations are used, after some manipulations, Eq.\eqref{eq:moment_based_rule} becomes
\begin{equation}\label{eq:moment_based_rule_well_conditioned}
    \tilde{\bsym{f}}_{UK} = - \bsym{\omega}_{UK} + \mathbb{M}\bsym{\omega}_{K} + \mathbb{A}_m \bsym{m}^\text{eq}_{UK} + \tilde{\rho}(t; \bsym{x}_N)\mathbb{A}_m \bsym{m}^\text{eq}_{UK} + \mathbb{M} \cdot \tilde{\bsym{f}}_{K} ,
\end{equation}
with all 9 elements of the equilibrium moments vector $\bsym{m}^\text{eq}$ defined as
\renewcommand{\arraystretch}{1.5}
\begin{equation}
    \begin{array}{rcl}
        m_{00}^\text{eq} &=& 1\\
        m_{10}^\text{eq} &=& V_0\\
        m_{01}^\text{eq} &=& V_1\\
        m_{11}^\text{eq} &=& V_0V_1\\
        m_{20}^\text{eq} &=& c_s^2+V_0\\
        m_{02}^\text{eq} &=& c_s^2+V_1\\
        m_{21}^\text{eq} &=& V_1(c_s^2+V_0)\\
        m_{12}^\text{eq} &=& V_0(c_s^2+V_1)\\
        m_{22}^\text{eq} &=& c_s^4 + c_s^2(V_0^2+V_1^2)+V_0^2+V_1^2.\\
    \end{array}
\end{equation}\renewcommand{\arraystretch}{1}
with the value of velocity components taken from the previous timestep. We note that in order to avoid the loss of significant digits (we use \verb|double| precision in all OLBM simulations) during the evaluation of Eq.~\eqref{eq:moment_based_rule_well_conditioned}, we initialize each element of vector $\tilde{\bsym{f}}_{UK}$ with the corresponding $-\omega_k$ value and perform summation of the other terms in the order specified in Eq.~\eqref{eq:moment_based_rule_well_conditioned}. This is because the first three terms in the sum on the right-hand side of Eq.~\eqref{eq:moment_based_rule_well_conditioned} have in general much larger elements than the other two (due to the multiplication of the latter by the deviations $\tilde{\rho}$ and $\tilde{f}_k$).

For the choice of the set of unknown distributions, $UK$, we divide the full $2\pi$ angle in to eight equal segments centered at angles $k\pi/4$, $k=0,\dots,7$. Then, each boundary node is assigned to one segment depending on its position along the perimeter of the cylinder. The bisector angle for each segment defines the local wall normal assumed for all the boundary nodes of that segment. Eq.~\eqref{eq:unknown_vdf_set_rule} is then used to determine the sets $UK$ and $K$ based on the assumed wall orientation within each segment. See Fig.~\ref{fig:UK_and_K}, subplot c) for the graphical interpretation.

Furthermore, we assume that the density at the wall node is unknown and thus its value at a boundary node should result from the evaluation of Eq.~\eqref{eq:moment_based_rule_well_conditioned}. This no-slip implementation will be referred to as MB. To explicitly enforce zero normal gradient of the density at a boundary, we introduce what will be further called the $f_0$-correction. Namely, after the MB no-slip is imposed in each boundary node, we set $f_0$ to the value of
\begin{equation}
    \tilde{f_0}(t+1; \bsym{x}_B) = \tilde{\rho}(t+1;\bsym{x}_N) - \sum\limits_{k=1}^{q-1} \tilde{f_k}(t+1; \bsym{x}_B).
\end{equation}
We will refer to the MB no-slip with this correction as MB$_0$. Finally, we note that in our implementation we pre-compute the matrices $\mathbb{M}$ and $\mathbb{A}_m$ for each of 8 combinations of the sets $K$ and $UK$.

\subsection{Hydrodynamic coefficients and stresses on a cylinder}

We use the pressure coefficient $C_P(\bsym{x})$ to represent the fluid pressure at point $\bsym{x}$ in non-dimensional form

\begin{equation}\label{eq:cp}
	C_P (\bsym{x}) = \dfrac{P_\infty - P(\bsym{x})}{\frac{1}{2}\rho_\infty V_\infty^2}.
\end{equation}
The calculation of the drag coefficient of the cylinder involves the integration of stresses around the cylinder's perimeter $\partial \Omega_w$
\begin{equation}\label{eq:drag_recipie}
	C_D = \dfrac{F_{D,P} + F_{D,WSS}}{\dfrac{1}{2}\rho_\infty V_\infty} = 
	\dfrac{\left(-\displaystyle\int\limits_{\bsym{x}: \> r=r_\text{in}} \!\!\! d\bsym{x} \> \bsym{n}P +
	\mu\int\limits_{\bsym{x}: \> r=r_\text{in}} \!\!\! d\bsym{x} \> \bsym{n} \cdot \nabla\bsym{V}\right) \cdot \hat{\bsym{e}}_0}{\dfrac{1}{2}\rho_\infty V_\infty},
\end{equation}
where $F_{D,P}$ and $F_{D,WSS}$ denote the pressure and shear stress contributions to the drag force. We use the same RBFFD method to calculate the gradient of the macroscopic velocity field, as for the interpolation (i.e. $\mathcal{L} = \partial/\partial x, \> \partial/\partial y$ in Eq.~\eqref{eq:rbffd_general_fd}). Taking the analytical stress tensor according to Eq.~\eqref{eq:stokes_analytical}, one obtains the true value of the drag force
\begin{equation}\label{eq:drag_recipie_true}
	C_{D,\text{true}} =
	\dfrac{\left(-\displaystyle\int\limits_{\theta = 0}^{2\pi} d\theta \> P\cos{\theta} +
	\mu\int\limits_{\theta = 0}^{2\pi} d\theta \> \partial_r V_\theta\sin{\theta}\right)_{r=r_\text{in}}}{\dfrac{1}{2}\rho_\infty V_\infty}
\end{equation}

\subsection{Solution error norms and steady state criteria used in this work}

We introduce the relative error of the drag coefficient $\text{err}(C_D)$ defined in a standard manner
\begin{equation}
	\text{err}(C_D) = \dfrac{|C_D - C_{D,\text{true}}|}{C_{D,\text{true}}}.
\end{equation}
To measure the error of a scalar field $\alpha(\bsym{x})$ given at discrete points, we calculate the normalized length of the vector of the values of $\alpha$
\begin{equation}\label{eq:L2_norm}
	L_2 = \dfrac{\sqrt{\sum\limits_{i=1}^{N}(\alpha_i - \alpha_{i,\text{true}})^2}}{\sqrt{\sum\limits_{i=1}^{N}\alpha_{i,\text{true}}^2}}.
\end{equation}
In case of the velocity magnitude, to avoid division by zero, we normalize the length of the error vector by its size
\begin{equation}\label{eq:length_norm}
	\langle L_2 \rangle = \dfrac{\sqrt{\sum\limits_{i=1}^{N}(\alpha_i - \alpha_{i,\text{true}})^2}}{N}.
\end{equation}

The mean order of convergence of quantity $\alpha$, $\langle p \rangle_{\alpha,\beta}$, is calculated as the arithmetic mean of the orders of convergence measured between the subsequently refined discretizations. The subscript $\beta$ denotes over which subset of the domain the error is calculated, such that $\beta = \Omega$ denotes the whole domain and $\beta = \partial\Omega$ denotes the cylinder's surface.

To define whether the CFD solution converged to a steady state, we monitor the mean $V_0$ and the mean absolute value of the pressure deviation $|\delta \rho|$ taken every $\Delta n_t=10^4$ timesteps
\begin{equation}
	\langle \alpha \rangle^t = \dfrac{1}{N}\sum\limits_{i=1}^{N}\alpha_i(t),
	\quad
	\alpha \in \{V_0, |\delta \rho|\},
\end{equation}
where $\alpha_i(t)$ denotes the value of $\alpha_i$ at timestep $t$. We calculate the relative root mean square of the most recent $n_{mon}$ values of each monitor
\begin{equation}
	\begin{array}{rcl}
		\overline{\langle \alpha \rangle^t} &=& \dfrac{1}{n_{mon}}\displaystyle\sum\limits_{i=0}^{n_{mon}-1} \langle \alpha \rangle^{t - i\Delta n_t},\\
		&&\\
		\text{RMS}_{\alpha}^t &=&
		\dfrac{1}{\overline{\langle \alpha \rangle^t}}
		\dfrac{\sqrt{\displaystyle\sum\limits_{i=1}^{N}\left(\langle \alpha \rangle^t -\overline{\langle \alpha \rangle^t}\right)^2}}{\sqrt{n_{mon}-1}}.
	\end{array}
\end{equation}
We choose $n_{mon}$ such that $n_{mon}\delta t = 0.03$ and terminate the calculations if both of the $\text{RMS}_{\alpha}^t$ values fall below $10^{-3}$.
	
\section{Results}\label{sec:results}

\subsection{Behavior of macroscopic fields at vanishing Reynolds number}\label{sec:macro_small_Re}

In the following Sections we investigate the behavior of the density and velocity fields in the limit of vanishing $Re$. The analysis of the convergence to the analytical solutions, Eqs.~\eqref{eq:stokes_analytical}, are given in Section 2 of the Supplementary Materials. Here, we show only the results obtained with $N_\theta=180$. For each implementation of the no-slip boundary we consider two discretization strategies described in Section~\ref{sec:discretization}, and several values of the relaxation parameter $\tau \in \{1.0, 0.75, 0.53125\}$. We set the Reynolds number $Re=0.01$. Since on the cylinder the true velocity is zero, we use the normalized length of the error vector, Eq.~\eqref{eq:length_norm}, as the measure of the velocity field error. We show the results only for the setups that did not give \verb|NaN| or infinite values in the solution. We note that the velocity values presented in the Figures are expressed in physical units, so in order to obtain the values in lattice units, one should divide them by the velocity conversion factor $C_V = \delta x/\delta t = 400$.

\subsubsection{Qualitative analysis of the hydrodynamic fields in the whole domain}

\begin{figure}[ht!]
	\centering
	\includegraphics[width=.45\linewidth]{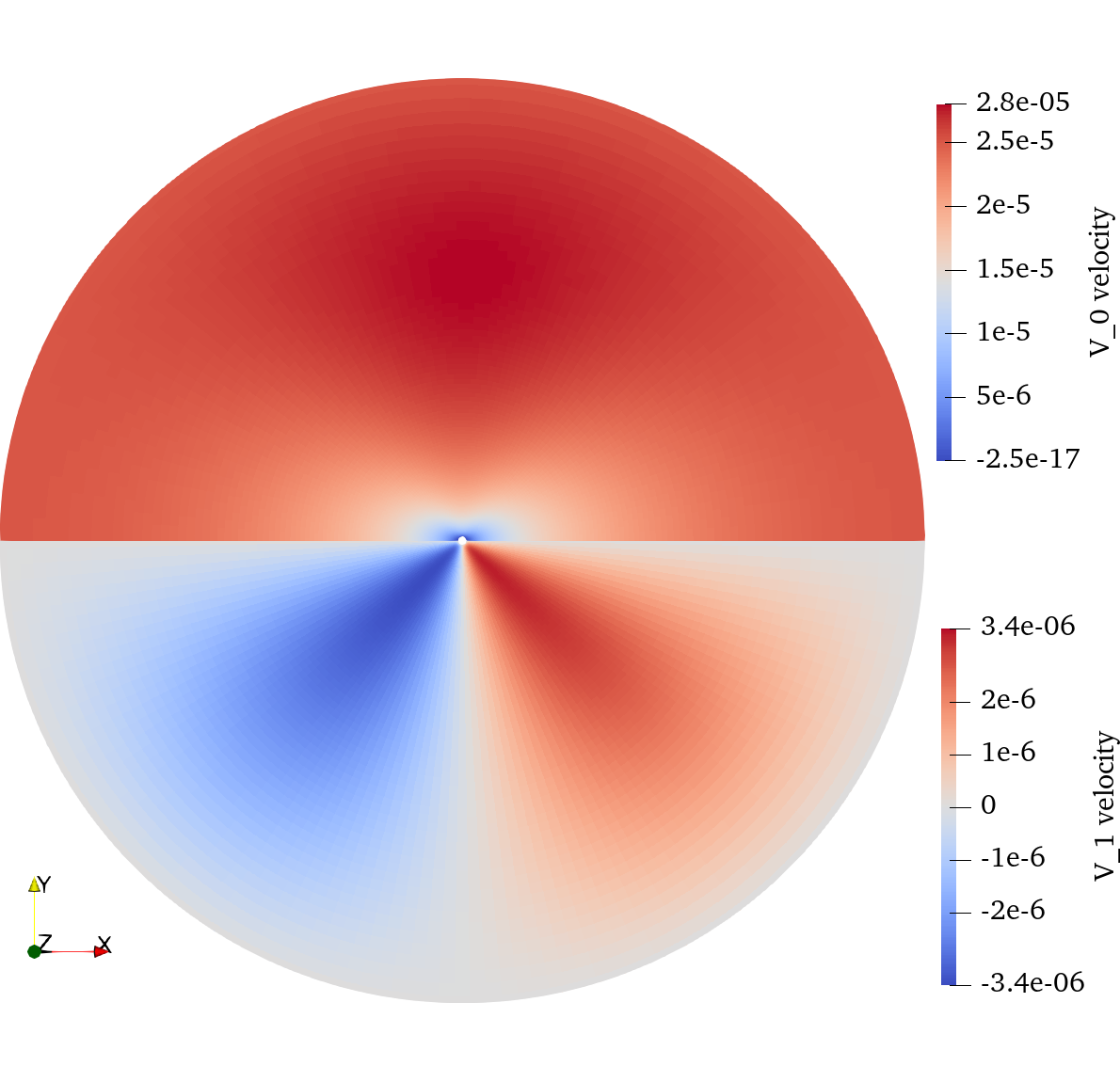}
	\includegraphics[width=.45\linewidth]{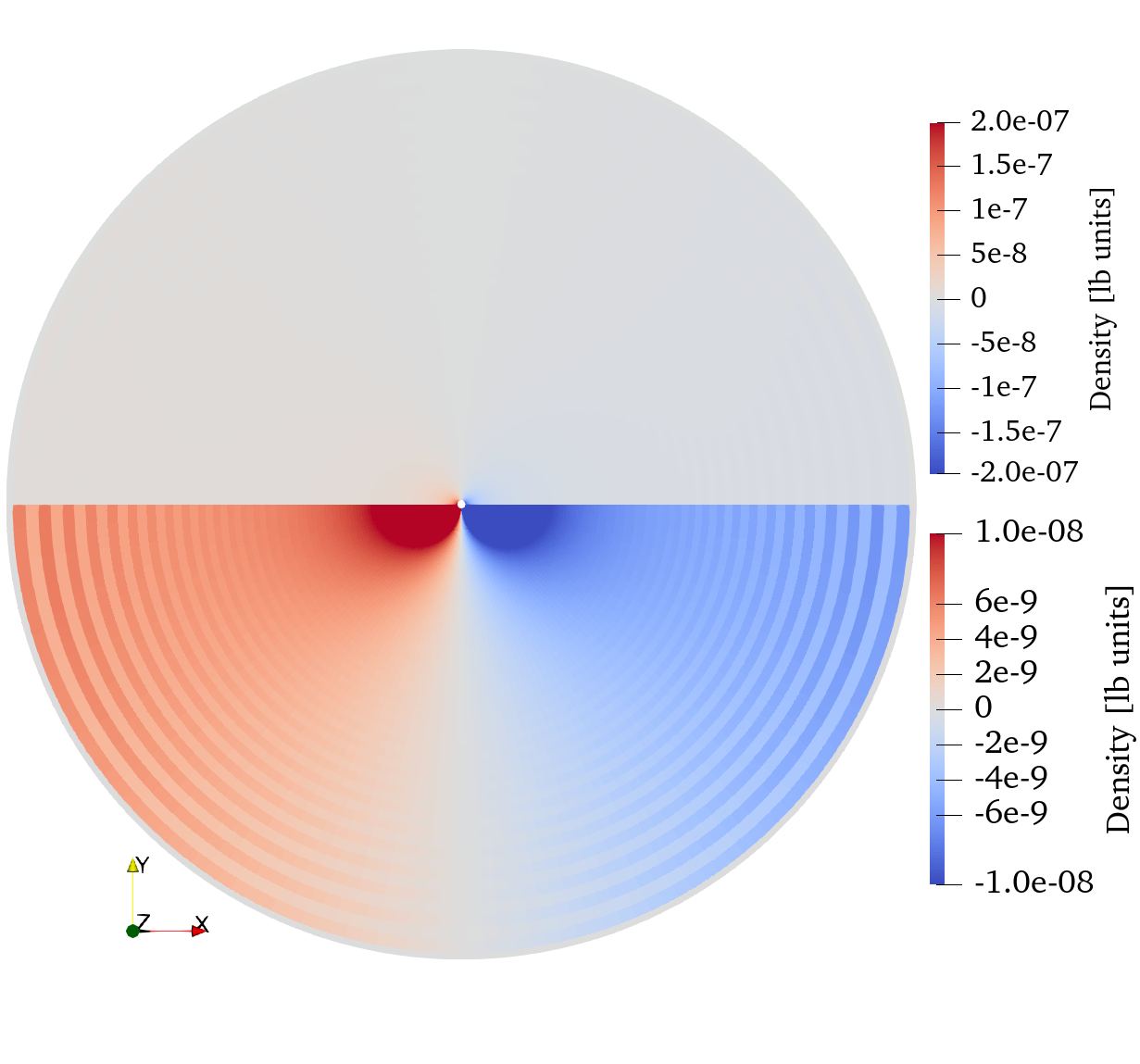}
	\caption{{\itshape Left}: velocity for non-equilibrium interpolation no-slip and structured discretization for $\tau=1$ and $N_\theta=180$. The upper half shows $V_0$ field, while the bottom half shows $V_1$. {\itshape Right}: density variation for the same setup. The colormap in the upper half is rescaled to the whole range of the data, in the bottom half -- to the range closer to the far-field density variation, $\rho_\infty-1=0$.}
	\label{fig:hydro_map}
\end{figure}

Fig.~\ref{fig:hydro_map} shows the velocity and the density fields in the whole domain obtained with non-equilibrium extrapolation boundary condition on the cylinder's wall. One sees that while the velocity field is smooth and free from oscillations, the density field exhibits standing waves propagating from the outer boundary, of amplitude about two orders of magnitude smaller than the density at the stagnation/wake points. Such waves are visible regardless of the no-slip implementation used, thus we assume that they are not related to this part of the algorithm. Furthermore, due to the damping of their amplitude towards the cylinder, we conclude that it doesn't significantly impact the results concerning the further analysis of the no-slip implementations.

\subsubsection{Analysis of drag coefficient and its components at vanishing Reynolds number}\label{sec:cd_small_Re}

Tables~\ref{tab:stokes_cd_summary_ogrid} and \ref{tab:stokes_cd_summary_hybrid} show the values of the drag coefficient $C_D$ defined in Eq.~\eqref{eq:drag_recipie}, and its components from the pressure and the wall shear stress (WSS) obtained with each of the studied setups. The values are accompanied by their error relative to the theoretical values, which are presented in the bottommost rows for each no-slip implementation. The two rightmost columns of each table also show the $L_2$ norms, Eq.~\eqref{eq:L2_norm}, of the WSS and pressure coefficient profiles on the cylinder's surface.

The relative errors of $C_D$ values obtained with OLBM are generally similar between the corresponding cases with regular and hybrid discretization, varying between about $2\%$ for non-equilibrium extrapolation on regular discretization to several tens of percents for IBB on both node sets. This indicates that the presence of scattered nodes does not impact the accuracy of the solution obtained with a given no-slip realization much. It can even have a slightly stabilizing effect on the simulation, as for the hybrid cases, non-equilibrium extrapolation and both bouncebacks managed to give stable solutions for $\tau=0.53125$. This can be attributed to the fact that the ring of scattered nodes acts as a filter to any spurious oscillations amplified on the regular part of the grid. Looking at each no-slip implementation separately, one sees that decreasing the relaxation time $\tau$ generally introduces larger errors to the drag coefficient and its components. Such behavior was previously observed for BGK collision with SBB, e.g., in~\cite{Pan06}, where the unit relaxation time gave the smallest errors of the permeability in a porous medium. The error of the pressure component of the drag does not exceed $5\%$ but for the lowest viscosity IBB on hybrid discretization. This is in contrast to the error of the WSS component of $C_D$, which reaches the values about twice as large as the error of the total drag. This indicates that the error of $C_D$ is dominated by the WSS part. Suprisingly, the $L_2$ errors of the pressure are often significantly higher than those of WSS (cf., e.g., IBB). This suggests that certain sing-dependent cancellations occur upon the integration of the pressure component of $C_D$, which does not occur in the squared integrand of $L_2$ error. Let us investigate this matter in more details in the further Sections.


\newcommand{\tablecaptionend}{at $N_\theta=180$ refinement on regular discretization. `P' denotes the drag component from the pressure and `WSS' -- from the wall shear stress. Hyphens denote diverged cases, `err' and $L_2$ errors are given in percent.}

\newcommand{\tablecaptionstart}{The values and relative errors of the drag coefficients and their components obtained with}

\newcommand{\tablecaption}[1]{The values and relative errors of the drag coefficients and their components obtained with various no-slip implemmentations at $N_\theta=180$ refinement on #1 discretizations. `P' denotes the drag component from the pressure and `WSS' -- from the wall shear stress. Hyphens denote diverged cases, `err' and $L_2$ errors are given in percent.}

\newcommand{\tableheader}{$\tau$ & $C_D$ & err$(C_D)$ & $C_{D,P}$ & err$(C_{D,P})$ & $C_{D,WSS}$ & err$(C_{D,WSS})$ &$L_{2,P}$ & $L_{2,WSS}$ \\}

\newcommand{\tableheaderX}[1]{\begin{tabular}{lrcrcrcrr}
	\multicolumn{9}{c}{#1} \\
	\hline
	$\tau$ & $C_D$ & err$(C_D)$ & $C_{D,P}$ & err$(C_{D,P})$ & $C_{D,WSS}$ & err$(C_{D,WSS})$ &$L_{2,P}$ & $L_{2,WSS}$ \\
	\hline
	\hline}

\newcommand{\tablefooterX}{\hline\end{tabular}\vspace{0.5cm}}

\begin{table}[!ht]
		\centering


\tableheaderX{NEE}
        $1.0$ & $660.88$ & $1.34$ & $330.62$ & $0.07$ & $330.27$ & $2.72$ & $0.53$ & $2.78$ \\
        $0.75$ & $656.04$ & $2.07$ & $332.17$ & $0.55$ & $323.86$ & $4.61$ & $0.79$ & $4.64$ \\
        $0.53$ &  --  &  --  &  --  &  --  &  --  &  --  &  --  &  --  \\
        \hline 
        True & $669.88$ &  & $330.37$ &  & $339.52$ & & &  \\
	\tablefooterX

		\tableheaderX{SBB}
		$1.0$ & $615.13$ & $8.17$ & $333.97$ & $1.09$ & $281.16$ & $17.19$ & $17.37$ & $19.06$ \\
		$0.75$ & $597.71$ & $10.77$ & $338.27$ & $2.39$ & $259.44$ & $23.59$ & $26.88$ & $25.05$ \\
		$0.53$ &  --  &  --  &  --  &  --  &  --  &  --  &  --  &  --  \\
		\hline 
		True & $669.88$ &  & $330.37$ &  & $339.52$ & & &  \\
	\tablefooterX

		\tableheaderX{IBB}
		$1.0$ & $565.54$ & $15.58$ & $324.2$ & $1.87$ & $241.35$ & $28.91$ & $50.12$ & $33.86$ \\
		$0.75$ & $538.01$ & $19.69$ & $326.28$ & $1.24$ & $211.74$ & $37.64$ & $90.18$ & $43.24$ \\
		$0.53$ &  --  &  --  &  --  &  --  &  --  &  --  &  --  &  --  \\
		\hline 
		True & $669.88$ &  & $330.37$ &  & $339.52$ & & &  \\
	\tablefooterX

		\tableheaderX{MB}
        $1.0$ & $670.71$ & $0.12$ & $340.32$ & $3.01$ & $330.39$ & $2.69$ & $25.11$ & $2.83$ \\
        $0.75$ & $627.5$ & $6.33$ & $337.68$ & $2.21$ & $289.82$ & $14.64$ & $20.25$ & $16.31$ \\
        $0.53$ &  --  &  --  &  --  &  --  &  --  &  --  &  --  &  --  \\
        \hline 
        True & $669.88$ &  & $330.37$ &  & $339.52$ & & &  \\
	\tablefooterX

		\tableheaderX{MB$_0$}
        $1.0$ & $660.88$ & $1.34$ & $330.62$ & $0.07$ & $330.27$ & $2.72$ & $0.53$ & $2.78$ \\
        $0.75$ & $623.03$ & $6.99$ & $331.29$ & $0.28$ & $291.74$ & $14.07$ & $7.73$ & $15.62$ \\
        $0.53$ &  --  &  --  &  --  &  --  &  --  &  --  &  --  &  --  \\
        \hline 
        True & $669.88$ &  & $330.37$ &  & $339.52$ & & &  \\
	\tablefooterX
	\caption{\tablecaption{regular}}
	\label{tab:stokes_cd_summary_ogrid}
\end{table}


\renewcommand{\tablecaptionend}{at $N_\theta=180$ refinement on hybrid discretization. `P' denotes the drag component from the pressure and `WSS' -- from the wall shear stress. Hyphens denote diverged cases, `err' and $L_2$ errors are given in percent.}

\begin{table}[!ht]
	\centering
		\tableheaderX{NEE}
        $1.0$ & $645.51$ & $3.64$ & $319.53$ & $3.28$ & $325.98$ & $3.99$ & $3.36$ & $4.02$ \\
        $0.75$ & $629.45$ & $6.04$ & $317.96$ & $3.76$ & $311.49$ & $8.25$ & $3.88$ & $8.27$ \\
        $0.53$ &  --  &  --  &  --  &  --  &  --  &  --  &  --  &  --  \\
        \hline 
        True & $669.88$ &  & $330.37$ &  & $339.52$ & & &  \\
		\tablefooterX

		\tableheaderX{SBB}
        $1.0$ & $612.52$ & $8.56$ & $332.3$ & $0.58$ & $280.23$ & $17.46$ & $16.98$ & $19.1$ \\
        $0.75$ & $595.45$ & $11.11$ & $335.5$ & $1.55$ & $259.95$ & $23.43$ & $25.7$ & $24.63$ \\
        $0.53$ & $542.67$ & $18.99$ & $335.16$ & $1.45$ & $207.51$ & $38.88$ & $48.23$ & $39.15$ \\
		\hline 
		True & $669.88$ &  & $330.37$ &  & $339.52$ & & &  \\
		\tablefooterX

		\tableheaderX{IBB}
        $1.0$ & $564.71$ & $15.7$ & $322.57$ & $2.36$ & $242.14$ & $28.68$ & $50.45$ & $33.3$ \\
        $0.75$ & $539.04$ & $19.53$ & $323.45$ & $2.09$ & $215.59$ & $36.5$ & $93.16$ & $41.67$ \\
        $0.53$ & $442.29$ & $33.97$ & $262.89$ & $20.43$ & $179.4$ & $47.16$ & $541.32$ & $54.08$ \\
        \hline 
        True & $669.88$ &  & $330.37$ &  & $339.52$ & & &  \\
		\tablefooterX

		\tableheaderX{MB}
        $1.0$ & $661.12$ & $1.31$ & $338.21$ & $2.37$ & $322.91$ & $4.89$ & $19.54$ & $5.48$ \\
        $0.75$ & $634.0$ & $5.36$ & $336.26$ & $1.78$ & $297.74$ & $12.3$ & $19.1$ & $13.23$ \\
        $0.53$ &  --  &  --  &  --  &  --  &  --  &  --  &  --  &  --  \\
        \hline 
        True & $669.88$ &  & $330.37$ &  & $339.52$ & & &  \\
		\tablefooterX

		\tableheaderX{MB$_0$}
        $1.0$ & $658.57$ & $1.69$ & $333.04$ & $0.81$ & $325.53$ & $4.12$ & $2.28$ & $4.19$ \\
        $0.75$ & $629.4$ & $6.04$ & $331.99$ & $0.49$ & $297.41$ & $12.4$ & $21.8$ & $13.29$ \\
        $0.53$ &  --  &  --  &  --  &  --  &  --  &  --  &  --  &  --  \\
        \hline 
        True & $669.88$ &  & $330.37$ &  & $339.52$ & & &  \\
		\tablefooterX
	\caption{\tablecaption{hybrid}}
	\label{tab:stokes_cd_summary_hybrid}
\end{table}

\subsubsection{Pressure coefficient}\label{sec:pressure_small_Re}

\begin{figure}[ht!]
	\centering
	\includegraphics[width=\linewidth]{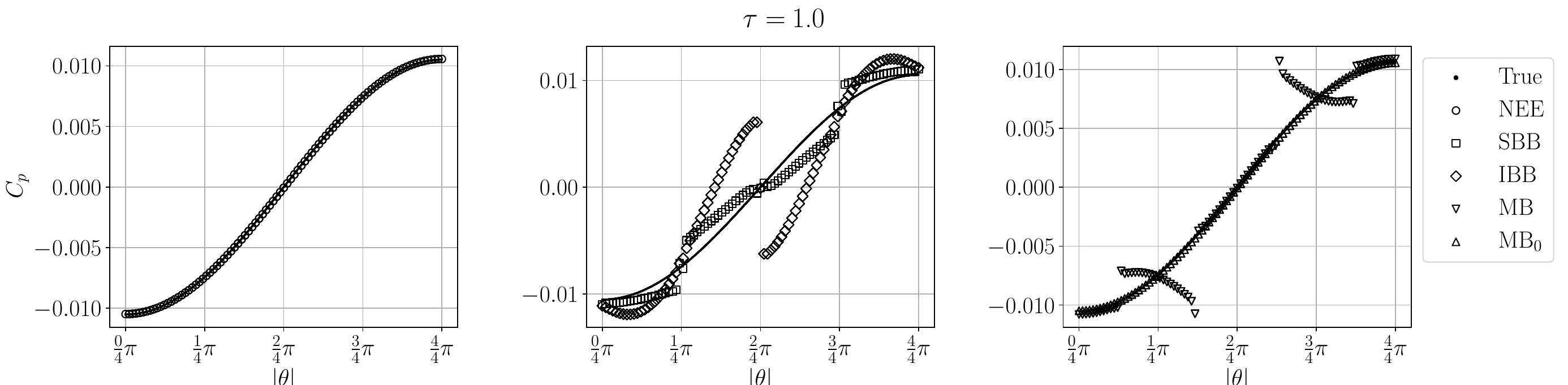}
	\includegraphics[width=\linewidth]{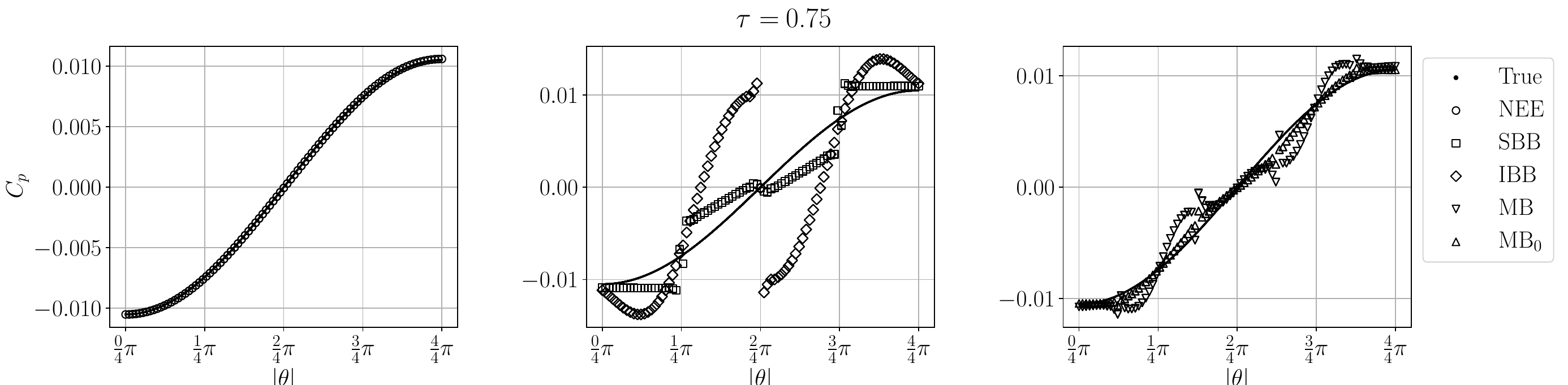}
	\caption{Profiles of the pressure coefficient on the cylinder's surface for various no-slip implementations and regular discretization for various $\tau$.}
	\label{fig:cp_allBCs_ogrid}
\end{figure}

The analysis of the pressure coefficient profiles will be conducted starting from the setups with regular discretizations presented in Fig.~\ref{fig:cp_allBCs_ogrid} and for $\tau=1$. For non-equilibrium extrapolation -- leftmost column in Fig.~\ref{fig:cp_allBCs_ogrid} -- the pressure coefficient profiles on the cylinder are perfectly aligned with the analytical solution. Overwriting all the populations constrained with the first order Neumann boundary condition for the density, along with structured discretization right at the cylinder's surface, allows for a good control over the pressure on the cylinder.

The bounceback-based implementations are shown in the middle column of Fig.~\ref{fig:cp_allBCs_ogrid}. For SBB, the obtained profiles only roughly follow the theoretical shape, with piecewise linear segments spanned between the subsequent angles $k\pi/4$, $k=1,2,3$, at which discontinuities are visible. Those angles are the ones corresponding to the direction of the discrete velocities in the used D2Q9 model. It suggests that the reason for the appearance of those discontinuities are the abrupt changes in the set of the populations undergoing bounceback. For IBB, the $C_p$ profiles have sinusoidal shape, piecewise continuous on segments spanning $\pi/2$ angles. In this case, the discontinuities in the profiles visible at angle $\pi/2$ have much larger amplitudes than any of the discontinuities in the SBB case. Interestingly, the values of $C_p$ at the angles $k\pi/4$ are in a good agreement with the theoretical predictions.

The moment-based no-slip implementations are shown in the rightmost column in Fig.~\ref{fig:cp_allBCs_ogrid}.The plain implementation gives huge errors away from angles $k\pi/2$, $k=0,1,2$. Discontinuities at the angles $k\pi/4 \pm \pi/8$, $k=0,1,2,3$ are present, which can relate to the strategy for determining the missing populations described in Section~\ref{sssec:moments_bc}. The version of the moments-based no-slip with $f_0$-correction follows the theoretical profile much closer. 

The decrease of viscosity have different implications for every type of the no-slip implementations. It has a practically invisible impact on the non-equilibrium extrapolation method. In the case of bounceback-based approaches, the amplitude of the discontinuities in the $C_p$ profile increases with decreasing $\tau$. The non-corrected moments-based no-slip exhibits discontinuities at similar angles as wiht $\tau=1$. The $f_0$-corrected version diverges visibly from the theoretical profile. It also develops discontinuities at angles similar to those observed in the non-corrected version of this boundary condition. It suggests that even though one has a direct control over the density at the boundary nodes in MB$_0$, the copying of its values from the neighboring nodes can introduce significant errors.

\begin{figure}[ht!]
	\centering
	\includegraphics[width=\linewidth]{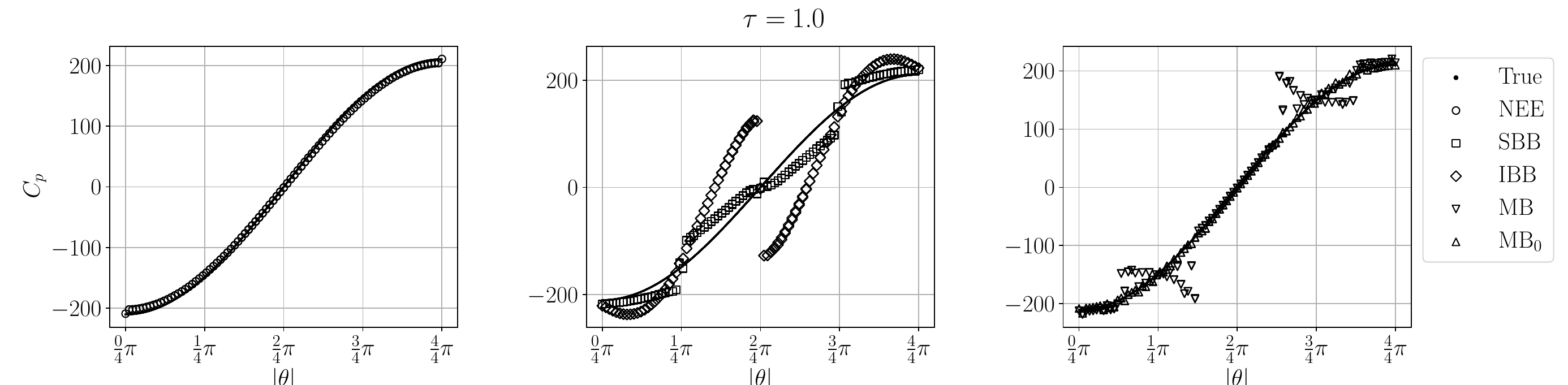}
	\includegraphics[width=\linewidth]{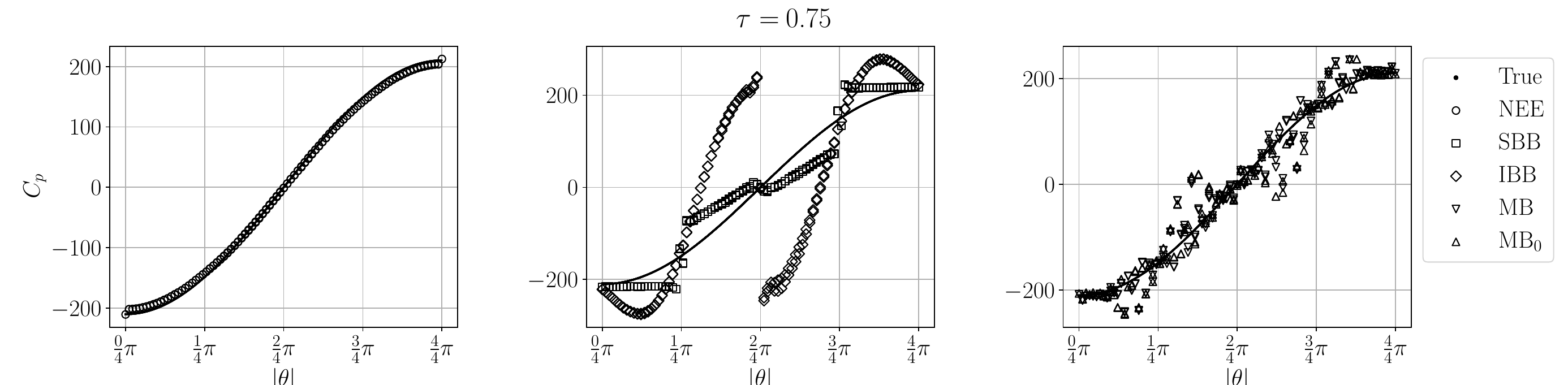}
	\includegraphics[width=0.4\linewidth]{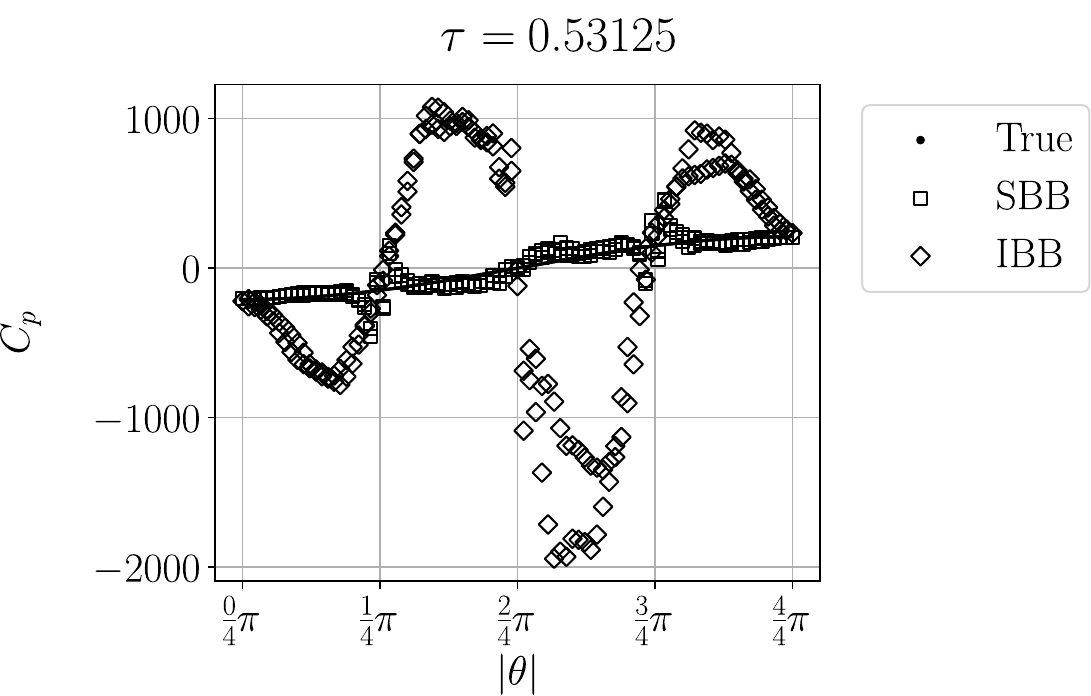}
	\caption{Profiles of the pressure coefficient on the cylinder's surface for various no-slip implementations and hybrid discretization for various $\tau$. The subplots with no data points but for the true solution correspond to the diverged cases.}
	\label{fig:cp_allBCs_hybrid}
\end{figure}

On hybrid discretizations, Fig.~\ref{fig:cp_allBCs_hybrid}, the $C_p$ profiles obtained with every no-slip implementation is similar to those from the regular discretizations. One visible difference is the slight noise added to the pressure profiles. The positions of the points within interpolation stencils are no longer symmetric between the upper and the lower half of the domain, thus slight differences in the solutions obtained therein occur. Further, as stated in Section~\ref{sec:cd_small_Re}, the lesser regularity of the nodes layout may have a stabilizing effect on the solution, by preventing the spurious oscillations from propagating freely through the domain~\cite{KOLARPOZUN2026} and preventing the local interpolation matrices from becoming ill-conditioned~\cite{Flyer2016}. This is observed in $\tau=0.53125$ case with SBB and IBB not diverging, although reaching extremely high errors.

The surprising fact of one order of magnitude difference between the pressure part of the drag coefficient, $C_{D,P}$, and the error of the pressure profile, $L_{2,P}$, can be analyzed in terms of the symmetry of the pressure profiles about the angle $\pi/2$. The setups where this occurs are all the studied ones on the regular discretization, and from the hybrid discretizations:
\begin{itemize}
	\item all no-slips for $\tau=1$,
	\item NEE and both bounce-backs for $\tau=0.75$.
\end{itemize}
One can represent the profiles of the pressure obtained numerically on the surface of the cylinder as the sum of the theoretical value from Eq.~\eqref{eq:stokes_analytical} and the error denoted by $\delta P$
\begin{equation}\label{eq:hydro_on_cylinder_with_error}
	P^\text{num}(r_\text{in},\theta) = P(r_\text{in},\theta) + \delta P(r_\text{in},\theta) \\
\end{equation}
From the setups for which the error of $C_{D,P}$ is small, one can assume that the errors of the pressure are approximately anti-symmetric about the $\theta = \pi/2$ angle
\begin{equation}\label{eq:hydro_on_cylinder_error_symmetric}
	\delta P(\theta) \approx -\delta P(\pi-\theta)
\end{equation}
In turn for the numerical pressure to give the approximately correct component of the drag coefficient, the following integral must hold
\begin{equation}
	\int\limits_0^{\pi/2} d\theta \>
	\cos{\theta} \delta P(\theta)
	\approx
	0
\end{equation}
which means that the pressure errors must cancel out over the lee and windward half of the cylinder. Despite the largest errors $\delta P$ occurring near the angle $\pi/2$, the cosine in the integrand diminishes their impact on the overall pressure drag component.

\subsubsection{Velocity}\label{sec:velocity_small_Re}

\begin{figure}[ht!]
	\centering
	\includegraphics[width=\linewidth]{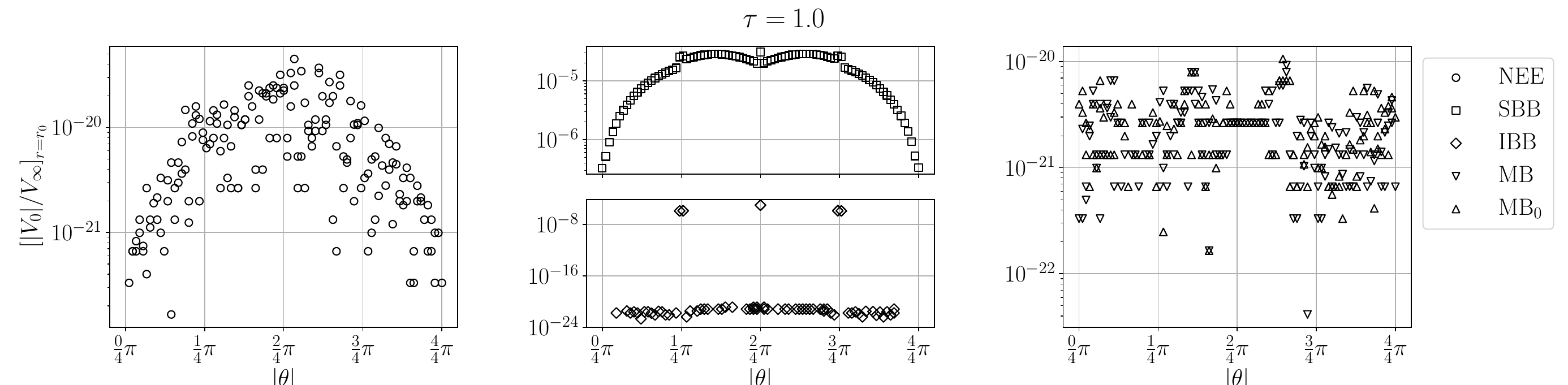}
	\includegraphics[width=\linewidth]{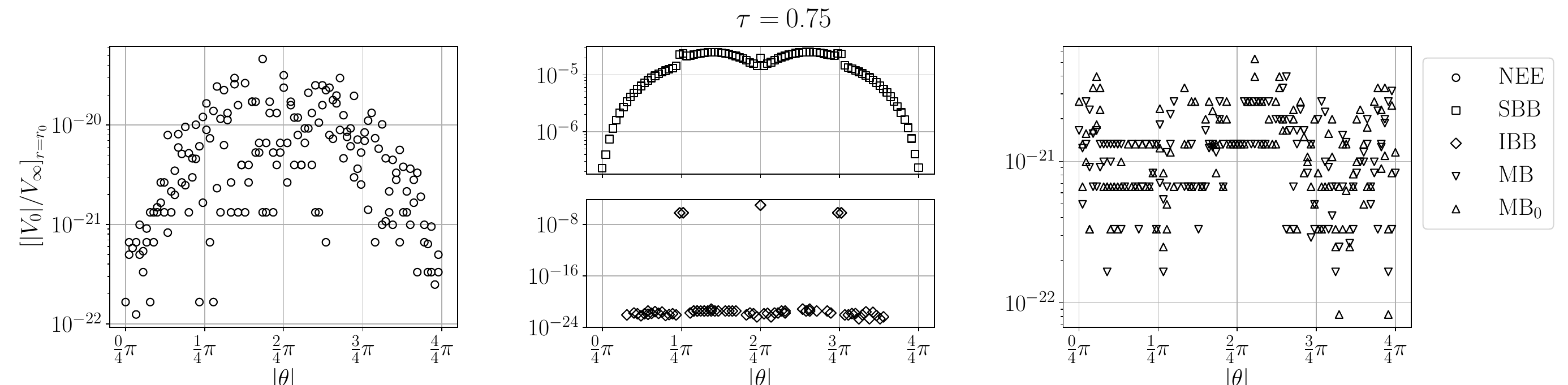}
	\caption{Profiles of the $x_0$-component of the velocity on the cylinder's surface for various no-slip implementations and regular discretization for various $\tau$.}
	\label{fig:v0_allBCs_ogrid}
\end{figure}

Next, we analyze the velocity profiles on the cylinder's surface, starting from the setups with regular discretizations, Fig.~\ref{fig:v0_allBCs_ogrid}. When the no-slip on the cylinder is implemented with the non-equilibrium extrapolation or any of the moments-based methods -- leftmost and rightmost columns in Fig.~\ref{fig:v0_allBCs_ogrid}, respectively -- the velocity profiles on the cylinder are on the order of numerical zero. This is a direct implication of the fact that in those no-slip implementations on has a direct control over the velocity imposed at the boundary nodes.

The bounceback-based implementations are shown in the middle column in Fig.~\ref{fig:v0_allBCs_ogrid}. With SBB, the velocity values on the cylinder exhibit discontinuities. The locations of the discontinuities are the $k\pi/4$ angles, $k=1,2,3$, similarly to the pressure profiles and the reason for the appearance of the discontinuities is most probably the same as previously. The magnitude of the $V_0$ velocity on the cylinder is significantly above zero, which is an expected result since in standard LBM the simple bounceback places the wall along the cut links, not at the nodes~\cite{CORNUBERT1991241}. With IBB implementation, the discontinuities in the velocity profiles on the cylinder are visible at angles $k\pi/4$, $k=1,2,3$, but apart from those points, the velocity is zero on the boundary. This is an expected results, as for this no-slip implementation on the lattice, the interpolation is meant to put the wall (i.e., the zero-velocity level set) in the desired place, controlled by the interpolation coefficients. In our case this is exactly at the boundary nodes. The reason for the discontinuities at the multiplicities of $\pi/4$ angle might be similar to SBB. We note that near the angles $0$ and $\pi$ the velocity value for IBB is very close to zero and thus is not visible on a logarithmic scale. This may be linked to the lack of shear at those points and thus -- little non-equilibrium part of VDF, see Section~\ref{sec:wss_small_Re}. 

On regular grids, decreasing the relaxation time from $1$ to $0.75$ does not have any visible impact on NEE, IBB, MB, and MB$_0$ setups. A slight decrease of the $V_0$ magnitude for SBB is observed. Decreasing $\tau$ even further, to the value of $0.53125$, made all the regular discretization setups unstable.

\begin{figure}[ht!]
	\centering
	\includegraphics[width=\linewidth]{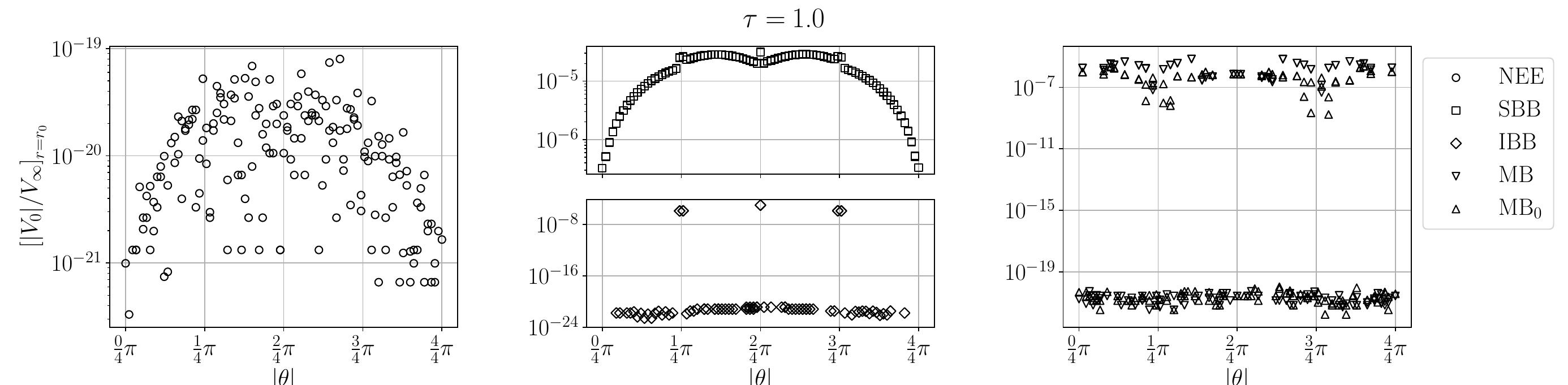}
	\includegraphics[width=\linewidth]{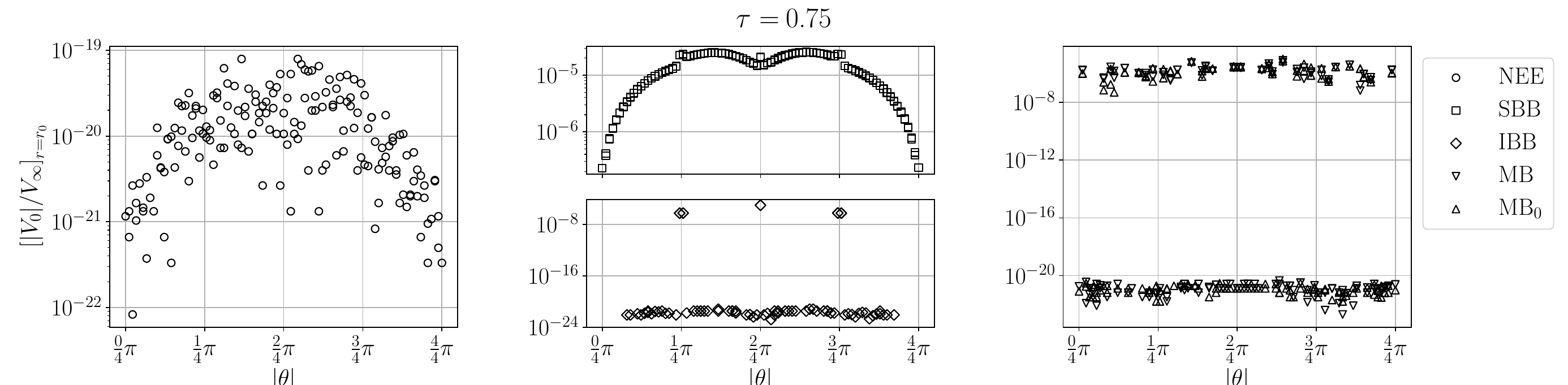}
	\includegraphics[width=0.4\linewidth]{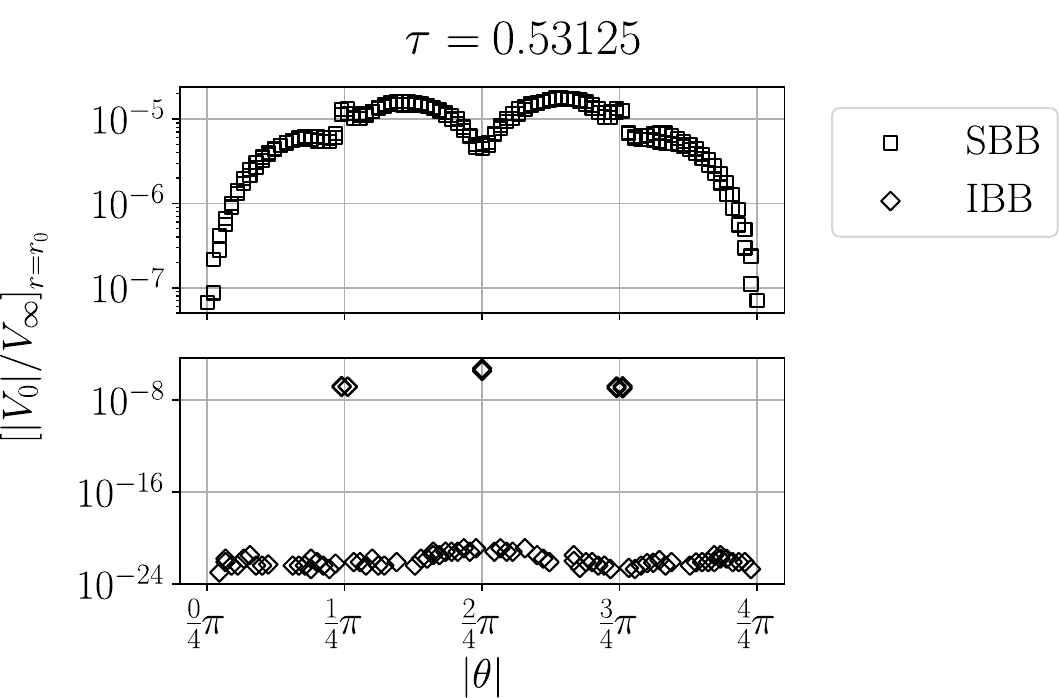}
	\caption{Profiles of the $x_0$-component of the velocity on the cylinder's surface for various no-slip implementations and hybrid discretization for various $\tau$. The subplots with no data points but for the true solution correspond to the diverged cases.}
	\label{fig:v0_allBCs_hybrid}
\end{figure}

Let us now move to the setups with hybrid discretizations shown in Fig.~\ref{fig:v0_allBCs_hybrid}. The behavior of the setups with NEE is similar to the one obtained on regular discretization, with a slight increase of the $V_0$ magnitude for both relaxation time values. In the case of SBB, one again sees a decrease of the error with the decrease of $\tau$, however the smoothness of the velocity profile deteriorates visibly between $\tau=0.75$ and $0.53125$. IBB behaves similarly to the regular discretization setup too. When moments-based no-slips are considered, the velocity at certain points increases far above numerical zero compared to the regular grid. The values in the remaining boundary nodes are mostly below $10^{-20}$, similarly to the regular grid case. We note again that both moments-based implementations and NEE went unstable for $\tau=0.53125$.

\subsubsection{Wall shear stress}\label{sec:wss_small_Re}

\begin{figure}[ht!]
	\centering
	\includegraphics[width=\linewidth]{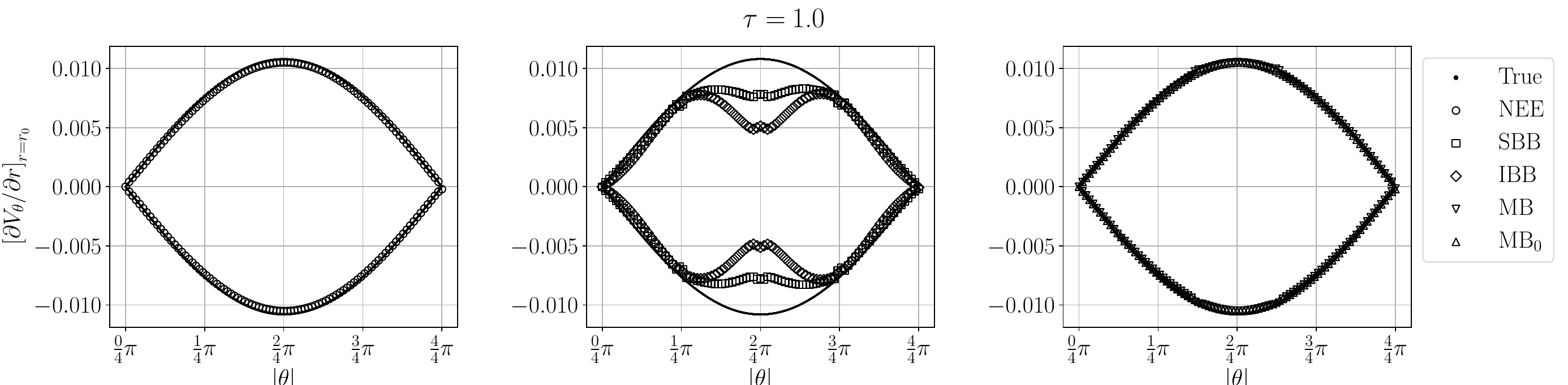}
	\includegraphics[width=\linewidth]{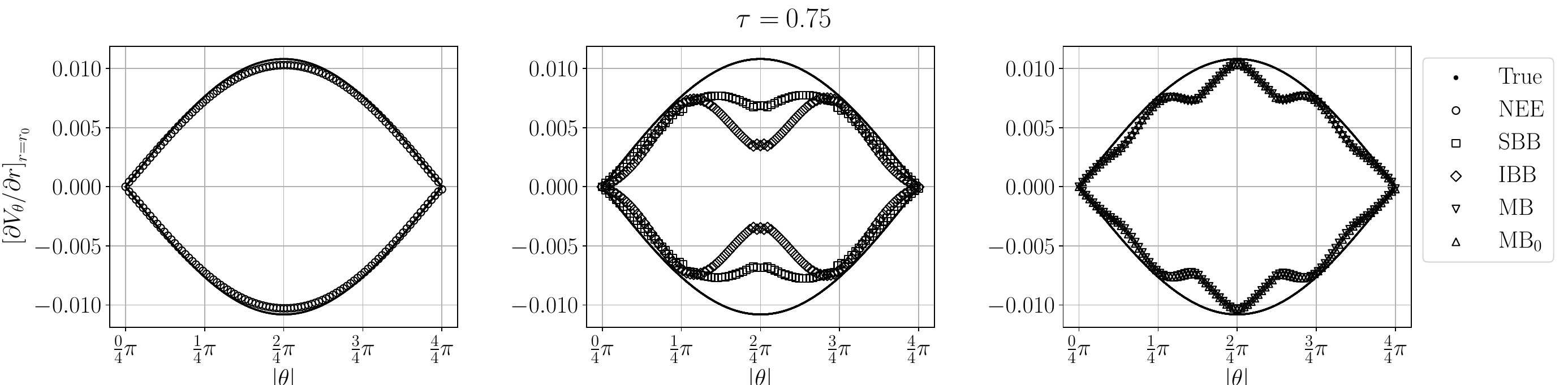}
	\caption{Profiles of the wall shear stress on the cylinder's surface for various no-slip implementations and regular discretization for various $\tau$.}
	\label{fig:wss_allBCs_ogrid}
\end{figure}

The profiles of the wall shear stress, $\partial_r V_\theta$ in Eq.~\eqref{eq:drag_recipie}, are shown for regular discretizations in Fig.~\ref{fig:wss_allBCs_ogrid}. For all no-slip implementations, the WSS profiles are symmetric about the $\pi/2$ angle. At the wake, $\theta=0$, and at the stagnation point, $\theta = \pi$, the shear is correctly predicted to be zero. NEE gives its profiles of correct shape, with the magnitude decreasing slightly when $\tau$ gets lower. Both bouncebacks exhibit a fair compliance with the theoretical profile in the vicinity of the stagnation and wake points. In the range $\theta \in [\pi/4; 3\pi/4]$, a significant error is visible, with several local extrema in the WSS profile. The extrema are present for both studied relaxation times, but the magnitude of the error increases with the decreasing viscosity. The moments-based implementations give a good compliance with the theory for $\tau=1$, however for the lower value of the relaxation time the error grows significantly, even for the angles closer to the wake and stagnation points compared to the bouncebacks. For the moment-based no-slips the error obtains a different profile than as with bouncebacks, however MB and MB$_0$ results are compliant with one another for both values of $\tau$. For all the no-slips, the biggest errors occur on the segments of the perimeter where WSS is the highest. In contrast to the pressure profiles, no significant discontinuities are visible at angles $k\pi/4$.

The presence of the scattered nodes does not introduce any new behavior to WSS profiles, apart from the slight noise visible for moment-based schemes. The error of WSS for each no-slip is the largest for the lowest stable $\tau$. The symmetry of the profile with respect to $\theta=\pi/2$ is broken for the lowest viscosity for both bouncebacks. We note again that for $\tau=0.51325$, the setups with NEE, MB< and MB$_0$ went unstable.

\begin{figure}[ht!]
	\centering
	\includegraphics[width=\linewidth]{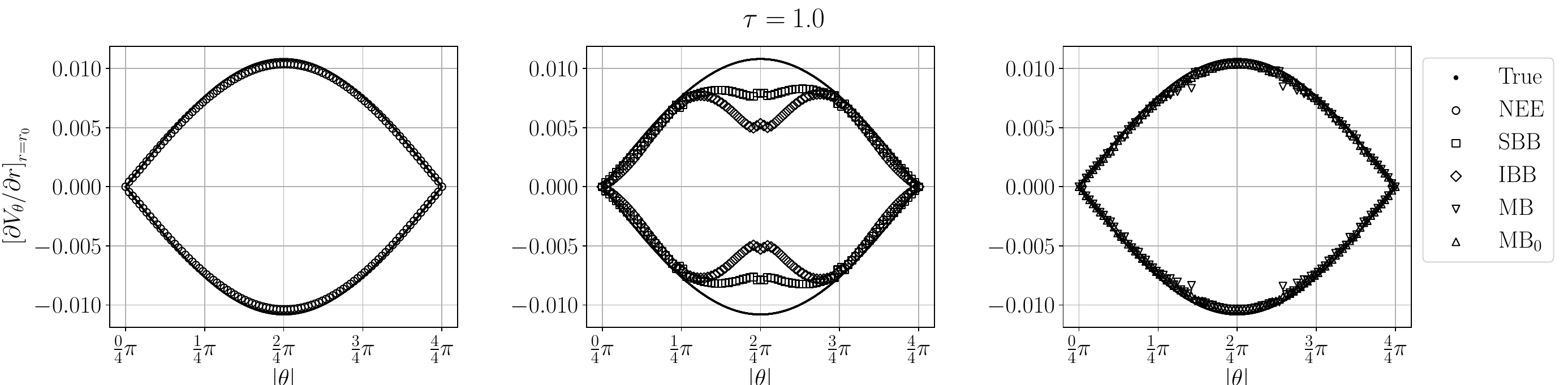}
	\includegraphics[width=\linewidth]{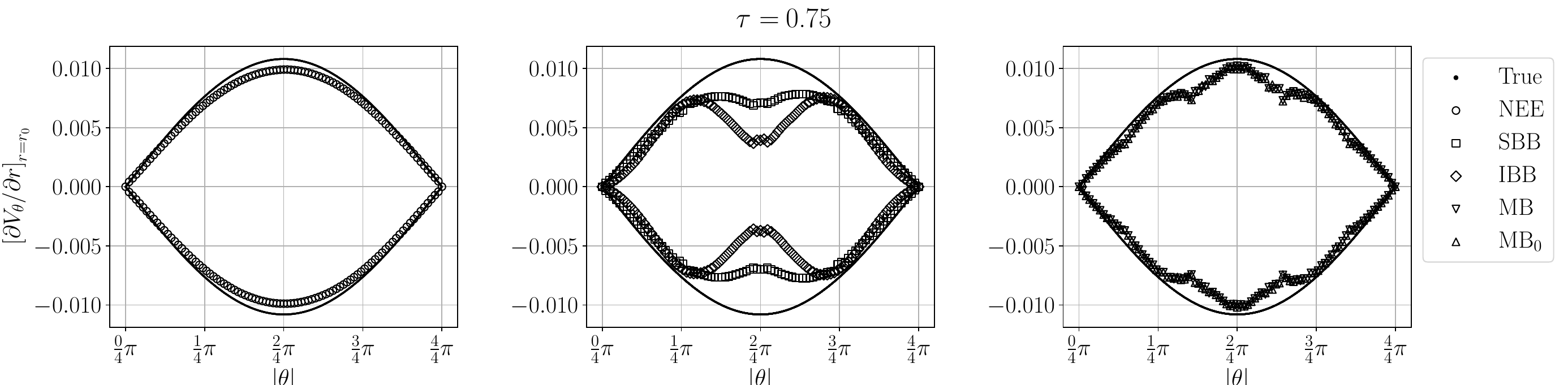}
	\includegraphics[width=0.4\linewidth]{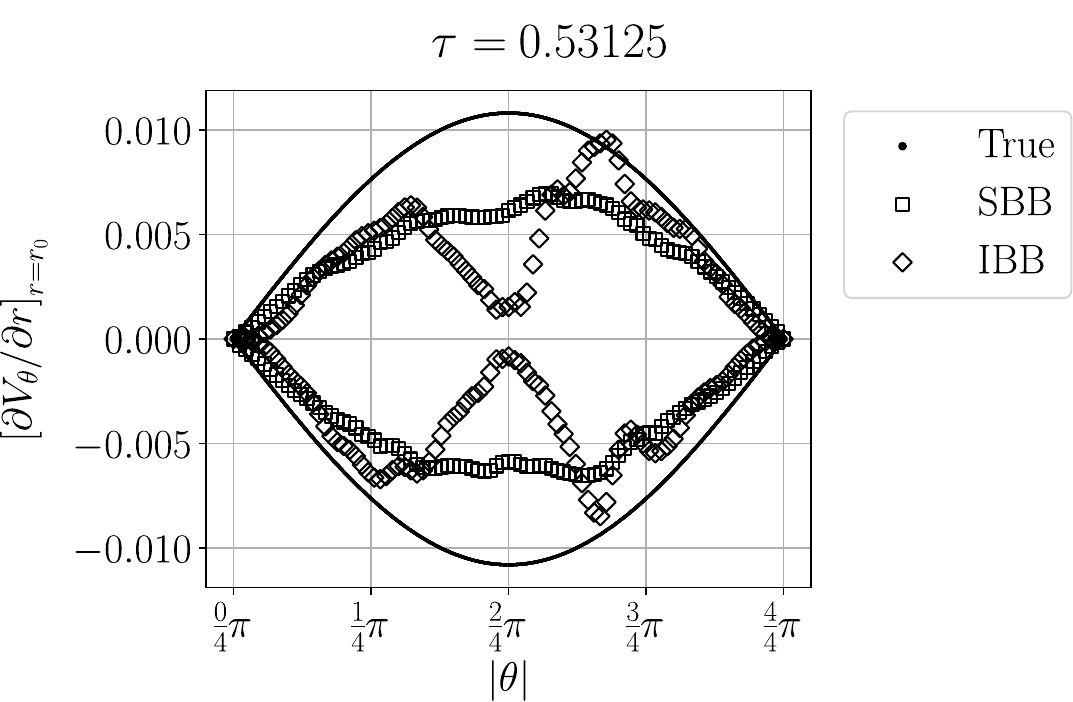}
	\caption{Profiles of the wall shear stress on the cylinder's surface for various no-slip implementations and hybrid discretization for various $\tau$. The subplots with no data points but for the true solution correspond to the diverged cases.}
	\label{fig:wss_allBCs_hybrid}
\end{figure}

\subsubsection{Practical implications -- deformation of an elastic cylinder}

\begin{figure}[ht!]
	\centering
	\includegraphics[height=.6\linewidth]{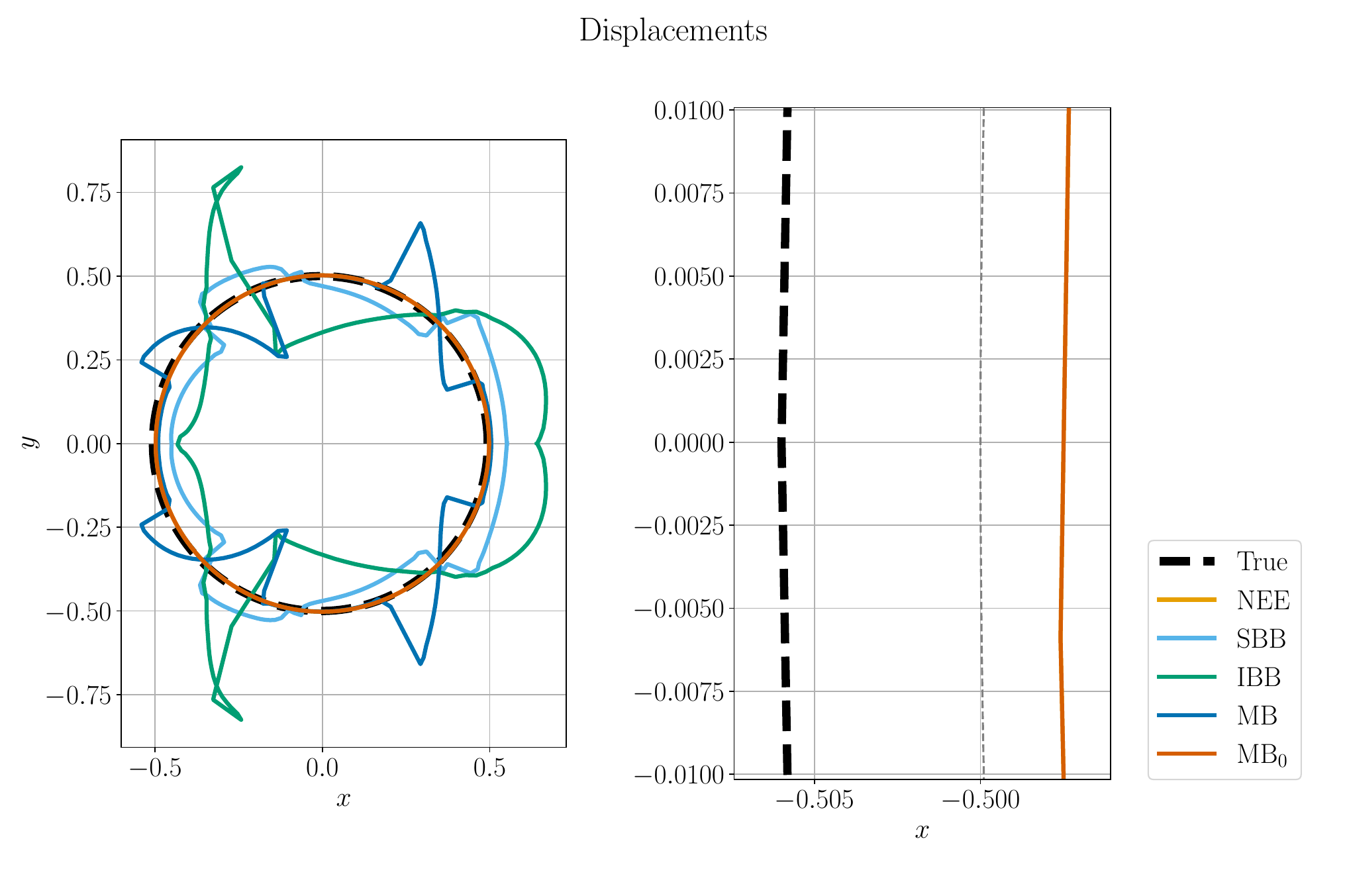}
	\caption{Shape of the deformed spring-mass system consisting of $N_\theta=180$ points, subject to hydrodynamic stresses obtained on regular discretizations, after time $t_{ms}=10^5$. The left subplot shows all the considered systems, while the right subplot shows the magnification near the stagnation point to better discern NEE, MB$_0$, the true solution, and the original cylinder shape (gray dashed line).}
	\label{fig:deformed_cylinder_ogrid}
\end{figure}

The presented results point directly at the possible errors/inconsistencies that simulations of more complex systems performed with the presented method can inherently bear. As the profiles of pressure and WSS on the walls can suffer significantly from the incorrect choice of the no-slip implementation, in studies involving fluid-structure interaction (FSI), such as those concerning deformable porous media~\cite{Azizi2023} or transport of deformable objects in fluid~\cite{Braesel2024}, special care should be given to the proper choice of the no-slip implementation. This is true even though the calculation of WSS in the off-lattice setting is much more straightforward than in the standard LBM~\cite{Matyka2013}. In Fig.~\ref{fig:deformed_cylinder_ogrid} we present the shape of the cylindrical mass-spring system, deformed according to the pressure and wall shear stress on the cylinder obtained numerically with each of the studied no-slips and the stresses calculated from the analytical solution (see Appendix~\ref{app:mass_spring_details} for a detailed description of the problem). One sees that the discrepancies in the shape of the deformed cylinder can be comparable to the radius of the cylinder itself, e.g., in case of IBB or MB. Slightly smaller maximal errors are visible in case of SBB. NEE and MB$_0$ give shapes very similar to each other and to the shape under analytical stresses. This suggests that the actual, fully coupled FSI simulations obtained with various presented no-slip implementations, could give significantly different solutions to the hydrodynamic fields and the obstacle shape.

\section{Discussion}

The results discussed in Sections~\ref{sec:pressure_small_Re}--\ref{sec:wss_small_Re} suggest that in meshfree LBM, when the discretization points can be placed exactly on the boundaries which are not tessellated by a square lattice, the use of certain, classical for LBM, no-slip implementations may not be a good choice. Even though for some setups the values of the obtained drag coefficient are within a few percent relative error with respect to the analytical solution, the relative discrepancy of the hydrodynamic stresses at individual points on the cylinder's surface may be on the order of its maximal value there (e.g., profile of $C_p$ with MB no-slip or WSS profile with IBB on regular discretization at $\tau=1$).

A general tendency of the increase of the solution error and eventually the loss of stability with the decrease of viscosity is a phenomenon known from the standard LBM. In the current meshfree model, however, the loss of stability occurs much sooner than one would expect in setups with an exact streaming. We believe that this observation will be true also for off-grid LBMs using other numerical schemes for advection, however a deeper investigation of that matter is beyond the scope of this work.

The no-slip implementations relying on the identification of the missing populations and overwriting only those populations (i.e., all but the non-equilibrium extrapolation) suffer from the decoupling of space and velocity discretization. The abrupt changes of the sets of known and unknown distributions introduce discontinuities to the pressure and velocity profiles on the cylinder's surface. Their angular positions correspond clearly to the angles where the change of the elements of the known distributions set occurs. This is true even for the MB$_0$. Despite imposing the Neumann condition for the density explicitly and without approximation (due to the copy-pasting of the target density from the neighbor point $\bsym{x}_N$), it fails to recover the correct $C_P$ profile at relatively high viscosity. We expect that also other strategies for the choice of the known/unknown populations at the boundary nodes, e.g., based on the sign of the dot product between the lattice direction and the local wall normal vector, would not solve this problem.

On the other hand, overwriting all populations at the boundary node in the NEE gives much more accurate and stable velocity and pressure profiles on the boundary. The fact that it also calculates the profile of WSS with satisfying accuracy suggests that NEE does not spoil the solution in the non-equilibrium part of VDF since~\cite{LATT2006165}
\begin{equation}
    f_k - f_k^\text{eq} = f_k^\text{neq} \approx f_k^{(1)} \sim
    (\bsym{e}_k\bsym{e}_k - c_s^2\bsym{1}) : \nabla (\rho \bsym{V}),
\end{equation}
where $\bsym{1}$ denotes the unit matrix and colon denotes the tensor contraction. The use of boundary-compliant unstructured discretization, with each boundary node having its closest bulk neighbor $\bsym{x}_N$ placed exactly along the local wall normal vector,  allows to easily use NEE in situations when the boundary has a complex shape, not aligned with the discrete velocity directions. The extension of this boundary condition to three-dimensional cases is trivial, since the only information about the geometry required for NEE is the identification of the neighbor $\bsym{x}_N$, which can be done during the generation of the discretization. This copy-pasting of the density from the bulk should increase the stability of the numerical model, compared to the setup with the interpolation of the density to a `ghost' fluid node, or approximation of the density gradient at the boundaries.
	
The use of a ring of finite width of scattered points around the cylinder introduces noise to the velocity and pressure profiles. Its magnitude depends on the specific no-slip implementation and the relaxation time value, with the pressure profiles obtained with moment-based no-slips suffering the most from this noise. At the same time, the results obtained with bounceback no-slips suggest that having a subset of the domain discretized with scattered nodes may even have a slightly stabilizing effect on the simulation, with respect to the decrease of the viscosity. This phenomenon was observed before in the numerical solution to hyperbolic problems involving RBFFD approximation~\cite{KOLARPOZUN2026}.

\begin{figure}
    \centering
    \includegraphics[width=\linewidth]{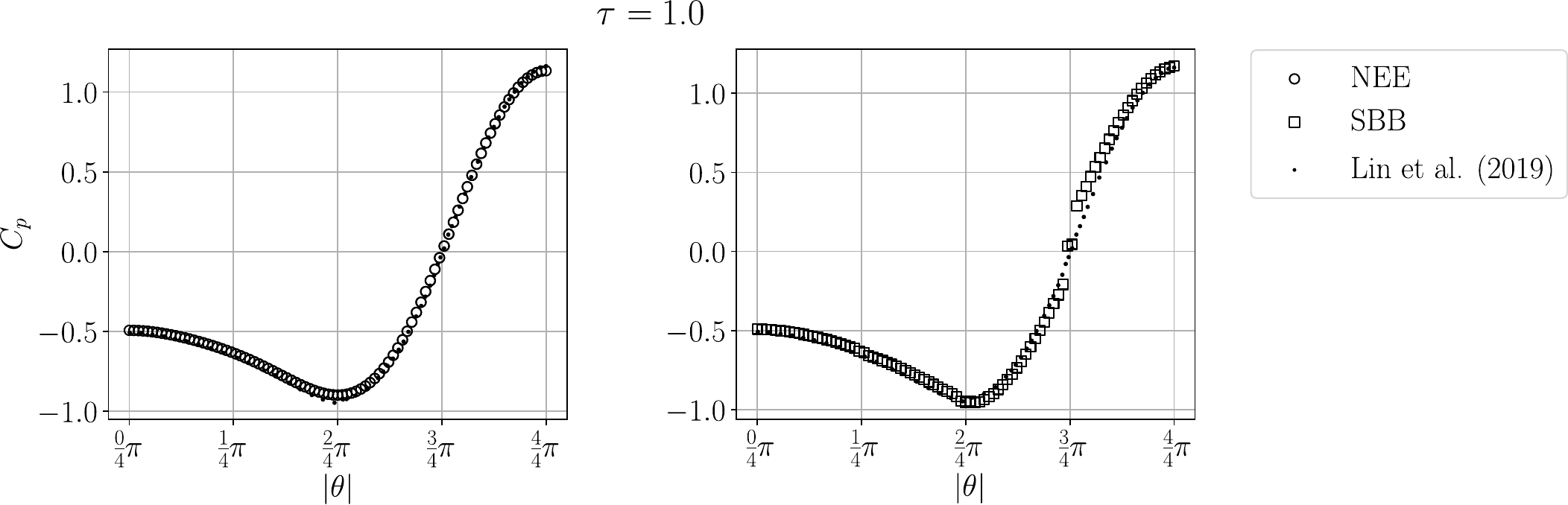}
    \caption{Results using the same NEE and SBB setups on regular discretizations at $N_\theta=180$ and $\tau=1$ as described in the Results of the present work, but for $Re=40$, achieved by increasing the far-field veloctiy. Small filled dots denote the results obtained by Lin and others~\cite{Lin2019}. We note that the values of the relaxation time, far-field velocity, streaming length (denoted as $dt$ in~\cite{Lin2019}), and the space discretization ($N_r$ and $N_\theta$) match the ones used in the referenced work. The reference data are reproduced manually from Fig.~7 of~\cite{Lin2019} using webplotdigitizer v4~\cite{WebPlotDigitizer}.}
    \label{fig:cp_at_higher_Re}
\end{figure}

Finally, the benchmark case for the vanishing Reynolds number, especially the outer boundary condition, was chosen such that it resembles the one used in previous works. For instance, the authors of~\cite{Lin2019} also use RBF-FD and semi-Lagrangian streaming along with SBB. Unfortunately, they consider higher Reynolds numbers, so a direct comparison at $Re \ll 1$ is not possible. Nevertheless, in Fig.~\ref{fig:cp_at_higher_Re}, we plot the results obtained with the current NEE and SBB setups for $Re=40$ on regular discretizations, along with the corresponding ones from~\cite{Lin2019} (cf. Fig.~7 in their work). One sees that NEE follows their values very closely, however our SBB results still exhibit a visible discontinuity at angle $3\pi/4$ and a smaller one at $\pi/2$ -- both smaller than the ones encountered in $Re \ll 1$ regime, cf. Fig.~\ref{fig:cp_allBCs_ogrid}. This suggests that the problems with high errors of the pressure field on the cylinder might be inherently linked to the flows in Stokes regime. Authors of~\cite{He1997} use Lagrange interpolation on a polar grid, along with SBB on the cylinder and imposing equilibrium distributions at the outer boundary. Again, an intriguing question is whether pressure discontinuities at $Re \ll 1$ would be visible in their results. Apart from that, none of those two works report problems with standing waves entering the domain from the outer boundary. Nevertheless, since NEE managed to obtain the results compliant with the analytical solution of the Stokes equation, we conclude that such setup can be used to highlight challenges faced by various implementations of the no-slip boundary condition. It can also provide stable solutions with satisfying accuracy, measured either with respect to the analytical solution (as in the present work) or compared with those obtained experimentally or using other numerical methods (as, e.g., in the works discussed at the beginning of this paragraph), but may not be suitable for a rigorous convergence study in the whole domain due to the standing waves propagating from the outer boundary (see right subplot of Fig.~\ref{fig:hydro_map} and the convergence plots of the pressure in the Supplementary Materials).

\section{Conclusions}

In this work we have implemented and benchmarked five realizations of the no-slip boundary condition in the meshfree framework of meshfree LBM -- the non-equilibrium extrapolation, simple and interpolated bounceback, and moment-based boundary with and without the novel correction of the stagnant population. In the model test case -- the unbounded flow around a cylinder at $Re \ll 1$ -- we investigated the profiles of pressure, one of the velocity components and tangential velocity gradient on the solid wall, as well as the total drag force acting on the cylinder. We found that non-equilibrium extrapolation performs the best compared to the analytical solution, while other no-slips suffer from errors in the profiles, including discontinuities. Finally, we applied the stresses obtained with each no-slip to an elastic cylinder and allowed it to deform. We have found that the errors in the stress field have profound impact on the shape of the deformed cylinder. The presented results will allow for a more sensible choice of the implementations of the no-slip boundary condition in off-grid LBM and can be used as a starting point for the works on improvement of irregular boundaries handling in standard LBM augmented with body-fitted discretizations.

\section{Acknowledgments}
Funded by National Science Centre, Poland grant no. 2024/53/N/ST6/04212 and the Slovenian Research and Innovation Agency (ARIS) research core funding No. P2-0095. For the purpose of Open Access, the author has applied a CC-BY public copyright license to any Author Accepted Manuscript (AAM) version arising from this submission.

\appendix

\section{Analytical solution of the considered Stokes problem}\label{sec:stokes_analytical_solution}

\newcommand{\ez}[0]{\hat{\bsym{e}}_z}
\newcommand{\er}[0]{\hat{\bsym{e}}_r}
\newcommand{\et}[0]{\hat{\bsym{e}}_\theta}
\newcommand{\pt}[0]{\partial_\theta}
\newcommand{\pr}[0]{\partial_r}
\newcommand{\prr}[0]{\partial_{rr}}
\newcommand{\prt}[0]{\partial_{r\theta}}
\newcommand{\ptt}[0]{\partial_{\theta\theta}}
\newcommand{\vr}[0]{V_r}
\newcommand{\vt}[0]{V_\theta}

Taking the curl of the momentum part of the Stokes equation, Eq.~\eqref{eq:stokes_eq}, and noting that the curl of the potential field is alway zero, i.e., $\nabla \times \nabla P \equiv 0$, one arrives at the vorticity transport equation, $(\nabla^2\bsym{\omega})\cdot\ez=0$, which can eventually be transformed into a biharmonic equation for the streamfunction $\psi$
\begin{equation}\label{eq:biharmonic}
    \nabla \times (\nabla^2\bsym{V}) \cdot\ez =
    \nabla^2(\nabla \times \bsym{V}) \cdot\ez =
    \nabla^4\psi = 0.
\end{equation}
The unit vector $\ez$ is normal to the problem plane. In polar coordinates, $\psi \equiv \psi(r,\theta)$, the velocity can be written in terms of the derivatives of the streamfunction as
\begin{equation}
\renewcommand{\arraystretch}{2.5}
    \bsym{V} =
    \er V_r+
    \et V_\theta =
    \er\dfrac{1}{r}\dfrac{\partial \psi}{\partial \theta}+
    -\et\dfrac{\partial \psi}{\partial r}.
\end{equation}
where $\er$ and $\et$ denote the unit vectors in polar coordinates and $V_r$ and $V_\theta$ denote the corresponding components of the velocity. We assume $\ez = \er \times \et$. One can consider the streamfunction in the form of $\psi(r,\theta)=R(r)\Theta(\theta)$. In such case, the most general solution to biharmonic equation is
\begin{equation}\label{eq:separated_components}
    \begin{array}{rcl}
        R_\lambda(r) & = &
        A r^{2+\lambda} +
        B r^{2-\lambda} +
        C r^\lambda +
        D r^{-\lambda} \\
        \Theta_\lambda(\theta) & = &
        E_0 \sin{\lambda\theta} + 
        E_1 \cos{\lambda\theta}
    \end{array}
\end{equation}
where $\lambda \in \mathbb{R}$, since we do not want oscillatory behavior of $\psi$ along $r$. As the boundary condition at the outer radius is $\psi(r_\text{out},\theta)=V_\infty r \sin{\theta}$, only $\lambda=1$ is the valid wavenumber. Once $R$ and $\Theta$ with such chosen $\lambda$ are inserted in to the biharmonic equation, we note that $\Theta = -\Theta'' = \Theta^{(4)}$ and the characteristic equation for the radial component has a double root at $\lambda=1$ (cf. terms preceded by $B$ and $C$ in Eq.~\eqref{eq:separated_components}). Thus, the corresponding solution must be multiplied by $\ln{r}$ to keep the solution basis non-degenerate. Finally, after combining the coefficient $E_0$ into the coefficients preceding each radial term, we arrive at the following form of the solution
\begin{equation}\label{eq:biharmonic_general_solution}
    \begin{array}{rcl}
        R(r) & = & Ar^3 + Br\ln{r} + Cr + Dr^{-1}, \\
        \Theta(\theta) & = & \sin{\theta}.
    \end{array}
\end{equation}
We note that due to the finite radius of the outer boundary, $r_\text{out}$, the logarithm and the cube of $r$ do not cause the unbounded growth of the velocity (the so-called Stokes paradox). The no-slip boundary at the cylinder's surface requires
\begin{equation}\label{eq:biharmonic_bc_0}
    \renewcommand{\arraystretch}{1.5}
    \begin{array}{rcl}
        V_r(r_\text{in}) &=& \cos{\theta}\left(Ar_\text{in}^2 + B\ln{r_\text{in}} + C + D/r_\text{in}^2\right)=0\\
        V_\theta(r_\text{in}) &=& -\sin{\theta}\left(3Ar_\text{in}^2 + B(1+\ln{r_\text{in}}) + C - D/r_\text{in}^2\right)=0
    \end{array}
\end{equation}
while imposing a constant horizontal velocity $\bsym{V}_\infty = V_\infty\bsym{e}_x = V_\infty\cos{\theta}\er + V_\infty\sin{\theta}\et$ at the outer boundary gives
\begin{equation}\label{eq:biharmonic_bc_1}
    \renewcommand{\arraystretch}{1.5}
    \begin{array}{rcl}
        V_r(r_\text{out}) &=& \cos{\theta}\left(Ar_\text{out}^2 + B\ln{r_\text{out}} + C + D/r_\text{out}^2\right)=V_\infty \cos{\theta}\\
        V_\theta(r_\text{out}) &=& -\sin{\theta}\left(3Ar_\text{out}^2 + B(1+\ln{r_\text{out}}) + C - D/r_\text{out}^2\right)=-\dfrac{1}{r_\text{out}}V_\infty \sin{\theta}.
    \end{array}
\end{equation}
The solution of the linear system of equations defined by Eqs.~\eqref{eq:biharmonic_bc_0} and \eqref{eq:biharmonic_bc_1} gives the desired values of the constants $A$, $B$, $C$, and $D$, and thus the particular form of the velocity field. In this work, we use Julia programming language~\cite{Julia2017} to find the solution of the linear system, using its left-division operator (\textbackslash). The pressure can be now calculated by solving the momentum equation in polar coordinates
\begin{equation}
    \nabla P = \mu \nabla^2(\er V_r + \et V_\theta).
\end{equation}
Using the polar coordinates form of the gradient
\begin{equation}
     \nabla P = \left(\er \partial_r + \et \dfrac{1}{r}\partial_\theta\right)P
\end{equation}
and the vector Laplace operator
\begin{equation}
    \begin{aligned}
         \nabla^2\bsym{V} = \nabla(\nabla \cdot \bsym{V}) - \nabla \times (\nabla \times \bsym{V}) =
         \er\left(\prr\vr + \dfrac{1}{r^2}\ptt\vr + \dfrac{1}{r}\pr\vr - \dfrac{2}{r^2}\pt\vt - \dfrac{1}{r^2}\vr\right) +\\
         \et\left(\prr\vt + \dfrac{1}{r^2}\ptt\vt + \dfrac{1}{r}\pr\vt + \dfrac{2}{r^2}\pt\vr - \dfrac{1}{r^2}\vt\right)
     \end{aligned}
\end{equation}
one arrives at the partial derivatives of the pressure
\begin{subequations}
    \begin{align}
        \partial_rP &= \mu\cos{\theta}\left(8A + 2B/r^2\right)\label{eq:pressure_gradient_radial} \\
        \partial_\theta P &= \mu\sin{\theta}\left(-8Ar + 3B/r\right)=\mu\sin{\theta}g(r)\label{eq:pressure_gradient_angular}.
    \end{align}
\end{subequations}
To get the pressure at point $[r,\theta]$ we first integrate Eq.~\eqref{eq:pressure_gradient_angular} w.r.t the angular coordinate
\begin{equation}\label{eq:pressure_unknown_r_const}
    P = -\mu\cos{\theta}g(r) + \alpha(r).
\end{equation}
We next differentiate Eq.~\eqref{eq:pressure_unknown_r_const} w.r.t. to the radial coordinate to get
\begin{equation}
    \partial_r P = -\mu\cos{\theta}g'(r) + \alpha'(r)=
    \mu\cos{\theta}\left(8A + 3B/r^2\right) + \alpha'(r)
\end{equation}
and use it along with Eq.~\eqref{eq:pressure_gradient_radial} to obtain $\alpha'(r)$
\begin{equation}
    \alpha'(r) = -\mu\cos{\theta}B/r^2
\end{equation}
which yields
\begin{equation}
    \alpha(r) = \mu\cos{\theta}B/r + \alpha_0.
\end{equation}
Next, we substitute $\alpha(r)$ back into Eq.~\eqref{eq:pressure_unknown_r_const} to obtain the explicit formula for the pressure shown in Eq.~\eqref{eq:stokes_analytical}
\begin{equation}\label{eq:pressure_final_almost}
    P(r,\theta) = \mu\cos{\theta}\left(8Ar-2B/r\right) + \alpha_0.
\end{equation}
The constant $\alpha_0$ is set such that there is a constant pressure $P_\infty$ on the outer boundary, $P(r_\text{out},\theta) = P_\infty$, i.e.
\begin{equation}
    \alpha_0 = P_\infty - \mu\cos{\theta}\left(8Ar_\text{out} - 2B/r_\text{out}\right).
\end{equation}
This from of $\alpha_0$, inserted back to Eq.~\eqref{eq:pressure_final_almost}, after transforming the difference of $r$- and $r_\text{out}$- related terms into the integral from Eq.~\eqref{eq:pressure_final}, gives the pressure in the form presented therein.

\section{Details of the mass-spring system deformation problem}\label{app:mass_spring_details}

\begin{figure}[ht!]
	\centering
	\includegraphics[width=.3\linewidth]{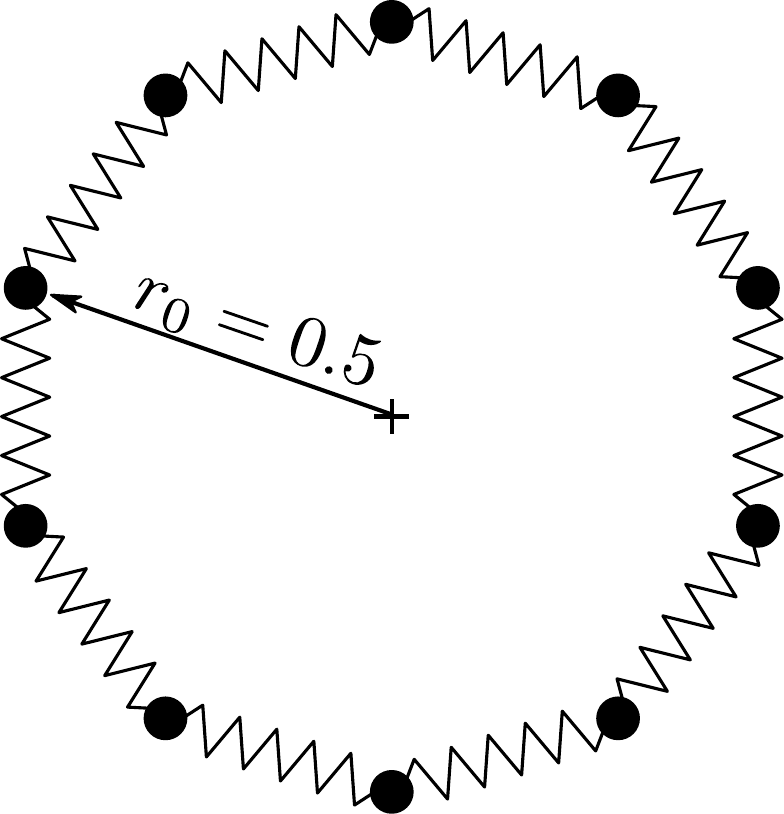}
	\caption{Schematic drawing of the considered mass and spring system. The two gray dots in the center of the polygon are the immobile nodes to which each of the perimeter nodes (black dots) is connected using a spring (shown for only one node for clarity). In the actual model we use $N_\theta=180$ nodes on the perimeter of the cylinder, here a smaller number is shown for clarity.}
	\label{fig:spring_mass_scheme}
\end{figure}

The considered mass spring system (see Fig.~\ref{fig:spring_mass_scheme}) is inspired by the reduced model of a droplet presented in~\cite{Matyka2022}. It consists of $N_\theta=180$ nodes $\bsym{x}_{sm,i}$ lying on the cylinder's surface, taken directly from the discretization used in the off-grid LBM simulations. Each node is of unit mass, i.e., $m=1$ and is connected with a linear spring to its two closest neighbors. All the springs in the system have stiffness $k=10^{-10}$ and their neutral lengths are their starting lengths, $L_{0,i}=2\pi r_\text{in}/N_\theta$. We introduce internal pressure dependent on the volume of the system via a simple equation of state
\begin{equation}
    P = \gamma(\mathcal{V}-\mathcal{V}_0)
\end{equation}
where $\mathcal{V}$ and $\mathcal{V}_0$ are the current and the initial (in the undeformed state) volume of the system. We set the proportionality constant $\gamma=10^{-6}$. The force from the internal pressure acting on node $\bsym{x}_{i,sm}$ is the mean of the pressure forces acting on the two edges it belongs to
\begin{equation}
    \bsym{F}_{P,i} = \dfrac{1}{2}P \Delta S\sum\limits_{j\in\mathcal{N}_i}\bsym{n}_{ij}
\end{equation}
where $\mathcal{N}_i$ is the set of indices of all the nodes that $\bsym{x}_{i,sm}$ is connected to and $\Delta S$ is the length of the surface element associated with each perimeter node. At each of the perimeter node we apply a force $\bsym{F}_{fl,i}$ coming from the hydrodynamic stresses at this node in the undeformed state
\begin{equation}
    \bsym{F}_{fl,i} = L_{0,i}\left(-\bsym{n}P + \bsym{n}\nabla\bsym{V}\right)_{
        \begin{subarray}{l}\bsym{x}=\bsym{x}_{sm,i}\\t=0\end{subarray}
    }.
\end{equation}
To mitigate infinite oscillations in the system we introduce also the damping force at each node, proportional to the node's velocity, i.e., $\bsym{F}_{fric,i}=-c\dot{\bsym{x}}_{sm,i}$ with the damping constant $c=10^{-4}$. We integrate the equations of motion of such system, i.e.
\begin{equation}
    m\ddot{\bsym{x}}_{sm,i} = \bsym{F}_{fl,i} + \displaystyle\sum\limits_{j\in\mathcal{N}_i}\bsym{F}_{k;ij}+
    \bsym{F}_{P,i}+
    \bsym{F}_{fric,i}
\end{equation}
where
\begin{equation}
    \bsym{F}_{k;ij} = k\left(1-\dfrac{L_{0,i}}{|\bsym{x}_{sm,ij}|}\right)\bsym{x}_{sm,ij}, \quad \bsym{x}_{sm,ij} =: \bsym{x}_{sm,i} - \bsym{x}_{sm,j}
\end{equation}
is the force acting on $\bsym{x}_{sm,i}$ form the spring connecting it to $\bsym{x}_{sm,i}$. Such defined system without any grounded points will float in space subjected to non-zero net force. Thus, in each timestep we calculate the center of mass of the system and extract it from the positions of each of the points, which is equivalent to the change of the frame of reference. We use explicit Euler method for time integration with the timestep $\Delta t_{ms}=1$ and we iterate the solution for $5\cdot 10^5$ timesteps.

\bibliography{sample}

@article{Nannelli1992,
  title = {The Lattice {{Boltzmann}} Equation on Irregular Lattices},
  author = {Nannelli, Francesca and Succi, Sauro},
  year = {1992},
  month = aug,
  journal = {Journal of Statistical Physics},
  volume = {68},
  number = {3-4},
  pages = {401--407},
  issn = {0022-4715, 1572-9613},
  doi = {10.1007/BF01341755},
  url = {http://link.springer.com/10.1007/BF01341755},
  urldate = {2023-11-20},
  langid = {english}
}

@article{Akkurt2022,
  title={An efficient edge based data structure for the compressible Reynolds-averaged Navier--Stokes equations on hybrid unstructured meshes},
  author={Akkurt, Semih and Sahin, Mehmet},
  journal={International Journal for Numerical Methods in Fluids},
  volume={94},
  number={1},
  pages={13--31},
  year={2022},
  publisher={Wiley Online Library}
}

@article{CORNUBERT1991241,
title = {A Knudsen layer theory for lattice gases},
journal = {Physica D: Nonlinear Phenomena},
volume = {47},
number = {1},
pages = {241-259},
year = {1991},
issn = {0167-2789},
doi = {https://doi.org/10.1016/0167-2789(91)90295-K},
url = {https://www.sciencedirect.com/science/article/pii/016727899190295K},
author = {R. Cornubert and D. d'Humières and D. Levermore}
}

@article{PM2023108074,
title = {Sensitivity analysis of geometric parameters on the aerodynamic performance of a multi-element airfoil},
journal = {Aerospace Science and Technology},
volume = {132},
pages = {108074},
year = {2023},
issn = {1270-9638},
doi = {https://doi.org/10.1016/j.ast.2022.108074},
url = {https://www.sciencedirect.com/science/article/pii/S1270963822007489},
author = {Mohamed Abubacker Siddique P.M. and L. Prince Raj}
}

@Article{Molins2021,
author={Molins, Sergi
and Soulaine, Cyprien
and Prasianakis, Nikolaos I.
and Abbasi, Aida
and Poncet, Philippe
and Ladd, Anthony J. C.
and Starchenko, Vitalii
and Roman, Sophie
and Trebotich, David
and Tchelepi, Hamdi A.
and Steefel, Carl I.},
title={Simulation of mineral dissolution at the pore scale with evolving fluid-solid interfaces: review of approaches and benchmark problem set},
journal={Computational Geosciences},
year={2021},
month={Aug},
day={01},
volume={25},
number={4},
pages={1285-1318},
issn={1573-1499},
doi={10.1007/s10596-019-09903-x},
url={https://doi.org/10.1007/s10596-019-09903-x}
}

@incollection{MALVE2019,
title = {Chapter 5 - Impact of the Fluid-Structure Interaction Modeling on the Human Vessel Hemodynamics},
editor = {Mohamed H. Doweidar},
booktitle = {Advances in Biomechanics and Tissue Regeneration},
publisher = {Academic Press},
pages = {79-93},
year = {2019},
isbn = {978-0-12-816390-0},
doi = {https://doi.org/10.1016/B978-0-12-816390-0.00005-4},
url = {https://www.sciencedirect.com/science/article/pii/B9780128163900000054},
author = {Mauro Malvè and Myriam Cilla and Estefanía Peña and Miguel Angel Martínez}
}

@article{Naqvi2026,
author = {Naqvi, Sahrish B. and Śnieżek, Damian and Strzelczyk, Dawid and Mądrala, Mariusz and Matyka, Maciej},
title = {Inertial Effects on Fluid Flow Through Natural Porous Media},
journal = {Water Resources Research},
volume = {62},
number = {1},
pages = {e2025WR041274},
doi = {https://doi.org/10.1029/2025WR041274},
url = {https://agupubs.onlinelibrary.wiley.com/doi/abs/10.1029/2025WR041274},
eprint = {https://agupubs.onlinelibrary.wiley.com/doi/pdf/10.1029/2025WR041274},
note = {e2025WR041274 2025WR041274},
year = {2026}
}

@article{Zadehgol2014a,
  title = {A Nodal Discontinuous {{Galerkin}} Lattice {{Boltzmann}} Method for Fluid Flow Problems},
  author = {Zadehgol, A. and Ashrafizaadeh, M. and Musavi, S.H.},
  year = {2014},
  month = dec,
  journal = {Computers \& Fluids},
  volume = {105},
  pages = {58--65},
  issn = {00457930},
  doi = {10.1016/j.compfluid.2014.09.015},
  url = {https://linkinghub.elsevier.com/retrieve/pii/S0045793014003521},
  urldate = {2023-11-20},
  langid = {english}
}

@article{Shan2006,
	title = {Kinetic Theory Representation of Hydrodynamics: A Way beyond the {{Navier}}--{{Stokes}} Equation},
	shorttitle = {Kinetic Theory Representation of Hydrodynamics},
	author = {Shan, Xiaowen and Yuan, Xue-Feng and Chen, Hudong},
	year = {2006},
	month = feb,
	journal = {Journal of Fluid Mechanics},
	volume = {550},
	number = {-1},
	pages = {413},
	issn = {0022-1120, 1469-7645},
	doi = {10.1017/S0022112005008153},
	url = {http://www.journals.cambridge.org/abstract_S0022112005008153},
	urldate = {2023-04-21},
	langid = {english}
}

@book{Succi2018,
	title = {The {{Lattice Boltzmann Equation}}: {{For Complex States}} of {{Flowing Matter}}},
	author = {Succi, S},
	year = {2018},
	publisher = {Oxford University Press},
	doi = {10.1093/oso/9780199592357.001.0001}
}

@book{Liu2005,
	title = {An {{Introduction}} to {{Meshfree Methods}} and {{Their Programming}}},
	author = {Liu, G. R. and Gu, Y. T.},
	year = {2005},
	publisher = {Springer-Verlag},
	address = {Berlin/Heidelberg},
	doi = {10.1007/1-4020-3468-7},
	isbn = {978-1-4020-3228-8}
}

@article{Rot2024,
	title = {Spatially Dependent Node Regularity in Meshless Approximation of Partial Differential Equations},
	author = {Rot, Miha and Jan{\v c}i{\v c}, Mitja and Kosec, Gregor},
	year = {2024},
	month = jul,
	journal = {Journal of Computational Science},
	volume = {79},
	pages = {102306},
	issn = {18777503},
	doi = {10.1016/j.jocs.2024.102306},
	url = {https://linkinghub.elsevier.com/retrieve/pii/S1877750324000991},
	urldate = {2024-05-21},
	langid = {english}
}

@article{Javed2013,
	title = {A {{Hybrid Mesh Free Local RBF- Cartesian FD Scheme}} for {{Incompressible Flow}} around {{Solid Bodies}}},
	author = {Javed, A and Djidjeli, K and Xing, J T and Cox, S J},
	year = {2013},
	volume = {7},
	number = {10},
	langid = {english}
}

@article{Ding2004,
	title = {Simulation of Incompressible Viscous Flows Past a Circular Cylinder by Hybrid {{FD}} Scheme and Meshless Least Square-Based Finite Difference Method},
	author = {Ding, H. and Shu, C. and Yeo, K.S. and Xu, D.},
	year = {2004},
	month = mar,
	journal = {Computer Methods in Applied Mechanics and Engineering},
	volume = {193},
	number = {9-11},
	pages = {727--744},
	issn = {00457825},
	doi = {10.1016/j.cma.2003.11.002},
	url = {https://linkinghub.elsevier.com/retrieve/pii/S0045782503005838},
	urldate = {2024-05-21},
	copyright = {https://www.elsevier.com/tdm/userlicense/1.0/},
	langid = {english}
}

@software{WebPlotDigitizer,
    author = {Ankit Rohatgi},
    title = {WebPlotDigitizer},
    url = {https://automeris.io},
    version = {4},
}

@article{Lyu2023,
    author = {Lyu, Changhao and Liu, Peiqing and Hu, Tianxiang and Geng, Xin and Qu, Qiuling and Sun, Tao and Akkermans, Rinie A. D.},
    title = {Hybrid method for wall local refinement in lattice Boltzmann method simulation},
    journal = {Physics of Fluids},
    volume = {35},
    number = {1},
    pages = {017103},
    year = {2023},
    month = {01},
    issn = {1070-6631},
    doi = {10.1063/5.0130467},
    url = {https://doi.org/10.1063/5.0130467},
    eprint = {https://pubs.aip.org/aip/pof/article-pdf/doi/10.1063/5.0130467/16675962/017103_1_online.pdf},
}

@article{Zamolo2019,
author = {R. Zamolo and E. Nobile},
title = {Solution of incompressible fluid flow problems with heat transfer by means of an efficient RBF-FD meshless approach},
journal = {Numerical Heat Transfer, Part B: Fundamentals},
volume = {75},
number = {1},
pages = {19--42},
year = {2019},
publisher = {Taylor \& Francis},
doi = {10.1080/10407790.2019.1580048},
URL = {https://doi.org/10.1080/10407790.2019.1580048},
eprint = { https://doi.org/10.1080/10407790.2019.1580048
}
}

@article{KOLARPOZUN2026,
title = {An RBF-based method for computational electromagnetics with reduced numerical dispersion},
journal = {Computer Methods in Applied Mechanics and Engineering},
volume = {454},
pages = {118865},
year = {2026},
issn = {0045-7825},
doi = {https://doi.org/10.1016/j.cma.2026.118865},
url = {https://www.sciencedirect.com/science/article/pii/S0045782526001398},
author = {Andrej Kolar-Požun and Gregor Kosec}
}

@online{Strzelczyk2024a,
  title = {On H-Refined Meshless Solution to {{Navier-Stokes}} Problem in Porous Media: Comparing Meshless {{Lattice Boltzman Method}} with {{ACM RBF-FD}} Approach},
  shorttitle = {On H-Refined Meshless Solution to {{Navier-Stokes}} Problem in Porous Media},
  author = {Strzelczyk, Dawid and Rot, Miha and Kosec, Gregor and Matyka, Maciej},
  date = {2024-04-22},
  eprint = {2404.14195},
  eprinttype = {arXiv},
  eprintclass = {physics},
  url = {http://arxiv.org/abs/2404.14195},
  urldate = {2024-05-25},
  langid = {english},
  pubstate = {prepublished}
}

@article{Pribec2021,
  title = {A {{Strong-Form Off-Lattice Boltzmann Method}} for {{Irregular Point Clouds}}},
  author = {Pribec, Ivan and Becker, Thomas and Fattahi, Ehsan},
  date = {2021-09},
  journaltitle = {Symmetry},
  volume = {13},
  number = {10},
  pages = {1802},
  issn = {2073-8994},
  doi = {10.3390/sym13101802},
  url = {https://www.mdpi.com/2073-8994/13/10/1802},
  urldate = {2023-04-21},
  langid = {english}
}

@article{Maidenberg2026,
  title = {A {{Meshless Discrete Boltzmann Solver With}} the {{Radial Basis Function Finite Difference Scheme}} for {{Fluid Flows With Dirichlet Boundary Conditions}}},
  author = {Maidenberg, Amandine R. and Chen, Leitao},
  date = {2026-02-01},
  journaltitle = {Journal of Fluids Engineering},
  volume = {148},
  number = {2},
  pages = {021501},
  issn = {0098-2202, 1528-901X},
  doi = {10.1115/1.4069638},
  url = {https://asmedigitalcollection.asme.org/fluidsengineering/article/148/2/021501/1221895/A-Meshless-Discrete-Boltzmann-Solver-With-the},
  urldate = {2026-09-15},
  langid = {english}
}

@article{Hu2024,
  title = {A Modified Lattice {{Boltzmann}} Approach Based on Radial Basis Function Approximation for the Non‐uniform Rectangular Mesh},
  author = {Hu, X. and Bergadà, J. M. and Li, D. and Sang, W. M. and An, B.},
  date = {2024-11},
  journaltitle = {International Journal for Numerical Methods in Fluids},
  shortjournal = {Numerical Methods in Fluids},
  volume = {96},
  number = {11},
  pages = {1695--1714},
  issn = {0271-2091, 1097-0363},
  doi = {10.1002/fld.5318},
  url = {https://onlinelibrary.wiley.com/doi/10.1002/fld.5318},
  urldate = {2026-09-15},
  langid = {english}
}

@article{Skordos1993,
  title = {Initial and Boundary Conditions for the Lattice {{Boltzmann}} Method},
  author = {Skordos, P. A.},
  date = {1993-12-01},
  journaltitle = {Physical Review E},
  shortjournal = {Phys. Rev. E},
  volume = {48},
  number = {6},
  pages = {4823--4842},
  issn = {1063-651X, 1095-3787},
  doi = {10.1103/PhysRevE.48.4823},
  url = {https://link.aps.org/doi/10.1103/PhysRevE.48.4823},
  urldate = {2024-05-26},
  langid = {english}
}

@thesis{Latt2007,
  type = {phdthesis},
  title = {Hydrodynamic {{Limit}} of {{Lattice Boltzmann Equations}}},
  author = {Latt, Jonas},
  date = {2007},
  doi = {10.13097/archive-ouverte/unige:464}
}

@article{Lee2003,
	title = {An {{Eulerian Description}} of the {{Streaming Process}} in the {{Lattice Boltzmann Equation}}},
	author = {Lee, Taehun and Lin, Ching Long},
	year = {2003},
	journal = {Journal of Computational Physics},
	volume = {185},
	number = {2},
	pages = {445--471},
	issn = {00219991},
	doi = {10.1016/S0021-9991(02)00065-7}
}

@article{Pan06,
	title={An evaluation of lattice Boltzmann schemes for porous medium flow simulation},
	author={Pan, Chongxun and Luo, Li-Shi and Miller, Cass T},
	journal={Computers \& fluids},
	volume={35},
	number={8-9},
	pages={898--909},
	year={2006},
	publisher={Elsevier}
}

@article{Min2011,
  title = {A Spectral-Element Discontinuous {{Galerkin}} Lattice {{Boltzmann}} Method for Nearly Incompressible Flows},
  author = {Min, Misun and Lee, Taehun},
  year = {2011},
  month = jan,
  journal = {Journal of Computational Physics},
  volume = {230},
  number = {1},
  pages = {245--259},
  issn = {00219991},
  doi = {10.1016/j.jcp.2010.09.024},
  url = {https://linkinghub.elsevier.com/retrieve/pii/S0021999110005279},
  urldate = {2023-11-20},
  langid = {english}
}

@book{Kruger2017,
	title = {The {{Lattice Boltzmann Method}}: {{Principles}} and {{Practice}}},
	shorttitle = {The {{Lattice Boltzmann Method}}},
	author = {Kr{\"u}ger, Timm and Kusumaatmaja, Halim and Kuzmin, Alexandr and Shardt, Orest and Silva, Goncalo and Viggen, Erlend Magnus},
	year = {2017},
	series = {Graduate {{Texts}} in {{Physics}}},
	publisher = {{Springer International Publishing}},
	address = {{Cham}},
	doi = {10.1007/978-3-319-44649-3},
	isbn = {978-3-319-44647-9 978-3-319-44649-3}
}

@Article{Strzelczyk2025,
author={Strzelczyk, Dawid
and Rot, Miha
and Kosec, Gregor
and Matyka, Maciej},
title={Cross-validation of meshless Navier--Stokes solvers in porous media flows},
journal={Scientific Reports},
year={2025},
month={Sep},
day={30},
volume={15},
number={1},
pages={33837},
issn={2045-2322},
doi={10.1038/s41598-025-05272-x},
url={https://doi.org/10.1038/s41598-025-05272-x}
}

@article{Strzelczyk2024,
	title = {Study of the Convergence of the {{Meshless Lattice Boltzmann Method}} in {{Taylor}}\textendash{{Green}}, Annular Channel and a Porous Medium Flows},
	author = {Strzelczyk, Dawid and Matyka, Maciej},
	year = {2024},
	month = jan,
	journal = {Computers and Fluids},
	volume = {269},
	pages = {106122},
	issn = {00457930},
	doi = {10.1016/j.compfluid.2023.106122},
	url = {https://linkinghub.elsevier.com/retrieve/pii/S004579302300347X},
	urldate = {2023-11-22},
	langid = {english}
}

@article{Bhatnagar1954,
	title = {A {{Model}} for {{Collision Processes}} in {{Gases}}. {{I}}. {{Small Amplitude Processes}} in {{Charged}} and {{Neutral One-Component Systems}}},
	author = {Bhatnagar, P L and Gross, E P and Krook, M},
	year = {1954},
	month = may,
	journal = {Phys. Rev.},
	volume = {94},
	number = {3},
	pages = {511--525},
	publisher = {{American Physical Society}},
	doi = {10.1103/PhysRev.94.511}
}

@article{Julia2017,
    title={Julia: A fresh approach to numerical computing},
    author={Bezanson, Jeff and Edelman, Alan and Karpinski, Stefan and Shah, Viral B},
    journal={SIAM {R}eview},
    volume={59},
    number={1},
    pages={65--98},
    year={2017},
    publisher={SIAM},
    doi={10.1137/141000671},
    url={https://epubs.siam.org/doi/10.1137/141000671}
}

@article{Guo2002_nee,
author = {Guo, Zhaoli and Chu-Guang, Zheng and Shi, Baochang},
year = {2002},
month = {03},
pages = {366-374},
title = {Non-equilibrium extrapolation method for velocity and pressure boundary conditions in the lattice Boltzmann method},
volume = {11},
journal = {Chinese Physics},
doi = {10.1088/1009-1963/11/4/310}
}

@article{Zou1997,
  title = {On Pressure and Velocity Boundary Conditions for the Lattice {{Boltzmann BGK}} Model},
  author = {Zou, Qisu and He, Xiaoyi},
  year = {1997},
  month = jun,
  journal = {Physics of Fluids},
  volume = {9},
  number = {6},
  pages = {1591--1598},
  issn = {1070-6631, 1089-7666},
  doi = {10.1063/1.869307},
  url = {https://pubs.aip.org/pof/article/9/6/1591/260464/On-pressure-and-velocity-boundary-conditions-for},
  urldate = {2023-09-28},
  langid = {english}
}

@article{Ginzburg2003,
  title = {Multireflection boundary conditions for lattice Boltzmann models},
  author = {Ginzburg, Irina and d'Humi\`eres, Dominique},
  journal = {Phys. Rev. E},
  volume = {68},
  issue = {6},
  pages = {066614},
  numpages = {30},
  year = {2003},
  month = {Dec},
  publisher = {American Physical Society},
  doi = {10.1103/PhysRevE.68.066614},
  url = {https://link.aps.org/doi/10.1103/PhysRevE.68.066614}
}

@article{Krastins2020,
author = {Krastins, Ivars and Kao, Andrew and Pericleous, Koulis and Reis, Timothy},
title = {Moment-based boundary conditions for straight on-grid boundaries in three-dimensional lattice Boltzmann simulations},
journal = {International Journal for Numerical Methods in Fluids},
volume = {92},
number = {12},
pages = {1948-1974},
doi = {https://doi.org/10.1002/fld.4856},
url = {https://onlinelibrary.wiley.com/doi/abs/10.1002/fld.4856},
eprint = {https://onlinelibrary.wiley.com/doi/pdf/10.1002/fld.4856},
year = {2020}
}

@article{Eichler2024,
  title = {Investigation of Mesoscopic Boundary Conditions for Lattice {{Boltzmann}} Method in Laminar Flow Problems},
  author = {Eichler, Pavel and Fu{\v c}{\'i}k, Radek and Strachota, Pavel},
  year = {2024},
  month = nov,
  journal = {Computers \& Mathematics with Applications},
  volume = {173},
  pages = {87--101},
  issn = {08981221},
  doi = {10.1016/j.camwa.2024.08.009},
  url = {https://linkinghub.elsevier.com/retrieve/pii/S089812212400350X},
  urldate = {2025-10-08},
  langid = {english}
}

@article{Lehto2017,
author = {Lehto, Erik and Shankar, Varun and Wright, Grady},
year = {2017},
month = {09},
pages = {A2129-A2151},
title = {A Radial Basis Function (RBF) Compact Finite Difference (FD) Scheme for Reaction-Diffusion Equations on Surfaces},
volume = {39},
journal = {SIAM Journal on Scientific Computing},
doi = {10.1137/16M1095457}
}

@article{Shankar2015,
	title = {Augmenting the {{Immersed Boundary Method}} with {{Radial Basis Functions}} ({{RBFs}}) for the {{Modeling}} of {{Platelets}} in {{Hemodynamic Flows}}},
	author = {Shankar, Varun and Wright, Grady B. and Kirby, Robert M. and Fogelson, Aaron L.},
	year = {2015},
	journal = {International Journal for Numerical Methods in Fluids},
	volume = {79},
	number = {10},
	pages = {536--557},
	issn = {10970363},
	doi = {10.1002/fld.4061}
}

@article{Slak2019,
	title = {Adaptive {{Radial Basis Function}}{\textendash} {{Generated Finite Differences Method}} for {{Contact Problems}}},
	author = {Slak, Jure and Kosec, Gregor},
	year = {2019},
	month = aug,
	journal = {International Journal for Numerical Methods in Engineering},
	volume = {119},
	number = {7},
	pages = {661--686},
	issn = {0029-5981, 1097-0207},
	doi = {10.1002/nme.6067}
}

@article{Guo2002,
  title = {An {{Extrapolation Method}} for {{Boundary Conditions}} in {{Lattice Boltzmann Method}}},
  author = {Guo, Zhaoli and Zheng, Chuguang and Shi, Baochang},
  year = {2002},
  journal = {Physics of Fluids},
  volume = {14},
  number = {6},
  pages = {2007--2010},
  issn = {10706631},
  doi = {10.1063/1.1471914}
}

@book{Guo2013,
	title = {Lattice {{Boltzmann}} Method: And Its Applications in Engineering},
	shorttitle = {Lattice {{Boltzmann}} Method},
	author = {Guo, Zhaoli and Shu, Chang},
	year = {2013},
	series = {Advances in Computational Fluid Dynamics},
	number = {vol. 3},
	publisher = {{World Scientific}},
	address = {{Singapore}},
	isbn = {978-981-4508-29-2},
	langid = {english}
}

@article{Flyer2016,
  title = {On the {{Role}} of {{Polynomials}} in {{RBF-FD Approximations}}: {{I}}. {{Interpolation}} and {{Accuracy}}},
  author = {Flyer, Natasha and Fornberg, Bengt and Bayona, Victor and Barnett, Gregory A.},
  year = {2016},
  journal = {Journal of Computational Physics},
  volume = {321},
  pages = {21--38},
  issn = {10902716},
  doi = {10.1016/j.jcp.2016.05.026}
}

@article{He1996,
  title = {Some {{Progress}} in {{Lattice Boltzmann Method}}. {{Part I}}. {{Nonuniform Mesh Grids}}},
  author = {He, Xiaoyi and Luo, Li Shi and Dembo, Micah},
  year = {1996},
  journal = {Journal of Computational Physics},
  volume = {129},
  number = {2},
  pages = {357--363},
  issn = {00219991},
  doi = {10.1006/jcph.1996.0255}
}

@article{He1997c,
  title = {Some Progress in the Lattice {{Boltzmann}} Method: {{Reynolds}} Number Enhancement in Simulations},
  shorttitle = {Some Progress in the Lattice {{Boltzmann}} Method},
  author = {He, Xiaoyi and Luo, Li-Shi and Dembo, Micah},
  year = {1997},
  month = may,
  journal = {Physica A: Statistical Mechanics and its Applications},
  volume = {239},
  number = {1-3},
  pages = {276--285},
  issn = {03784371},
  doi = {10.1016/S0378-4371(96)00486-4},
  url = {https://linkinghub.elsevier.com/retrieve/pii/S0378437196004864},
  urldate = {2024-02-21},
  langid = {english}
}

@article{He1997,
  title = {Lattice {{Boltzmann Method}} on {{Curvilinear Coordinates System}}: {{Flow}} around a {{Circular Cylinder}}},
  author = {He, Xiaoyi and Doolen, Gary},
  year = {1997},
  journal = {Journal of Computational Physics},
  volume = {134},
  number = {2},
  pages = {306--315},
  issn = {00219991},
  doi = {10.1006/jcph.1997.5709}
}

@article{Wu2009,
title = {Implicit velocity correction-based immersed boundary-lattice Boltzmann method and its applications},
journal = {Journal of Computational Physics},
volume = {228},
number = {6},
pages = {1963-1979},
year = {2009},
issn = {0021-9991},
doi = {https://doi.org/10.1016/j.jcp.2008.11.019},
url = {https://www.sciencedirect.com/science/article/pii/S0021999108006116},
author = {J. Wu and C. Shu}
}

@article{Hejranfar2014a,
  title = {Implementation of a High-Order Compact Finite-Difference Lattice {{Boltzmann}} Method in Generalized Curvilinear Coordinates},
  author = {Hejranfar, Kazem and Ezzatneshan, Eslam},
  year = {2014},
  month = jun,
  journal = {Journal of Computational Physics},
  volume = {267},
  pages = {28--49},
  issn = {00219991},
  doi = {10.1016/j.jcp.2014.02.030},
  url = {https://linkinghub.elsevier.com/retrieve/pii/S0021999114001508},
  urldate = {2023-11-20},
  langid = {english}
}

@article{Hejranfar2015,
  title = {Chebyshev Collocation Spectral Lattice {{Boltzmann}} Method for Simulation of Low-Speed Flows},
  author = {Hejranfar, Kazem and Hajihassanpour, Mahya},
  year = {2015},
  month = jan,
  journal = {Physical Review E},
  volume = {91},
  number = {1},
  pages = {013301},
  issn = {1539-3755, 1550-2376},
  doi = {10.1103/PhysRevE.91.013301},
  url = {https://link.aps.org/doi/10.1103/PhysRevE.91.013301},
  urldate = {2023-11-20},
  langid = {english}
}

@article{He1997a,
  title = {Lattice Boltzmann method on a curvilinear coordinate system: Vortex shedding behind a circular cylinder},
  author = {He, Xiaoyi and Doolen, Gary D.},
  journal = {Phys. Rev. E},
  volume = {56},
  issue = {1},
  pages = {434--440},
  numpages = {0},
  year = {1997},
  month = {Jul},
  publisher = {American Physical Society},
  doi = {10.1103/PhysRevE.56.434},
  url = {https://link.aps.org/doi/10.1103/PhysRevE.56.434}
}

@article{Bouzidi2001,
    author = {Bouzidi, M’hamed and Firdaouss, Mouaouia and Lallemand, Pierre},
    title = {Momentum transfer of a Boltzmann-lattice fluid with boundaries},
    journal = {Physics of Fluids},
    volume = {13},
    number = {11},
    pages = {3452-3459},
    year = {2001},
    month = {11},
    issn = {1070-6631},
    doi = {10.1063/1.1399290},
    url = {https://doi.org/10.1063/1.1399290},
    eprint = {https://pubs.aip.org/aip/pof/article-pdf/13/11/3452/19042466/3452_1_online.pdf},
}

@article{Sanjeevi2018,
  title = {Choice of No-Slip Curved Boundary Condition for Lattice {{Boltzmann}} Simulations of High-{{Reynolds-number}} Flows},
  author = {Sanjeevi, Sathish K. P. and Zarghami, Ahad and Padding, Johan T.},
  year = {2018},
  month = apr,
  journal = {Physical Review E},
  volume = {97},
  number = {4},
  pages = {043305},
  issn = {2470-0045, 2470-0053},
  doi = {10.1103/PhysRevE.97.043305},
  url = {https://link.aps.org/doi/10.1103/PhysRevE.97.043305},
  urldate = {2024-06-11},
  langid = {english}
}

@article{Kramer2020,
  title = {Lattice {{Boltzmann}} Simulations on Irregular Grids: {{Introduction}} of the {{NATriuM}} Library},
  shorttitle = {Lattice {{Boltzmann}} Simulations on Irregular Grids},
  author = {Kr{\"a}mer, Andreas and Wilde, Dominik and K{\"u}llmer, Knut and Reith, Dirk and Foysi, Holger and Joppich, Wolfgang},
  year = {2020},
  month = jan,
  journal = {Computers \& Mathematics with Applications},
  volume = {79},
  number = {1},
  pages = {34--54},
  issn = {08981221},
  doi = {10.1016/j.camwa.2018.10.041},
  url = {https://linkinghub.elsevier.com/retrieve/pii/S0898122118306382},
  urldate = {2023-07-05},
  langid = {english}
}

@article{Musavi2016,
  title = {A Mesh-Free Lattice {{Boltzmann}} Solver for Flows in Complex Geometries},
  author = {Musavi, S. Hossein and Ashrafizaadeh, Mahmud},
  year = {2016},
  month = jun,
  journal = {International Journal of Heat and Fluid Flow},
  volume = {59},
  pages = {10--19},
  issn = {0142727X},
  doi = {10.1016/j.ijheatfluidflow.2016.01.006},
  url = {https://linkinghub.elsevier.com/retrieve/pii/S0142727X16300030},
  urldate = {2023-04-24},
  langid = {english}
}

@article{Matin2017,
  title = {Evaluation of the Finite Element Lattice {{Boltzmann}} Method for Binary Fluid Flows},
  author = {Matin, Rastin and Misztal, Marek Krzysztof and {Hern{\'a}ndez-Garc{\'i}a}, Anier and Mathiesen, Joachim},
  year = {2017},
  month = jul,
  journal = {Computers \& Mathematics with Applications},
  volume = {74},
  number = {2},
  pages = {281--291},
  issn = {08981221},
  doi = {10.1016/j.camwa.2017.04.027},
  url = {https://linkinghub.elsevier.com/retrieve/pii/S0898122117302584},
  urldate = {2024-01-02},
  langid = {english}
}

@article{Misztal2015,
  title = {Simulating {{Anomalous Dispersion}} in {{Porous Media Using}} the {{Unstructured Lattice Boltzmann Method}}},
  author = {Misztal, Marek K. and {Hernandez-Garcia}, Anier and Matin, Rastin and M{\"u}ter, Dirk and Jha, Diwaker and S{\o}rensen, Henning O. and Mathiesen, Joachim},
  year = {2015},
  journal = {Frontiers in Physics},
  volume = {3},
  number = {JUL},
  pages = {1--9},
  issn = {2296424X},
  doi = {10.3389/fphy.2015.00050}
}

@article{Ubertini2004,
  title = {Lattice {{Boltzmann}} Schemes without Coordinates},
  author = {Ubertini, S. and Succi, S. and Bella, G.},
  editor = {{Al--Ghoul}, M. and Boon, J. P. and Coveney, P. V.},
  year = {2004},
  month = aug,
  journal = {Philosophical Transactions of the Royal Society of London. Series A: Mathematical, Physical and Engineering Sciences},
  volume = {362},
  number = {1821},
  pages = {1763--1771},
  issn = {1364-503X, 1471-2962},
  doi = {10.1098/rsta.2004.1413},
  url = {https://royalsocietypublishing.org/doi/10.1098/rsta.2004.1413},
  urldate = {2023-10-18},
  langid = {english}
}

@article{Misztal2015a,
  title = {Detailed Analysis of the Lattice {{Boltzmann}} Method on Unstructured Grids},
  author = {Misztal, Marek Krzysztof and {Hernandez-Garcia}, Anier and Matin, Rastin and S{\o}rensen, Henning Osholm and Mathiesen, Joachim},
  year = {2015},
  month = sep,
  journal = {Journal of Computational Physics},
  volume = {297},
  pages = {316--339},
  issn = {00219991},
  doi = {10.1016/j.jcp.2015.05.019},
  url = {https://linkinghub.elsevier.com/retrieve/pii/S0021999115003538},
  urldate = {2023-10-13},
  langid = {english}
}

@article{Maidenberg2025,
    author = {Maidenberg, Amandine R. and Chen, Leitao},
    title = {A Meshless Discrete Boltzmann Solver With the Radial Basis Function Finite Difference Scheme for Fluid Flows With Dirichlet Boundary Conditions},
    journal = {Journal of Fluids Engineering},
    volume = {148},
    number = {2},
    pages = {021501},
    year = {2025},
    month = {10},
    issn = {0098-2202},
    doi = {10.1115/1.4069638},
    url = {https://doi.org/10.1115/1.4069638},
    eprint = {https://asmedigitalcollection.asme.org/fluidsengineering/article-pdf/148/2/021501/7534794/fe-25-1220.pdf},
}

@article{Lin2019,
  title = {A {{Mesh-Free Radial Basis Function}}\textendash{} {{Based Semi-Lagrangian Lattice Boltzmann Method}} for {{Incompressible Flows}}},
  author = {Lin, Xingjian and Wu, Jie and Zhang, Tongwei},
  year = {2019},
  journal = {International Journal for Numerical Methods in Fluids},
  volume = {91},
  number = {4},
  pages = {198--211},
  issn = {10970363},
  doi = {10.1002/fld.4749}
}

@book{Bertin2002,
publisher = {Prentice Hall},
booktitle = {Aerodynamics for engineers},
isbn = {0130646334},
year = {2002},
title = {Aerodynamics for engineers / John J. Bertin},
edition = {4th ed},
language = {eng},
address = {Upper Saddle River, NJ},
author = {Bertin, John J.},
lccn = {2001036583},
}

@article{LATT2006165,
title = {Lattice Boltzmann method with regularized pre-collision distribution functions},
journal = {Mathematics and Computers in Simulation},
volume = {72},
number = {2},
pages = {165-168},
year = {2006},
note = {Discrete Simulation of Fluid Dynamics in Complex Systems},
issn = {0378-4754},
doi = {https://doi.org/10.1016/j.matcom.2006.05.017},
url = {https://www.sciencedirect.com/science/article/pii/S0378475406001583},
author = {Jonas Latt and Bastien Chopard}
}

@Article{Matyka2022,
author={Matyka, Maciej},
title={Pushing Droplet Through a Porous Medium},
journal={Transport in Porous Media},
year={2022},
month={Aug},
day={01},
volume={144},
number={1},
pages={55-68},
issn={1573-1634},
doi={10.1007/s11242-021-01705-z},
url={https://doi.org/10.1007/s11242-021-01705-z}
}

@article{Matyka2013,
  title = {Wall Orientation and Shear Stress in the Lattice {{Boltzmann}} Model},
  author = {Matyka, Maciej and Koza, Zbigniew and Miros{\l}aw, {\L}ukasz},
  year = {2013},
  month = mar,
  journal = {Computers \& Fluids},
  volume = {73},
  pages = {115--123},
  issn = {00457930},
  doi = {10.1016/j.compfluid.2012.12.018},
  url = {https://linkinghub.elsevier.com/retrieve/pii/S0045793013000078},
  urldate = {2023-05-30},
  langid = {english}
}

@article{Braesel2024,
    author = {Bräsel, Berinike and Geiger, Matthias and Linkhorst, John and Wessling, Matthias},
    title = {Transport and clogging dynamics of flexible rods in pore constrictions},
    journal = {Soft Matter},
    volume = {20},
    number = {34},
    pages = {6767-6778},
    year = {2024},
    month = {09},
    issn = {1744-683X},
    doi = {10.1039/d4sm00734d},
    url = {https://doi.org/10.1039/d4sm00734d},
    eprint = {https://pubs.rsc.org/sm/article-pdf/20/34/6767/10078124/d4sm00734d.pdf},
}

@ARTICLE{Azizi2023,
AUTHOR={Azizi, Pedram  and Drobek, Christoph  and Budday, Silvia  and Seitz, Hermann },           
TITLE={Simulating the mechanical stimulation of cells on a porous hydrogel scaffold using an FSI model to predict cell differentiation},          
JOURNAL={Frontiers in Bioengineering and Biotechnology},          
VOLUME={Volume 11 - 2023},  
YEAR={2023},  
URL={https://www.frontiersin.org/journals/bioengineering-and-biotechnology/articles/10.3389/fbioe.2023.1249867},  
DOI={10.3389/fbioe.2023.1249867},  
ISSN={2296-4185}}

\end{document}


\centering \Huge Supplementary materials for the work {\itshape On challenges faced in modeling unbounded flows with meshless Lattice Boltzmann Method}

\vspace{0.75cm}
\centering
\large Dawid Strzelczyk$^{*,1,2}$, Pavel Eichler$^{3}$, Maciej Matyka$^{1,2}$ \normalsize

\vspace{0.25cm}

\noindent\textit{$^1$Faculty of Physics and Astronomy, University of Wrocław, pl. Maxa Borna 9, 50-204 Wrocław, Poland}

\noindent\textit{$^2$Jo\v{z}ef Stefan Institute, Scientific Computing Laboratory, Jamova cesta 39, 1000 Ljubljana, Slovenia}

\noindent\textit{$^3$ Department of Software Engineering, Faculty of Nuclear Sciences and Physical Engineering, Czech Technical University in Prague, Trojanova 13, Prague, 12000, Czech Republic}

\vspace{0.25cm}
\RaggedRight
$^*$ \texttt{dawid.strzelczyk@ijs.si}
\linebreak

\justifying

\section{Analytical derivation of the convergence rate for non-equilibrium bounceback in off-gird setup}\label{app:nee_convergence_analytical}

Following the asymptotic analysis presented in~\cite{Guo2002}, for a boundary node O and its adjacent node B from which $f^\text{neq}$ and $\rho$ is taken in NEE no-slip, one can assess that the non-equilibrium part of the VDF in the boundary node is approximated with order of accuracy
\begin{equation}
	f^\text{neq}_k(O) =
	\epsilon(f^{(1)}_k(O)) + \mathcal{O}(\epsilon^2) =
	\epsilon(f^{(1)}_k(B) + \mathcal{O}(h)) + \mathcal{O}(\epsilon^2) =
	f^\text{neq}_k(B) + \mathcal{O}(\epsilon h) + \mathcal{O}(\epsilon^2)
\end{equation}
which is because the space discretization parameter $h$ and the Chapman-Enskog expansion parameter $\epsilon$ can be varied independently in OLBM, so the two asymptotic terms after the last equation sign need to be distinguished (in contrast to standard LBM where $h \propto \epsilon$ always). At the same time, with the current approximation of the density in the boundary node, the leading order term of the approximation error has the form of
\begin{equation}
	\rho(O) - \rho(B) = (\bsym{n} \cdot \nabla \rho)h = \mathcal{O}(hM^2).
\end{equation}
where $\bsym{n}$ is the local boundary normal vector. As at the boundary we approximate the equilibrium part of VDF $f_k^0(O)$ by the equilibrium parametrized with zero velocity and the density taken from the neighbor B ($\boldsymbol{V}=\boldsymbol{0}$ and $\rho(B)$, respectively), its error scales linearly with space discretization parameter and quadraticaly with Mach number
\begin{equation}
	f^0_k(O) = \rho(O)s_k(\bsym{0}) =
	(\rho(B) + \mathcal{O}(hM^2)) s_k(\bsym{0}) =
	\rho(B) s_k(\bsym{0}) + \mathcal{O}(hM^2).
\end{equation}
In the present study we scale the space and velocity discretization proportionally, i.e., $h \propto \delta x \propto \epsilon$. Thus, at constant $Re$, the velocity (and so, the Mach number) scales linearly with $\delta x$ and $h$, i.e., $Ma \propto \delta x \propto h$. What follows is that the total error of the approximation of VDF in the boundary nodes scales as $\mathcal{O}(h^2) = \mathcal{O}(\delta x^2)$
\begin{equation}
	f_k(O) = f^\text{neq}(O) + f^0_k(O) =
	f^\text{neq}_k(B) + \rho(B) s_k(\bsym{0}) + \mathcal{O}(\epsilon h) + \mathcal{O}(\epsilon^2) + \mathcal{O}(hM^2),
\end{equation}
with the following sources of each asymptotic term
\begin{itemize}
	\item $\mathcal{O}(\epsilon h) + \mathcal{O}(\epsilon^2)$ -- approximation of non-equilibrium part in space and velocities,
	\item $\mathcal{O}(hM^2)$ -- approximation of equilibrium in space.
\end{itemize}
Considering the scenarios where only $h$ or $\delta x$ is scaled, the convergence will proceed with the corresponding order as long as the error associated with either space discretization or compressibility, respectively, dominates the solution~\cite{Strzelczyk2024,Kramer2020}.


\section{Convergence of the solution at vanishing Reynolds number}\label{app:convergence}

In this Supplementary Material we present the convergence of the solution to the analytical value obtained with the studied no-slip implementations on regular and hybrid discretizations, for various relaxation times. For each type of discretization we consider three levels of refinement: $N_{\theta,0}=45$, $N_{\theta,1}=90$, and $N_{\theta,2}=180$. The number of nodes along the radial coordinate and streaming distance is scaled proportionally, which gives the following sequence of the values of the latter: $\delta x_0 = 0.06$, $\delta x_1 = 0.03$, and $\delta x_2 = 0.015$. For the hybrid discretizations, the target internodal distance in the scattered part of the domain was scaled proportionally to $N_{\theta}$. We set the Reynolds number to $Re=0.01$. We perform the study for the range of relaxation times $\tau \in [0.51625;1.0]$. Since on the cylinder the true velocity is zero, we use the normalized length of the error vector $\langle L_2 \rangle$, Eq.\meanLTwoEq in the main text, as the measure of the velocity field error. We show the results only for the setups that did not diverge.

\subsection{Velocity}\label{app:velocity_convergence}

\begin{figure}[ht!]
	\includegraphics[width=\linewidth]{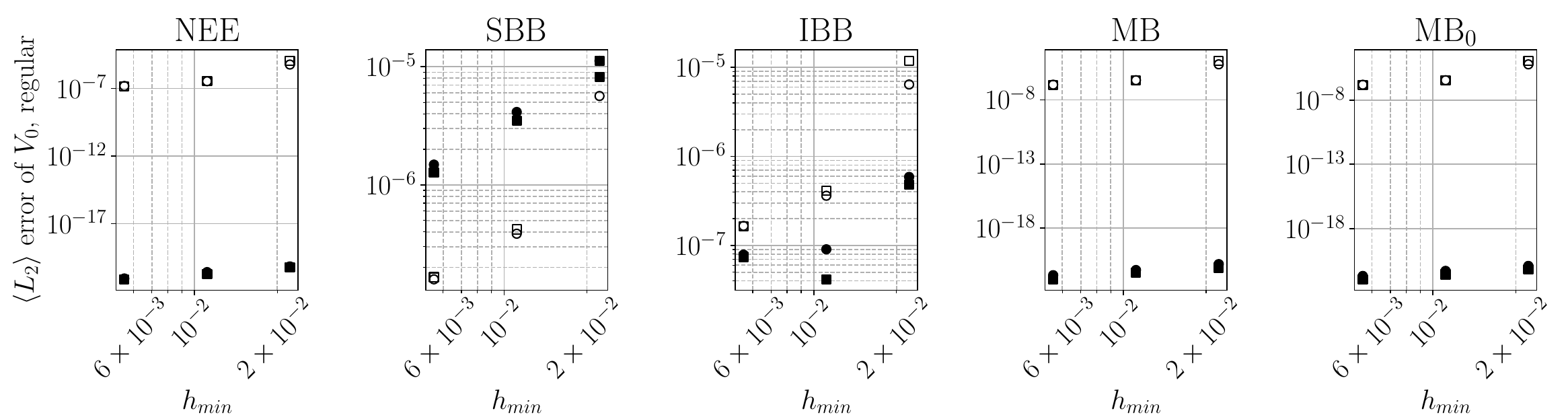}
    \includegraphics[width=\linewidth]{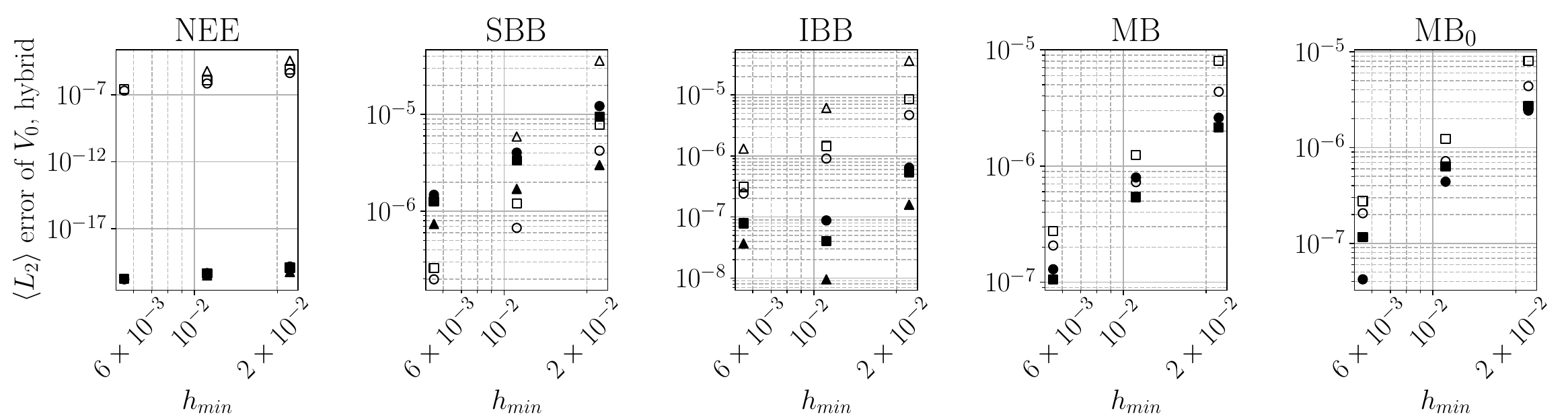}
	\caption{Error scaling of the $x_0$-component of the velocity obtained at regular ({\itshape top row}) and hybrid ({\itshape bottom row}) discretizations. The symbols $\bigcirc, \square, \triangle$ denote $\tau = 1,0.75,0.53125$, respectively. {\itshape Empty symbols}: whole domain, {\itshape filled symbols}: only cylinder's surface.}
	\label{fig:v0_convergence}
\end{figure}

Fig.~\ref{fig:v0_convergence} shows the $\langle L_2 \rangle$ errors of the $V_0$ velocity component. On regular discretizations, the values of the error calculated on the cylinder's surface are close to numerical zero for the implementations imposing the zero-velocity condition explicitly, namely NEE and both MB's. SBB exhibits a stable convergence on the cylinder's surface while IBB suffers from non-monotonic error behavior for $\tau=0.75$. For the error calculated in the whole domain, its values are confined within range $[10^{-8};\>10^{-5}]$ for all no-slips. The convergence is visible with slightly decreasing order at finer discretizations. On hybrid discretizations, the numerical zero is not achieved by moment-based no-slips -- most likely because of the points on the perimeter of the cylinder where the velocity is on the order of $10^{-6}$ (cf. Fig.~\velocityProfileHybridFig in the main text). Apart from that, the convergence plots look similarly to those obtained on regular discretizations.

\subsection{Pressure coefficient}\label{app:cp_convergence}

\begin{figure}[ht!]
	\includegraphics[width=\linewidth]{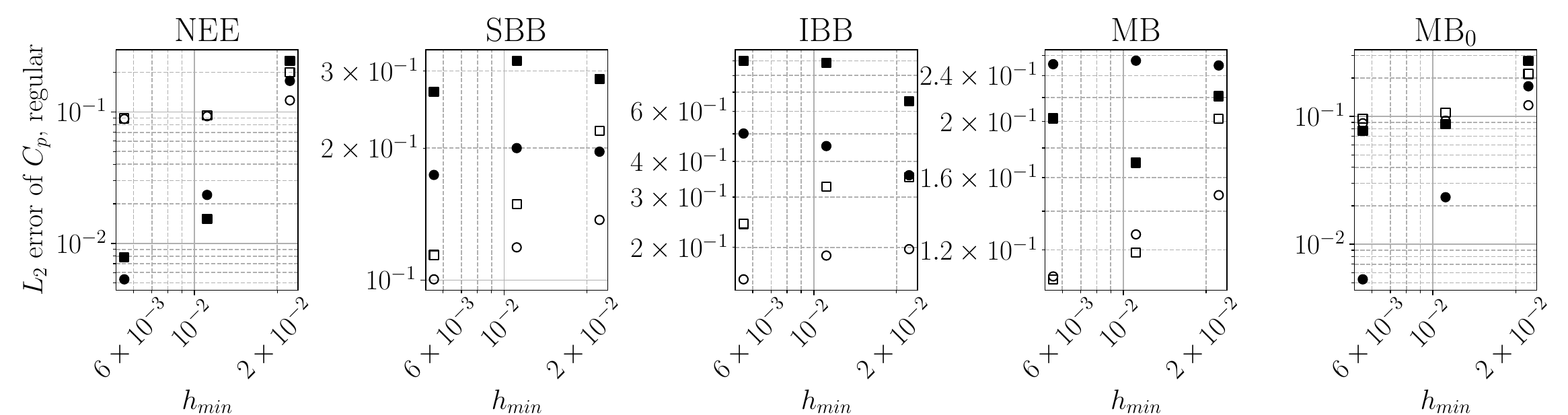}
    \includegraphics[width=\linewidth]{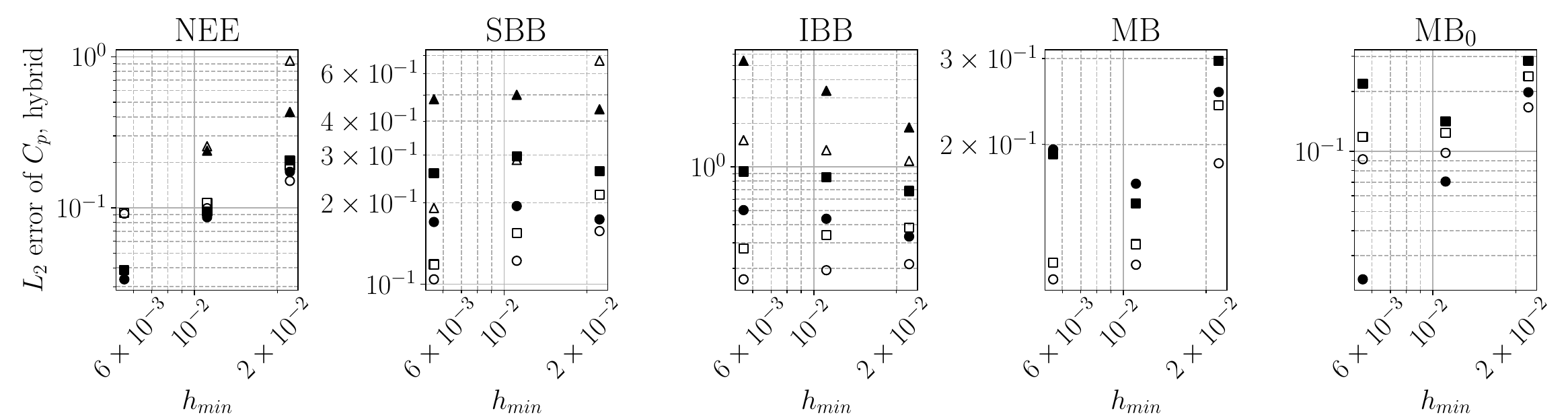}
	\caption{Error scaling of the pressure coefficient obtained at regular ({\itshape top row}) and hybrid ({\itshape bottom row}) discretizations. The symbols $\bigcirc, \square, \triangle$ denote $\tau = 1,0.75,0.53125$, respectively. {\itshape Empty symbols}: whole domain, {\itshape filled symbols}: only cylinder's surface.}
	\label{fig:cp_convergence}
\end{figure}

Fig.~\ref{fig:cp_convergence} shows the $L_2$ errors of the pressure coefficient field. On the cylinder's surface, its convergence reflects the problems with discontinuities discussed in Section~\sectionCp of the main text. On cylinder's surface with regular discretization, only NEE and MB$_0$ managed to keep the decreasing trend of the error, with MB$_0$'s convergence rate visibly decreasing for $\tau=0.75$. On the cylinder's surface with hybrid discretization, MB$_0$ at $\tau=0.75$ gave non-monotonic error behavior, while NEE, as the only out of all five no-slips, gave a stable, decreasing error for all refinements. The convergence of the solution in the whole domain halts at the value approximately 0.1, most probably caused by the standing waves entering the domain from the outer boundary (cf. Fig.~\hydroFieldsFig in the main text). 

\subsection{Analysis of the orders of convergence of the solutions}\label{app:convergence_order_study}

The mean orders of convergence $\langle p \rangle$ for the setups and quantities considered in Figs.~\ref{fig:v0_convergence} and \ref{fig:cp_convergence} are shown in Tables~\ref{tab:fneq_convergence}--\ref{tab:moments_f0_convergence}. The subscripts at the $\langle p \rangle$ symbols denote which quantity is considered (the velocity $V_0$ or the pressure coefficient $C_p$) and over what set the error was calculated (the whole domain -- $\Omega$ or the cylinder's surface -- $\partial \Omega$).

For NEE (Table~\ref{tab:fneq_convergence}), the mean orders of convergence for the velocity calculated from the whole domain are not far from the theoretical ones, derived in Section~\ref{app:nee_convergence_analytical} of the Supplementary Materials. With space and velocity discretizations refined proportionally, as in our study, the order of convergence should be equal to 2 or 3, depending on whether the error of the approximation of equilibrium or non-equilibrium part of VDF dominates in the solution. This applies to regular and hybrid discretizations, with the former having visibly higher orders of convergence than the latter, possibly due to larger interpolation errors originating from the scattered part of the discretization. For the error of the velocity calculated on the cylinder's surface, a stable convergence of the order about 1.4--1.5 is visible for all cases. The fact that the order is lower than in the case of the error taken from the whole domain might be explained by the limited floating-point precision. To do this, let us write the well-conditioned VDF as
\begin{equation}\label{eq:vdf_Ov}
    f_k-\omega_k =
    \omega_k(1+\delta\rho)(1+\mathcal{O}(V)) + f^\text{neq}_k - \omega_k =
    \omega_k\delta\rho + \omega_k(1+\delta\rho)\mathcal{O}(V) + f^\text{neq}_k,
\end{equation}
which describes the VDF in the bulk of the flow, used in the interpolated streaming within the boundary nodes' stencils. However, the interpolated streaming at the near-boundary nodes uses also the values of the well-conditioned VDF directly from the wall, where the velocity is zero, and so the well-conditioned VDF from Eq.~\eqref{eq:vdf_Ov} becomes
\begin{equation}\label{eq:vdf_Ov_wall}
    (f_k-\omega_k)|_{V=0} = \omega_k\delta\rho + f^\text{neq}_k.
\end{equation}
Now, let us compare the corresponding terms in the right-hand sides of Eqs.~\eqref{eq:vdf_Ov} and \eqref{eq:vdf_Ov_wall}. First, we note that the density variations $\delta \rho$ are constant along the wall normal due to the imposed macroscopic boundary conditions. Second, due to Eq.\fneqHydroApproxEq in the main text, we can assume that in the vicinity of the wall, the non-equilibrium part of VDF will be the largest exactly at the wall, since the shear stresses are the largest at the wall. Third, upon moving way from the wall, the fluid velocity increases rapidly from zero to a finite value. Thus, within the near-wall stencils, the nodes lying in the bulk will have the values of well-conditioned VDF orders of magnitude larger than those at the wall, due to the domination of the term $\omega_k(1+\delta\rho)\mathcal{O}(V)$. What follows is that during the computation of the linear combination of VDF values within near-wall stencils (Eq.~\rbffdLinearCombination in the main text), a loss of significant digits may occur in the near-wall nodes data. Upon the copying of the values of $\rho$ and $f_k^\text{neq}$ to the wall nodes, the round-off errors may be propagated back to the wall nodes, thus reducing the order of convergence of the velocity at the wall, in spite it being practically zero. As for the pressure coefficient, the orders of convergence calculated on the cylinder's surface are around the theoretical ones ($\sim 2.5$) for the regular discretization, and significantly lower on hybrid discretizations, similarly to the velocity described earlier. The fact that on hybrid setups the order of convergence of the pressure on the cylinder is lower than the theoretical (implied by the orders of the NEE and meshfree approximation), even though the values of error are far from the numerical precision, may be explained by the propagation of the errors from the standing waves at the outer boundary -- a phenomenon not included in the theoretical derivations of Section~\ref{app:nee_convergence_analytical} of the Supplementary Materials.

For SBB (Table~\ref{tab:sbb_convergence}), the theoretical order of convergence of the velocity on straight, lattice-aligned walls, should be equal to 1. On the other hand, as stated above, the order of convergence, theoretically, should not exceed the order of the interpolation scheme, equal to 2. In our SBB results, the mean order of convergence of velocity on the cylinder is confined between approximately 1 and 1.5, for both types of discretizations, which is compliant with those expectations. For the error taken over the whole domain, the velocity converges with orders above 2. This is larger than the order of the used meshfree approximation and suggests that the pressure waves at the outer boundary can prevent the solution from entering the asymptotic convergence regime yet. The pressure coefficient exhibits a very weak convergence for all studied cases of SBB -- both on the cylinder, where the errors are not monotonic with the refinement, and in the whole domain.

The velocity obtained with IBB (Table~\ref{tab:ibb_convergence}) gives similar mean orders of convergence as in the SBB setups, but its errors calculated on the cylinder are not monotonic with the refinement. The pressure exhibits similar orders of convergence as in the corresponding setups equipped with SBB no-slip. The divergence of the pressure coefficient on the cylinder is visible for all IBB setups, which further supports the hypothesis that the introduction of interpolation in IBB makes the boundary-condition less table and accurate in this benchmark case.

The orders of convergence of the velocity on the cylinder obtained with MB and MB$_0$ are similar to those obtained with NEE. The same is true for the error of the velocity calculated over the whole domain on regular discretizations. However, on hybrid node sets, the moment-based no-slips exhibit significantly higher orders of convergence than NEE. This may be caused by a different mechanism of the exchange of information between the wall nodes and their in-bulk neighbors than in NEE. The orders of convergence of the pressure for MB results are well below 1. For MB$_0$, the on-cylinder convergence at $\tau=1$ proceeds with the order above 1.5, which corresponds to the $C_P$ field compliant with the theory in Figs.~\pressureProfilesFigs in the main text. For higher relaxation rates, the orders fall below 1.

\begin{table}[!ht]
	\centering
	\begin{tabular}{ccc}
		\begin{tabular}{lrrrr}
			\multicolumn{5}{c}{NEE, regular} \\
			\hline
			\hline
			$\tau$ & $\langle p \rangle_{V_0,\Omega}$ & $\langle p \rangle_{V_0,\partial\Omega}$ & $\langle p \rangle_{C_p,\Omega}$ & $\langle p \rangle_{C_p,\partial\Omega}$ \\
			\hline
            1.0 & $2.635$ & $1.481$ & $0.235$ & $2.509$ \\
            0.75 & $3.135$ & $1.443$ & $0.579$ & $2.483$ \\
            0.53125 & -- & -- & -- & -- \\
		\end{tabular}
		& &
		\begin{tabular}{lrrrr}
			\multicolumn{5}{c}{NEE, hybrid} \\
			\hline
			\hline
			$\tau$ & $\langle p \rangle_{V_0,\Omega}$ & $\langle p \rangle_{V_0,\partial\Omega}$ & $\langle p \rangle_{C_p,\Omega}$ & $\langle p \rangle_{C_p,\partial\Omega}$ \\
			\hline
            1.0 & $2.200$ & $1.586$ & $0.360$ & $1.184$ \\
            0.75 & $2.465$ & $1.396$ & $0.520$ & $1.204$ \\
            0.53125 & -- & -- & -- & -- \\
		\end{tabular}
	\end{tabular}
	\caption{Assessed orders of convergence of the $x_0$-component of velocity, $V_0$, pressure coefficient $C_p$ for NEE no-slip on regular and hybrid discretization with various $\tau$'s. }
	\label{tab:fneq_convergence}
\end{table}

\begin{table}[!ht]
	\centering
	\begin{tabular}{ccc}
		\begin{tabular}{lrrrr}
			\multicolumn{5}{c}{SBB, regular} \\
			\hline
			\hline
			$\tau$ & $\langle p \rangle_{V_0,\Omega}$ & $\langle p \rangle_{V_0,\partial\Omega}$ & $\langle p \rangle_{C_p,\Omega}$ & $\langle p \rangle_{C_p,\partial\Omega}$ \\
			\hline
			1.0 & $2.579$ & $1.459$ & $0.225$ & $0.088$ \\
            0.75 & $3.045$ & $1.346$ & $0.472$ & $0.049$ \\
            0.53125 & -- & -- & -- & -- \\
		\end{tabular}
		& &
		\begin{tabular}{lrrrr}
			\multicolumn{5}{c}{SBB, hybrid} \\
			\hline
			\hline
			$\tau$ & $\langle p \rangle_{V_0,\Omega}$ & $\langle p \rangle_{V_0,\partial\Omega}$ & $\langle p \rangle_{C_p,\Omega}$ & $\langle p \rangle_{C_p,\partial\Omega}$ \\
			\hline
            1.0 & $2.208$ & $1.523$ & $0.296$ & $0.016$ \\
            0.75 & $2.461$ & $1.449$ & $0.429$ & $0.012$ \\
            0.53125 & $2.351$ & $1.011$ & $0.904$ & $-0.062$ \\
		\end{tabular}
	\end{tabular}
	\caption{Assessed orders of convergence of the $x_0$-component of velocity, $V_0$, pressure coefficient $C_p$ for SBB no-slip on regular and hybrid discretizations with various $\tau$'s. }
	\label{tab:sbb_convergence}
\end{table}

\begin{table}[!ht]
	\centering
	\begin{tabular}{ccc}
		\begin{tabular}{lrrrr}
			\multicolumn{5}{c}{IBB, regular} \\
			\hline
			\hline
			$\tau$ & $\langle p \rangle_{V_0,\Omega}$ & $\langle p \rangle_{V_0,\partial\Omega}$ & $\langle p \rangle_{C_p,\Omega}$ & $\langle p \rangle_{C_p,\partial\Omega}$ \\
			\hline
            1.0 & $2.640$ & $1.450$ & $0.176$ & $-0.243$ \\
            0.75 & $3.094$ & $1.349$ & $0.272$ & $-0.235$ \\
            0.53125 & -- & -- & -- & -- \\
		\end{tabular}
		& &
		\begin{tabular}{lrrrr}
			\multicolumn{5}{c}{IBB, hybrid} \\
			\hline
			\hline
			$\tau$ & $\langle p \rangle_{V_0,\Omega}$ & $\langle p \rangle_{V_0,\partial\Omega}$ & $\langle p \rangle_{C_p,\Omega}$ & $\langle p \rangle_{C_p,\partial\Omega}$ \\
			\hline
            1.0 & $2.135$ & $1.502$ & $0.174$ & $-0.302$ \\
            0.75 & $2.388$ & $1.394$ & $0.241$ & $-0.225$ \\
            0.53125 & $2.395$ & $1.055$ & $-0.237$ & $-0.765$ \\
		\end{tabular}
	\end{tabular}
	\caption{Assessed orders of convergence of the $x_0$-component of velocity, $V_0$, pressure coefficient $C_p$ for IBB no-slip on regular and hybrid discretizations with various $\tau$'s. }
	\label{tab:ibb_convergence}
\end{table}

\begin{table}[!ht]
	\centering
	\begin{tabular}{ccc}
		\begin{tabular}{lrrrr}
			\multicolumn{5}{c}{MB, regular} \\
			\hline
			\hline
			$\tau$ & $\langle p \rangle_{V_0,\Omega}$ & $\langle p \rangle_{V_0,\partial\Omega}$ & $\langle p \rangle_{C_p,\Omega}$ & $\langle p \rangle_{C_p,\partial\Omega}$ \\
			\hline
            1.0 & $2.639$ & $1.434$ & $0.233$ & $-0.004$ \\
            0.75 & $3.145$ & $1.506$ & $0.461$ & $0.063$ \\
            0.53125 & -- & -- & -- & -- \\
		\end{tabular}
		& &
		\begin{tabular}{lrrrr}
			\multicolumn{5}{c}{MB, hybrid} \\
			\hline
			\hline
			$\tau$ & $\langle p \rangle_{V_0,\Omega}$ & $\langle p \rangle_{V_0,\partial\Omega}$ & $\langle p \rangle_{C_p,\Omega}$ & $\langle p \rangle_{C_p,\partial\Omega}$ \\
			\hline
            1.0 & $2.199$ & $2.166$ & $0.395$ & $0.195$ \\
            0.75 & $2.434$ & $2.172$ & $0.536$ & $0.318$ \\
            0.53125 & -- & -- & -- & -- \\
		\end{tabular}
	\end{tabular}
	\caption{Assessed orders of convergence of the $x_0$-component of velocity, $V_0$, pressure coefficient $C_p$ for MB no-slip on regular and hybrid discretizations with various $\tau$'s.}
	\label{tab:moments_convergence}
\end{table}

\begin{table}[!ht]
	\centering
	\begin{tabular}{ccc}
		\begin{tabular}{lrrrr}
			\multicolumn{5}{c}{MB$_0$, regular} \\
			\hline
			\hline
			$\tau$ & $\langle p \rangle_{V_0,\Omega}$ & $\langle p \rangle_{V_0,\partial\Omega}$ & $\langle p \rangle_{C_p,\Omega}$ & $\langle p \rangle_{C_p,\partial\Omega}$ \\
			\hline
            1.0 & $2.635$ & $1.297$ & $0.235$ & $2.509$ \\
            0.75 & $3.143$ & $1.259$ & $0.581$ & $0.911$ \\
            0.53125 & -- & -- & -- & -- \\
		\end{tabular}
		& &
		\begin{tabular}{lrrrr}
			\multicolumn{5}{c}{MB$_0$, hybrid} \\
			\hline
			\hline
			$\tau$ & $\langle p \rangle_{V_0,\Omega}$ & $\langle p \rangle_{V_0,\partial\Omega}$ & $\langle p \rangle_{C_p,\Omega}$ & $\langle p \rangle_{C_p,\partial\Omega}$ \\
			\hline
            1.0 & $2.204$ & $2.926$ & $0.432$ & $1.559$ \\
            0.75 & $2.435$ & $2.280$ & $0.503$ & $0.194$ \\
            0.53125 & -- & -- & -- & -- \\
		\end{tabular}
	\end{tabular}
	\caption{Assessed orders of convergence of the $x_0$-component of velocity, $V_0$, pressure coefficient $C_p$ for MB$_0$ no-slip on regular and hybrid discretizations with various $\tau$'s.}
	\label{tab:moments_f0_convergence}
\end{table}


\bibliography{sample}